\documentclass[twocolumn,twocolappendix,usenatbib]{aastex631}

\graphicspath{{./}}
\usepackage{float}

\newcommand{\dphot}{\texttt{DOLPHOT}}
\newcommand{\cch}{Cnr-CMi-Hyd}
\newcommand{\W}{\ensuremath{\lambda}} 

\shorttitle{XMP}
\shortauthors{Breneman}

\usepackage{hyperref}
\AtBeginDocument{}%
\AtBeginDocument{}%
\AtBeginDocument{}%

\begin{document}

\title{
Leonessa: An Extremely Metal-poor Galaxy Undergoing Secular Chemical Evolution\footnote{Based on observations made with the NASA/ESA Hubble Space Telescope: Program HST-GO-16048; Green Bank Observatory, Robert C. Byrd Green Bank Telescope: Program GBT20B-468; and McDonald Observatory Hobby-Eberly Telescope.}}
\author[0009-0005-8590-0829]{Joseph A. Breneman}
\affiliation{Rutgers University, Department of Physics and Astronomy, 136 Frelinghuysen Road, Piscataway, NJ 08854, USA}
\author[0000-0001-5538-2614]{Kristen B. W. McQuinn}
\affiliation{Rutgers University, Department of Physics and Astronomy, 136 Frelinghuysen Road, Piscataway, NJ 08854, USA}
\affiliation{Space Telescope Science Institute, 3700 San Martin Drive, Baltimore, MD 21218, USA.}
\author[0000-0002-8815-536X]{Alexander Menchaca}
\affiliation{Department of Astronomy, The University of Texas at Austin, 2515 Speedway, Stop C1400, Austin, TX 78712, USA}
\author[0000-0002-4153-053X]{Danielle A. Berg}
\affiliation{Department of Astronomy, The University of Texas at Austin, 2515 Speedway, Stop C1400, Austin, TX 78712, USA}
\author[0000-0003-4122-7749]{O. Grace Telford}
\affiliation{Department of Astrophysical Sciences, Princeton University, 4 Ivy Lane, Princeton, NJ
08544, USA}
\affiliation{The Observatories of the Carnegie Institution for Science, 813 Santa Barbara Street,
Pasadena, CA 91101, USA}
\affiliation{Rutgers University, Department of Physics and Astronomy, 136 Frelinghuysen Road, Piscataway, NJ 08854, USA}
\author[0000-0002-8092-2077]{Max J. B. Newman}
\affiliation{Rutgers University, Department of Physics and Astronomy, 136 Frelinghuysen Road, Piscataway, NJ 08854, USA}
\author[0000-0001-8416-4093]{Andrew Dolphin}
\affiliation{Raytheon, 1151 E. Hermans Road, Tucson, AZ 85756, USA}
\affiliation{Steward Observatory, University of Arizona, 933 North Cherry Avenue, Tucson, AZ 85721, USA}
\author[0000-0003-2307-0629]{Gregory R. Zeimann}
\affiliation{Hobby Eberly Telescope, University of Texas, Austin, Austin, TX, 78712, USA}

\begin{abstract}
Extremely metal-poor (XMP) galaxies are systems with gas-phase oxygen abundances below $\sim$5\% Solar metallicity (12+log(O/H) $ \le 7.35$). These galaxies populate the metal-poor end of the mass-metallicity and luminosity-metallicity relations (MZR and LZR, respectively). Recent studies have found XMP galaxies in the nearby Universe to be outliers on the LZR, where they show enhanced luminosities relative to other galaxies of similar gas-phase oxygen abundance. Here, we present a study of the recently discovered XMP galaxy Leonessa and characterize the system’s properties using new imaging from the Hubble Space Telescope and spectra from the Green Bank Telescope and Hobby-Eberly Telescope. We use these observations to measure a tip of the red giant branch (TRGB) distance ($15.86 \pm 0.78$ Mpc) to Leonessa, the \ion{H}{1} gas mass, the gas-phase oxygen abundance, and N/O ratio. We find Leonessa is an isolated, gas rich (gas fraction $\mu = 0.69$), low-mass (log(M$_\star$/M$_\odot) = 6.12 \pm 0.08$), XMP (12+log(O/H) $= 7.32 \pm 0.04$), star-forming galaxy at a distance of $15.86 \pm 0.78$ Mpc. Our measurements show that Leonessa agrees with the MZR, but disagrees with the LZR; we conclude the LZR offset is due to recent star formation enhancing the system's luminosity. To investigate possible chemical evolution pathways for nearby XMP galaxies we also compile a comparison sample of 150 dwarf galaxies (53 XMP systems) taken from the literature with gas-phase metallicity measurements based on the direct method. We find evidence for an anti-correlation between gas-phase oxygen abundance and \mbox{\ion{H}{1}} gas-to-stellar mass ratios. We posit Leonessa is undergoing a chemical evolution pathway typical of field dwarf galaxies.
\end{abstract}


\section{Introduction}
\label{sec:intro}
Extremely metal poor (XMP) galaxies are defined by low gas-phase oxygen abundances (12+log(O/H) $\le 7.35$, or $Z_{\rm gas} \lesssim 0.05 \, Z_{\odot}$)\footnote{We adopt the value of 12+log(O/H) $ \le 7.35$ here, instead of a percentage of solar abundance, to define the XMP regime in accordance with \citet{mcquinn2020}} in their interstellar medium (ISM). Despite a relatively narrow range in oxygen abundance, nearby XMP galaxies exhibit a large range in luminosity ($-8$ to $-19$ mag in $g$-band) and stellar mass ($5 <$ log(M$_\star$/M$_{\odot}$) $< 9$); \cite[e.g.,][]{brown2008,papaderos2008,morales-luis2011,izotov2012,guseva2015,sanchez2016,hsyu2018,izotov2019,senchyna2019,kojima2020,miller2023,nishigaki2023}. Given their large luminosity and mass range at such low metallicity, nearby XMP galaxies provide intriguing laboratories for exploring chemical enrichment in the extremely low metallicity regime of galaxy formation and evolution that is scarcely populated due to observational constraints.

Empirical evidence from representative low-mass field galaxies\footnote{A field galaxy is defined here as a low-mass system that does not reside in a cluster, and while some are found to be interacting with other galaxies, they are not satellite systems or part of galaxy groups.} in the local universe has shown a linear relationship between the gas-phase oxygen abundance and galaxy luminosity, known as the luminosity-metallicity relation \cite[LZR;][]{skillman1989,lee2006a,berg2012}. Similarly, a more intrinsic relation is observed between a system's gas-phase oxygen abundance and the log of its stellar mass, referred to as the mass-metallicity relation \cite[MZR;][]{tremonti2004,ekta2010,berg2012}. 

The LZR exhibits a tight correlation between luminosity and gas-phase metallicity, spanning several orders of magnitude and approximately $1.5$ dex in 12+log(O/H) \citep{berg2012}. The MZR also displays a tight relationship between gas-phase metallicity and log of the stellar mass, ranging $\sim 3.5$ dex in log(M$_\star$/M$_{\odot}$) and $\sim 1$ dex in 12+log(O/H) \citep{berg2012}. Typical low-mass dwarf irregular (dIrr) field galaxies that agree with the LZR and MZR are suspected of undergoing a secular chemical evolution, marked by bursty and possibly inefficient star formation, and dominated by stellar feedback capable of expelling a majority of the metals synthesized in their stellar populations. However, at lower metallicity (12+log(O/H) $\lesssim 7.6$), the tight correlation in these scaling relations shows an increasing scatter, which continues down into the XMP regime \cite[e.g.,][]{mcquinn2020}. 

Indeed, most XMP galaxies appear as significant outliers from the LZR and MZR. The scatter is most significant on the LZR plane where almost all XMP galaxies disagree with the established trend. Several external and internal mechanisms are suspected to be responsible for outliers on the LZR and MZR planes, including: interactions with other systems, dilution through accretion of metal-poor gas, and stellar feedback \cite[e.g.,][]{strickland2009,ekta2010,filho2013,mcquinn2015,sanchez2016,chisholm2018,hsyu2017,mcquinn2019,mcquinn2020}.

A specific trait that can help characterize a galaxy's (past and present) star-forming capability is its gas content. One of the most remarkable characteristics of the majority of XMP outliers on the LZR is the contrast in their gas content relative to field dIrr galaxies. While a few XMP systems have gas-mass fractions ($\mu \equiv \, $M$_{\rm gas}/($M$_{\rm gas} + $M$_\star)$, where $\rm M_{gas} = 1.4 \times M_{HI}$ to account for the mass of helium) similar to average field dIrr galaxies \cite[$\mu \sim 0.60$;][equivalent to a gas-to-stellar mass ratio of M$_{\rm HI}/$M$_\star \gtrsim 1.0$]{geha2006}, the vast majority of XMP systems are extremely gas-rich, with $\mu \gtrsim 0.9$ (M$_{\rm HI}/$M$_\star \ge 10.0$). \citet{thuan2016} found that the mean gas-mass fractions for their XMP sample was $0.90 \pm 0.15$. Previous studies have also found XMP systems to be extremely gas-rich, with nearly metal-free (pristine) \mbox{\ion{H}{1}} components \citep{filho2013,filho2015}. Pristine gas envelopes could also be indicative of a recent cold mode accretion capable of driving gas-phase oxygen abundance values down through dilution \citep{keres2005,brooks2009}. The large gas reservoirs in XMP systems serve as the fuel for driving current and future starbursts that can significantly enhance galactic luminosity.

The extremely gas-rich characteristic of XMP systems in the nearby universe often correlate with a high surface brightness. Well-studied XMP systems like DDO 68/B, IZw18A/C, and SBS 0335-052W are high surface brightness systems and are representative of most XMP galaxies \citep{berg2012,izotov2019,izotov2018b}. They are extremely gas-rich (M$_{\rm HI}$/M$_\star \ge 10.0$) with extended \mbox{\ion{H}{1}} envelopes that are thought to be a possible source of dilution for gas-phase oxygen abundances \cite[e.g.,][]{ekta2008,ekta2010,filho2013}. The current intense star formation (SF) they harbor, likely triggered by a recent interaction or ongoing merger, can enhance their galactic luminosity by several magnitudes \citep{izotov2018b}. The enhanced luminosity from SF and dilution of the ISM from in-falling gas are thought to both contribute to the system's deviation from the LZR and MZR. 

While most XMP galaxies disagree with the LZR and MZR, there are two galaxies \cite[Leo A and Leo P;][respectively]{berg2012,skillman2013} that agree with the trends. These systems are isolated low-mass, low-luminosity, star-forming galaxies. Leo A and Leo P are gas-rich galaxies, although they are considerably less gas-rich than most XMP systems, with gas-mass fractions that align more with typical field dIrr galaxies. Leo A and Leo P also display flat stellar age-metallicity gradients and low effective yields,\footnote{$y_{\rm eff} = Z_{\rm gas}/$ln$\left(1/\mu\right)$. A low effective yield (log$\left( y_{\rm eff} \right) < -2.4$) is characterized by a system having metal yields that are less than those of true yields. (log$\left(y_{\rm true}\right) = -2.4$ indicates a closed-box model yield for all of a system's stellar populations, i.e., the galaxy has retained 100\% of the metal yields created in its stars). } both of which indicate that they have lost a majority of their metals to strong outflows \citep{cole2007,mcquinn2015,mcquinn2015Z}. Leo A and Leo P are unique as the only two known XMP galaxies that have gas-phase oxygen abundance measurements that align with the LZR, and have characteristics that indicate a chemical evolution pathway more typical of a field dIrr system, i.e., driven by secular evolution. 

A recently discovered XMP galaxy, Sloan Digital Sky Survey (SDSS) J100512.15+372201.5 (nicknamed here: Leonessa), has also been identified as agreeing with the LZR, although the results relied on an approximate, velocity-based distance estimate \citep{senchyna2019}. Leonessa was identified as a blue star-forming galaxy in the SDSS catalog. Follow-up Multiple-Mirror Telescope (MMT) spectral line observations revealed the system to be very metal-poor \citep[12+log(O/H) $= 7.25 \pm 0.22$;][]{senchyna2019}. The initial luminosity (based on a local flow distance and SDSS photometry) for Leonessa showed it to agree with the LZR \citep[Distance $= 2.7$ Mpc, luminosity: M$_i = -7.77$;][]{senchyna2019}; making the system a prime candidate for more detailed analysis.

Here, we present results obtained from follow-up observations of Leonessa. We introduce Hubble Space Telescope (HST) observations of the resolved stellar populations in Leonessa, from which we derive a robust distance using the TRGB method. We also present an updated gas-phase oxygen abundance (12+log(O/H)) and nitrogen-to-oxygen (N/O) abundance ratio measured using the direct method and emission lines from optical spectra obtained with the Hobby-Eberly Telescope (HET). Finally, we present an \mbox{\ion{H}{1}} mass calculated from integrated flux measurements of the \mbox{\ion{H}{1}} 21cm line detected in L-band radio spectral observations obtained from the Green Bank Telescope (GBT). 

To place Leonessa in context with the growing number of XMP galaxies, we compile a sample of 53 XMP galaxies from the literature with robust oxygen abundance measurements determined using the direct method. We present the sample of XMP galaxies alongside a sample of 97 low-mass, slightly more metal-enriched, low-luminosity galaxies in the nearby universe including 44 galaxies known to be located in voids, adding gas-mass values where reported (see \S~\ref{sec:props_leonessa_comp} for a description of the sample). We compare Leonessa's properties, along with the comparison sample to the LZR, MZR, and the log(N/O) vs 12+log(O/H) (N/O-O/H) trends (taken from \citet{berg2012,berg2019}) for field dIrr galaxies in order to explore different possible chemical evolution pathways.

This paper is organized as follows. In \S~\ref{sec:obs} we present our observations from HST, HET, and GBT. In \S~\ref{sec:hst_meas} we discuss the photometry performed on the HST data, measure the TRGB magnitude, and define the environment surrounding Leonessa. In \S~\ref{sec:atom_neb} we discuss the measurement of the nebular and radio emission lines, and subsequent elemental abundances. In \S~\ref{sec:props_leonessa_comp} we introduce the comparison sample. In \S~\ref{sec:gal_evo_trend} we report our findings for Leonessa, and investigate galaxy evolution trends. In \S~\ref{sec:dis} we investigate potential chemical evolution pathways that are likely for Leonessa. Finally, in \S~\ref{sec:conc} we summarize our conclusions.

\begin{table}
\caption{Properties of Leonessa}
\begin{tabular}{lr}
\hline
\hline
Parameter & Value\\
\hline
R.A. (J2000) & 10:05:12.154 \\
Dec. (J2000) & +37:22:01.55 \\
$z$ (redshift) & 0.0016\\
12+log(O/H) & 7.32 \(\pm 0.04\) \\
log(N/O) & \(-1.41 \pm  0.2\) \\
A\textsubscript{V}\textsuperscript{1} (mag) & 0.045\\
50\% (F606W; mag) & 29.15\\
50\% (F814W; mag) & 27.91\\
m\textsubscript{g}\textsuperscript{2} (mag) & $19.09 \pm 0.02$\\  
m\textsubscript{r}\textsuperscript{2} (mag) & $19.06 \pm 0.03$\\  
m\textsubscript{i}\textsuperscript{2} (mag) & $19.38 \pm 0.05$\\  
M\textsubscript{g} (mag) & \(-11.91 \pm 0.11\)\\
F814W$_{0,TRGB}$ (mag) & $26.95 \pm 0.10$\\
Distance Modulus (mag)& \(31.00 \pm 0.10\)\\
Distance (Mpc) & \(15.86 \pm 0.78\)\\
P.A. ($^{\circ}$) & $160$\\
$e$ (eccentricity) & $0.70$\\
SGX (Mpc) & $6.22 \pm 0.30$\\
SGY (Mpc) & $13.48 \pm 0.65$\\
SGZ (Mpc) & $-5.59 \pm 0.27$\\
$v_{\rm Helio}$ (km s$^{-1}$) & 493.80\\
S\textsubscript{HI} (mJy km s\textsuperscript{-1}) & $35.51 \pm 8.08$ \\
M\textsubscript{HI} $($M$_{\odot})$& $(2.10 \pm 0.56) \times 10^{6}$\\
M$_\star \ ($M$_{\odot})$ & $(1.31 \pm 0.24) \times 10^{6}$\\
M$_{\rm HI}$/M$_\star$ & $1.61 \pm 0.52$\\
$\mu$ & $0.69$\\
\hline
\end{tabular}
\label{tab:properties}
\tablecomments{The measured properties for Leonessa. A\textsubscript{V} is the Galactic extinction. 50\% completeness limits are provided for the F606W and F814W filters. F814W$_{0,TRGB}$ is the extinction corrected TRGB magnitude in the F814W filter. SGX, SGY, SGZ are the supergalactic coordinates for Leonessa. $v$\textsubscript{Helio} is the heliocentric (radial) velocity. S\textsubscript{HI} is the \mbox{\ion{H}{1}} 21cm line integrated flux measurement. $\mu$ is the gas-mass fraction.\\
References: \\
(1) \citet{schlafly2011} \\ (2) \citet{ahumada2020} \\ }
\end{table}

\section{Observations and Data Reduction}
\label{sec:obs}
We obtained observations of Leonessa using the HST, HET, and GBT. Here, we present those observations and the techniques used for reducing the data.

\subsection{HST Observations}
\label{ssec:hst}
HST observations (GO-16048; PI McQuinn) for Leonessa were obtained between 2020, April 12th and 15th, with the Advanced Camera for Surveys/Wide Field Channel (ACS/WFC) instrument \citep{ford1998}. The galaxy was observed for 12 orbits split between the F606W and F814W filters, resulting in exposure times of $14835$ s and $17423$ s, respectively. The HST images are available through the Mikulski Archive for Space Telescope (MAST) portal on the STScI website.\footnote{All of the HST data used in this paper can be found in MAST: \dataset[http://dx.doi.org/10.17909/f4q0-p798]{http://dx.doi.org/10.17909/f4q0-p798}} 

To reduce and combine the raw data, and prepare for running photometry, we began by retrieving the flat-fielded and charge transfer corrected {\tt flc.fits} files output from ACS/WFC instrument calibration pipeline. We then combined the {\tt flc.fits} image files using AstroDrizzle \cite[\textit{HST DrizzlePac v3.2.1} suite;][]{drizzlepac}, aligning them based on their World Coordinate System (WCS). AstroDrizzle made corrections for sky background, identified cosmic rays, and created a distortion-corrected FITS (drizzled) image file for each filter.

\begin{figure*}[!htb]
\epsscale{1.15}
\plotone{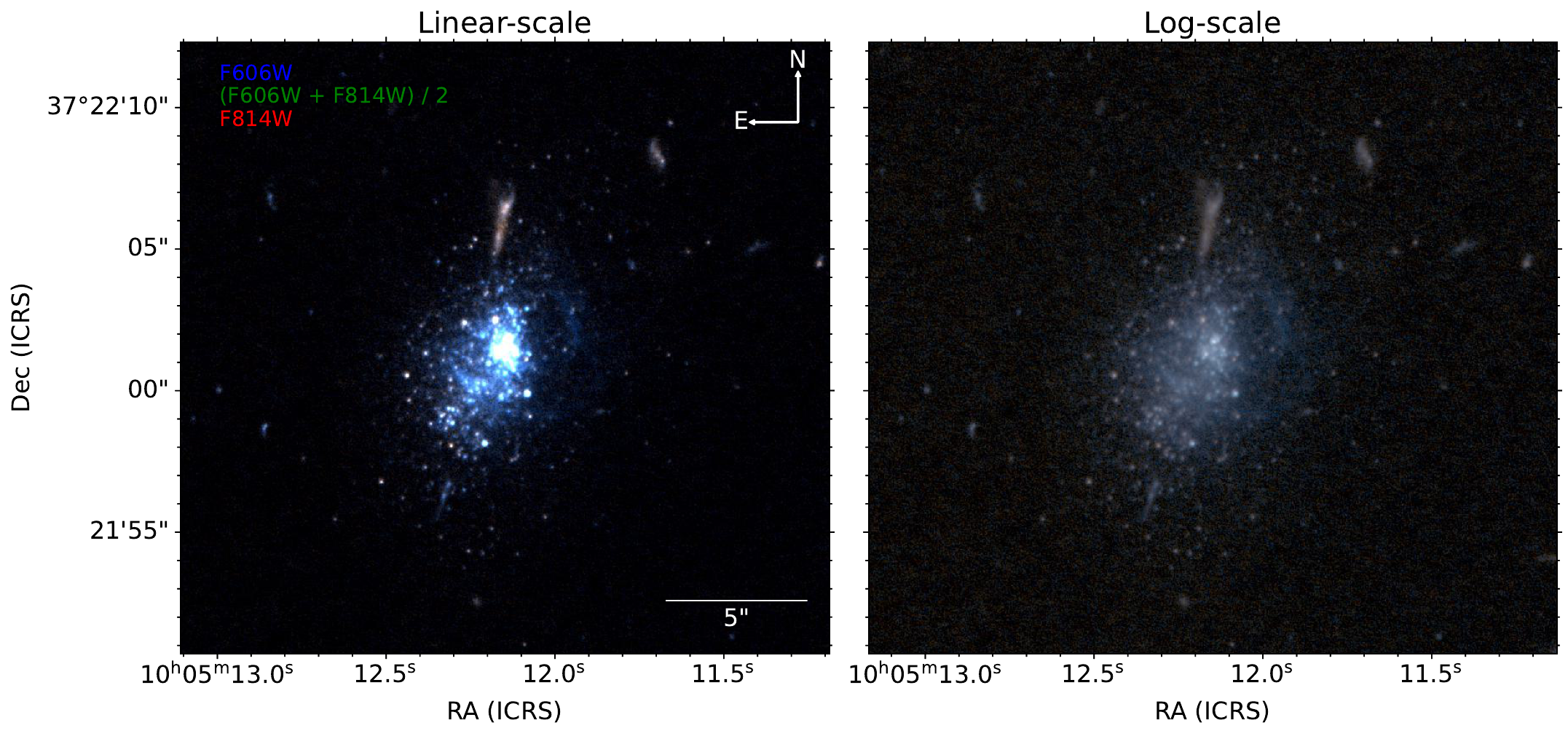}
\caption{Color images (North: up; East: left) of the galaxy Leonessa created using the {\tt APLpy} package and data from the HST F606W and F814W filters. {\it Left}: This image is created using a linear scale that highlights the intensity of the compact central star-forming region of Leonessa, likely harboring a population of young massive stars. A curve-like structure is traced out by possible nebular emission coming from the regions to the NW and SE of the central star-formation. {\it Right}: This image is created using a logarithmic scale that highlights a population of older, redder stars that extend to larger radial distances. The log scaling also diminishes the saturation caused by the bright stars in the central region.}
\label{fig:rgb}
\end{figure*}

\autoref{fig:rgb} displays color images created from the drizzled images. Data from the F606W and F814W images were used to generate the blue and red colors, respectively. The green color was constructed by averaging the F606W and F814W images. In the left panel of \autoref{fig:rgb} we have applied a linear scaling to emphasize the high-surface brightness regions. There are clear signs of recent SF in Leonessa, indicated by the population of photometrically recovered upper main sequence (MS) stars located in the crowded central region of the galaxy (see \S~\ref{sssec:phot_cat}). To the NW and SE of the central star-forming region, we see some curve-like structure that could possibly be nebular emission, although this is not verifiable with broad-band imaging alone. There is an extended object near the northern edge of the galaxy on the sky that we interpret as a background system due to its redder color; this object was omitted from photometric analysis. In the right panel of \autoref{fig:rgb} we use a logarithmic scaling in order to highlight the low-surface brightness aspects of Leonessa which accentuates the older, redder stellar populations extending out to greater radial distances.

\subsection{HET Optical Spectroscopy Observations}
\label{ssec:het_deets}
Optical spectroscopic observations were taken in clear to moderate cirrus conditions between 2021, November 3rd and 11th, using the Low-Resolution Spectrograph 2 
\citep[LRS2;][]{chonis2014,chonis2016}
on the $\sim$10-meter HET \citep{ramsey1998,hill2021}. 
We used the blue unit, LRS2-B, which has an UV arm and an ``orange" arm covering 
3700--4700~\AA\ and 4600--7000~\AA, respectively, 
with a usable spectral range 3700--7000~\AA. 
The spectral resolution, $R\sim1910$ in the UV and $R\sim1140$ in the orange, 
corresponds to a full-width-half-max (FWHM) $\sim160$ km s$^{-1}$ and $\sim260$ km s$^{-1}$, respectively. 
We obtained 12$\times$2400-s exposures.
Seeing was variable, with an average of 1\farcs57, but the LRS2 is an integral field 
spectrograph (IFS) with a view (FoV) of $12\arcsec\times6\arcsec$, allowing us to fully 
capture all of the light in our integrated extraction.

The raw LRS2 data were initially processed with 
\texttt{Panacea}\footnote{\url{https://github.com/grzeimann/Panacea}}, 
which carries out bias subtraction, dark subtraction, fiber tracing, fiber wavelength 
evaluation, fiber extraction, fiber-to-fiber normalization, source detection, source
extraction, and flux calibration for each channel. 
The absolute flux calibration comes from default response curves and measures of the 
mirror illumination, as well as the exposure throughput from guider images.
Leonessa is a compact star-forming galaxy (see \autoref{fig:rgb}) 
with a small enough footprint on the LRS2 FoV to allow for ample sky fibers within 
a given observation.
Sky fibers were identified by hand to be more than $6\arcsec$ away from the center
of Leonessa and median-combined to create a master sky spectrum for each exposure.
The master sky spectrum was then subtracted from all 280 fibers, followed by a
 principal component analysis (PCA) sky residual subtraction. 
 We note that [\ion{N}{2}] \W6548 is significantly affected by bright sky line emission in many exposures and, therefore,
 should not be trusted; [\ion{N}{2}] \W6584 should be used on its own.

\begin{figure*}
    \centering
    \includegraphics[width=0.9\linewidth]{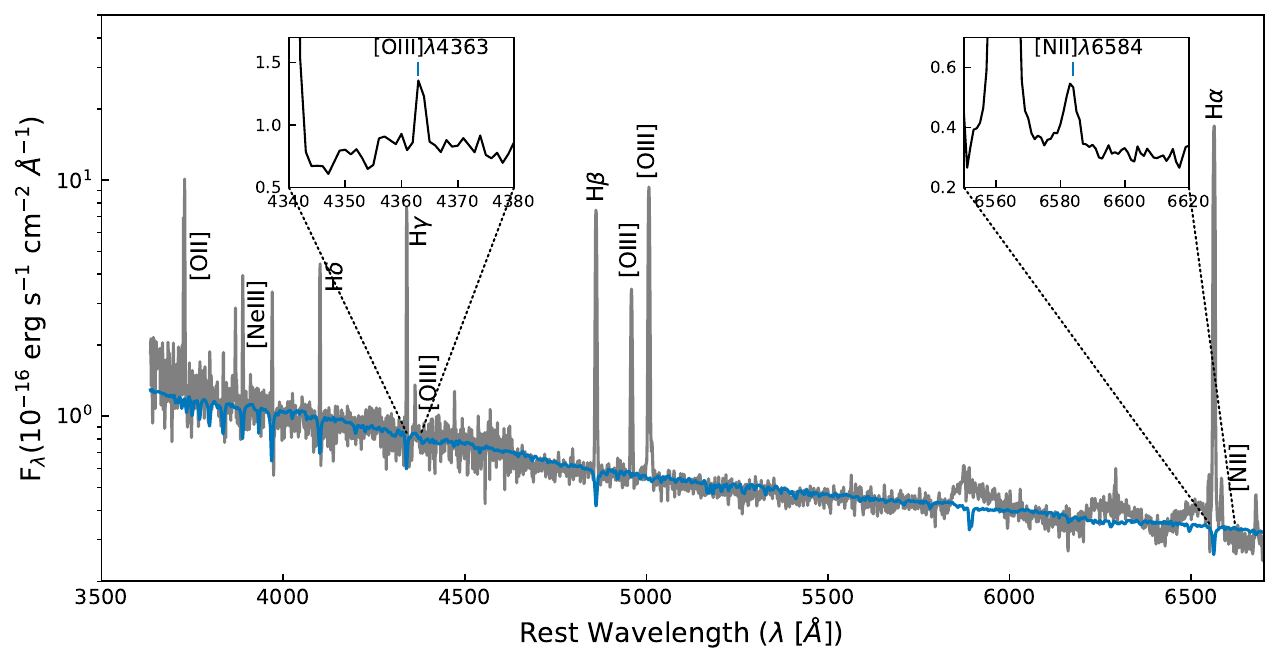}
    \caption{Emission spectrum obtained from the HET. Important emission lines are labeled. We add zoomed-in subplots to emphasize the [\ion{O}{3}] \W4363 line, used to measure the gas-phase oxygen abundance using the direct method, and the [\ion{N}{2}] \W6584 line, used for measuring the N abundance and subsequent N/O ratio for Leonessa. After \texttt{STARLIGHT} was used to fit and subtract the continuum (synthetic spectrum generated and represented here as the blue spectrum) the \texttt{python} package \texttt{lmfit} was used to measure emission line strengths (see \S~\ref{sec:neb_meas}).}
    \label{fig:HET_spec}
\end{figure*}

Each spectrum was then extracted using a weighted-extraction limited to a
3\farcs5 aperture centered on a 2D Gaussian model fit to the [\ion{O}{3}] \W5007 
emission image.
Individual extracted spectra were normalized to one another by the strength of the
integrated [\ion{O}{3}] \W5007 flux.
Finally, the 12 exposures were combined using a variance-weighted average,
which down-weights observations taken in poor conditions. 
Standard error propagation was used throughout.
The final spectrum is shown \autoref{fig:HET_spec}, where the observed spectrum is represented in grey with the synthetic spectrum overlaid in blue. We add two zoomed-in regions of the spectrum in the subplots, focusing on the [\ion{O}{3}] \W4363 and the [\ion{N}{2}] \W6584 lines used for measuring gas-phase abundance ratios.

\subsection{GBT 21cm Radio Spectroscopy Observations}
\label{ssec:GBT}
GBT radio observations in the \mbox{\mbox{\ion{H}{1}}} 21cm frequency range were obtained on 2021, February 3rd (GBT20B-468; PI: McQuinn) using the Robert C. Byrd Green Bank Telescope (100m single-dish). The aim of these observations was to establish an atomic gas mass for Leonessa. The galaxy was observed in the L-band at a redshifted frequency of $1418.42$ MHz calculated using a redshift of $z=0.0014$ \citep{senchyna2019} and an \mbox{\ion{H}{1}} 21cm resting line frequency of $1420.40575$ MHz. Position switching (On/Off) observations were performed in 5 min intervals. We used a bandwidth of $23.44$ MHz across 8192 channels, a spectral resolution of $2.9$ kHz, and a beam width of  $9.159\arcmin$.

\begin{figure}[!htb]
\epsscale{1.2}
\plotone{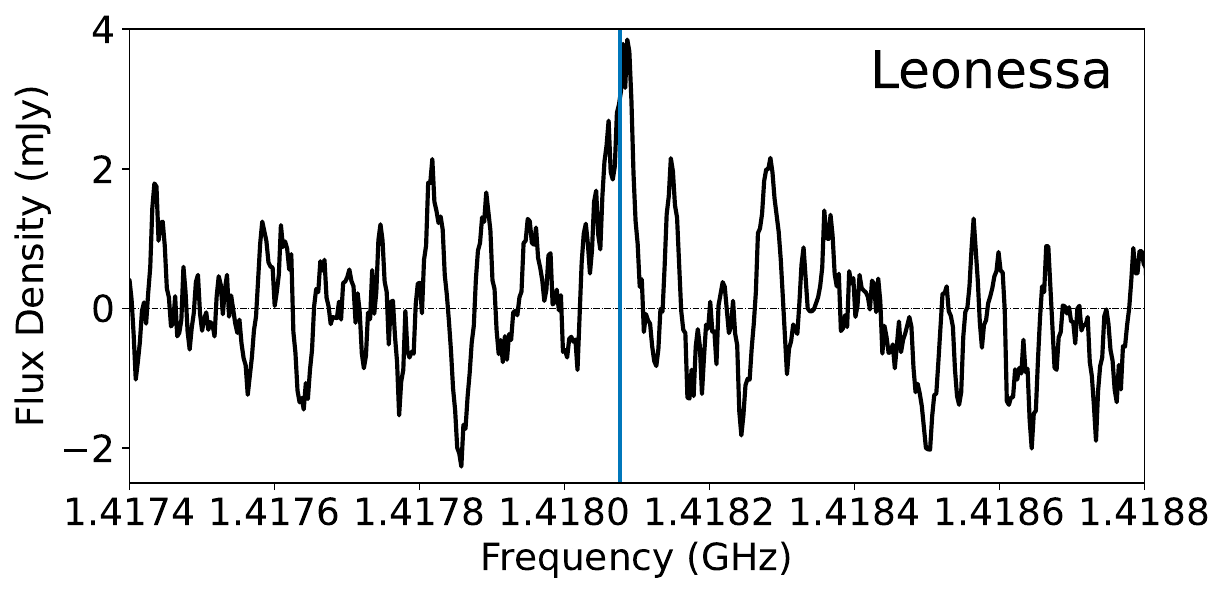}
\caption{Spectrum of the \mbox{\ion{H}{1}} 21cm line from the 100 meter single dish Green Bank Telescope at the Green Bank Observatory. The detected line is indicated with a blue vertical line, located at a frequency of 1.41808 GHz, and redshift $z=0.00163$ \cite[a $-72.65$ km s$^{-1}$ offset from the previously reported redshift $z=0.0014$;][]{senchyna2019}, possible evidence for a dynamical separation between the atomic and ionized gas. The 21cm line is recovered with a $4.4\sigma$ significance; it has a sharply peaked Gaussian profile, indicating a low velocity dispersion in the HI disc.}
\label{fig:hi_21}
\end{figure}

The GBT data was reduced to identify and measure the 21cm line. We first combined the observations and averaged the spectra, including both polarizations. Before reducing the combined spectra, we scaled the observed flux up by 1.2 in accordance with noise diode calibration offset values for the On/Off observing mode presented in \citet{goddy2020}. The 21cm line detection was identified, a continuum was fit, and the baseline subtracted using {\texttt{GBTIDL}}. We then smoothed the data with the final 21cm spectrum for Leonessa appearing in \autoref{fig:hi_21}, with the central frequency highlighted by the vertical blue line at 1.41808 GHz corresponding to a redshift $z=0.00163$. The original redshift ($z=0.0014$) was established using optical spectra, the offset between the optical and radio derived redshifts equates to a velocity difference of $\Delta v \sim 68.85$ km s$^{-1}$. This could be indirect evidence of inflows or outflows; detailed \mbox{\ion{H}{1}} spatial mapping would give us better insight into the mechanisms responsible for this discrepancy.

\section{Resolved Stellar Population Analysis}
\label{sec:hst_meas}
\subsection{Photometry of Resolved Stars}
\label{sssec:phot_cat}
\label{sec:ast}
Photometry was carried out on the HST images using \dphot\ \citep{dolphin2000,dolphin2016}. \dphot\ is a point-spread function (PSF) fitting photometry software package with HST specific modules and PSFs. Through an iterative process, \dphot\ uses the F606W drizzled image as a reference and identifies the central location of flux peaks. These locations are then used to fit and measure the peaks and sky in the individual {\tt flc.fits} files using well-established PSF models. 

\dphot\ is run using an extensive list of user-specified parameters. For Leonessa, we adopted the parameter values from \citet{williams2014} Table 2 (WFC/All). \citet{williams2014} performed extensive testing, resulting in optimized parameter values for \dphot\ in crowded stellar fields. 

After the photometry was run, quality and spatial cuts were performed on the output data to create our final stellar catalog of well-recovered sources in the galaxy. These cuts were implemented to reject non-stellar objects, poorly-recovered sources, and background and foreground contamination. Quality cuts were made using metrics output by \dphot. Specifically, we removed sources with signal to noise ratio (SNR) values $< 5$, poorly-fit sources with object flag values that indicate a non-stellar detection, and sharpness and crowding values where the object was too broad/sharp ($\rm Sharp^2 \ge 0.0625$), or had magnitudes significantly impacted by a neighboring source ($\rm Crowd \ge 0.3$). Sharpness and crowding value limits were applied on a per filter basis.

Leonessa's spatial size is small in comparison to the field of view of the HST observations, and despite quality cuts, background and foreground sources from the greater field of view are still present in the source catalog. To reduce the impact from this contamination we performed a spatial cut, excluding sources outside the vicinity of Leonessa. To define the boundary beyond which sources were removed from the catalog, nested annuli centered on the SDSS R.A. and Decl. position for Leonessa  were extended until additional annuli added more contaminating sources than high-confidence target sources. Contamination was identified by sampling and comparing sources outside the vicinity of the target to sources within Leonessa on a color-magnitude diagram (CMD). The final ellipse containing the stellar populations within Leonessa that we used to construct our high-fidelity catalog has a semi-major axis of $a=16.45\arcsec$, a position angle (P.A.) of $160^{\circ}$ east of north, and an eccentricity of $e=0.70$ (see \autoref{tab:properties}).

We measured the completeness of the photometry by running artificial star tests (ASTs) in the region of the chip surrounding Leonessa. In order to recreate the compact characteristics present in the HST imaging of Leonessa when carrying out the ASTs, we first create a detection list of the recovered sources in the region of the detector chip where Leonessa is located. We then use this list as a reference for generating artificial stars while running the ASTs. Approximately $300$k artificial stars were placed into each of the raw {\tt flc.fits} files. \dphot\ was then re-run with artificial stars individually inserted and recovered from the images. We applied the same quality cuts to ASTs that were used for creating the photometric catalog. The $50\%$ completeness magnitudes for the F606W and the F814W filters are $29.15$ and $27.91$ mag, respectively (see \autoref{tab:properties}).

\subsection{Color-Magnitude Diagram}
\begin{figure}[!htb]
\plotone{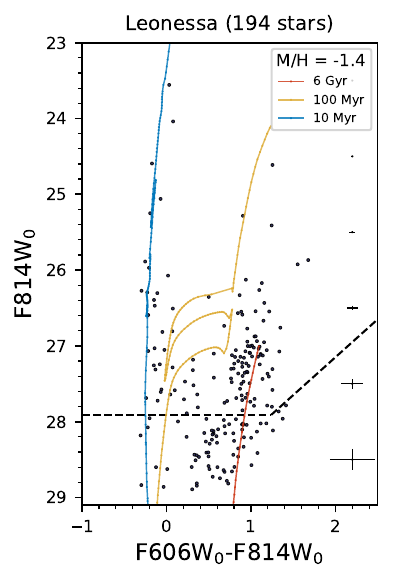}
\caption{Color-Magnitude Diagram of all sources that lie within the spatial region of our target and that pass our quality cuts. CMD is overlaid with PARSEC isochrones tracing stellar ages of $10$ Myr (blue line), $\sim 100$ Myr (yellow line), and $\sim 6$ Gyr (red line), with a metallicity of [M/H]$= -1.4$, and using our TRGB distance measurement from \S~\ref{sec:trgb}. Representative errorbars are shown to the right of the data. The dashed lines represent the 50\% completeness magnitudes for F606W and F814W filters (see \autoref{tab:properties}).}
\label{fig:cmd_iso}
\end{figure}

\autoref{fig:cmd_iso} presents a CMD of the high-fidelity photometric catalog for Leonessa, which contains 194 stars. The MS stars populate the CMD along a vertical sequence from a color (F606W-F814W) of approximately $-0.5$ to $0.0$. The red giant branch (RGB) is well-defined at a color of F606W$-$F814W~$\sim~1.0$ and ascends vertically until reaching the TRGB. F814W magnitudes were binned ($\Delta$bin $= 1$ mag) and the average errors in magnitude and color were determined for each bin. These errors are represented in \autoref{fig:cmd_iso} by bars placed at the center of their respective F814W magnitude bin. We applied reddening corrections to account for Galactic extinction along the line of sight to the galaxy. Dust extinction in the ACS bandpass is based on dust maps from \citet{schlegal1998}. The calibrated magnitude corrections are $0.041$ mag and $0.025$ mag for the F606W and F814W filters, respectively \citep{schlafly2011}. The subscript of `0' on the axis labels in \autoref{fig:cmd_iso} indicate that the magnitudes have been extinction corrected.

In \autoref{fig:cmd_iso} we also overplot stellar isochrones. The isochrones are generated from the PAdova and TRieste Stellar Evolution Code (\texttt{PARSEC}) 1.2 stellar evolution models \citep{bressan2012}, with ages of $10$ Myr (blue line), $\sim 100$ Myr (yellow line), and $\sim 6$ Gyr (red line). Approximately half of the MS stars that we recovered in the HST photometry for Leonessa are likely O- and B-class stars (stars above the $\sim 100$ Myr MS turn off; e.g., stars with magnitudes F814W $\lesssim 27$). All isochrones have a metallicity of [M/H] = -1.4; for comparison, this is the approximate equivalent of the gas-phase oxygen abundance of Leonessa (measured below; see \S~\ref{sec:neb_meas}) converted into a [M/H]\footnote{[M/H] $=$ log((Z/X)/0.0207), with (Z/X)$_\odot = 0.0207$} value \citep{bressan2012}.

\subsection{TRGB Distance}
\label{sec:trgb}
Resolving older stars that populate the RGB in the CMD enables us to determine the distance to Leonessa using the TRGB distance indicator method. Low-mass stars \cite[$\rm M_\star \lesssim 1.6 \, M_{\odot}$ for metal-rich stars; $\rm M_\star \lesssim 1 \, M_{\odot}$ for metal-poor stars;][]{lee1993} ascending the RGB have H-shell burning around an inert degenerate \ion{He}{0}core. The subsequent transition to triple-$\alpha$ fusion in the cores of these stars initiates their abrupt migration off of the RGB, corresponding to an intrinsic luminosity that can be used as a standard candle distance indicator. The magnitude of this luminosity defines the TRGB, and results in a discontinuity in the luminosity function (LF) of the stars, and thus a measurable ``edge" on the CMD. 

Multiple methods have been developed to measure and calibrate the TRGB magnitude. The first estimates of the TRGB were visually determined from the CMD using \textit{I}-band data \citep{hoessel1982, reid1987, vanderydt1991, lee_freedman_madore1993}, but had unconfirmed reliability. \citet{lee1993}, and \citet{madore1995} devised a more robust way of measuring the TRGB that used binned magnitudes weighted by stellar count to create a LF that was then convolved with an edge-detection (Sobel) kernel. These methods emphasized the \textit{edge} in the LF created by the scarcity of stars populating magnitudes brighter than the TRGB in the CMD. The simple sobel methods give a more precise measurement of the TRGB than visual determinations, although they are affected by bin size and magnitude range selection. \citet{sakai1996} implemented a Gaussian-smoothed LF and an adapted edge-detection filter that was not reliant on bin-width and magnitude range, giving a TRGB measurement free from binning or magnitude selection bias. 

Due to the paucity of resolved stellar populations recovered from our photometry we chose a smoothed LF Sobel response approach to measure the TRGB magnitude for Leonessa \citep{sakai1996,makarov2006}. Specifically, we use a GLOESS \cite[Gaussian-windowed, Locally Weighted Scatterplot Smoothing,][]{persson2004, hatt2017, beaton2019} algorithm on the constructed F814W filter LF. This non-parametric smoothing method becomes relevant when there is a low number of photometrically recovered point sources to bin \citep{freedman2019}. We bin the point sources, weighted by their F814W magnitudes, with a bin-width of $0.01$ mag. The smoothed LF is then constructed by computing a weighted least-squares regression on every bin. The bin weights ($w(n)$) are defined according to:
\begin{equation}
\label{eq:gauss}
w(n)=e^{-\frac{M-M_n}{2\sigma^2}}
\end{equation}
where $M$ includes all bins, $M_n$ denotes the $n^{th}$ bin, and $\sigma$ is a smoothing scale.

\begin{figure}[!htb]
\epsscale{1.25}
\plotone{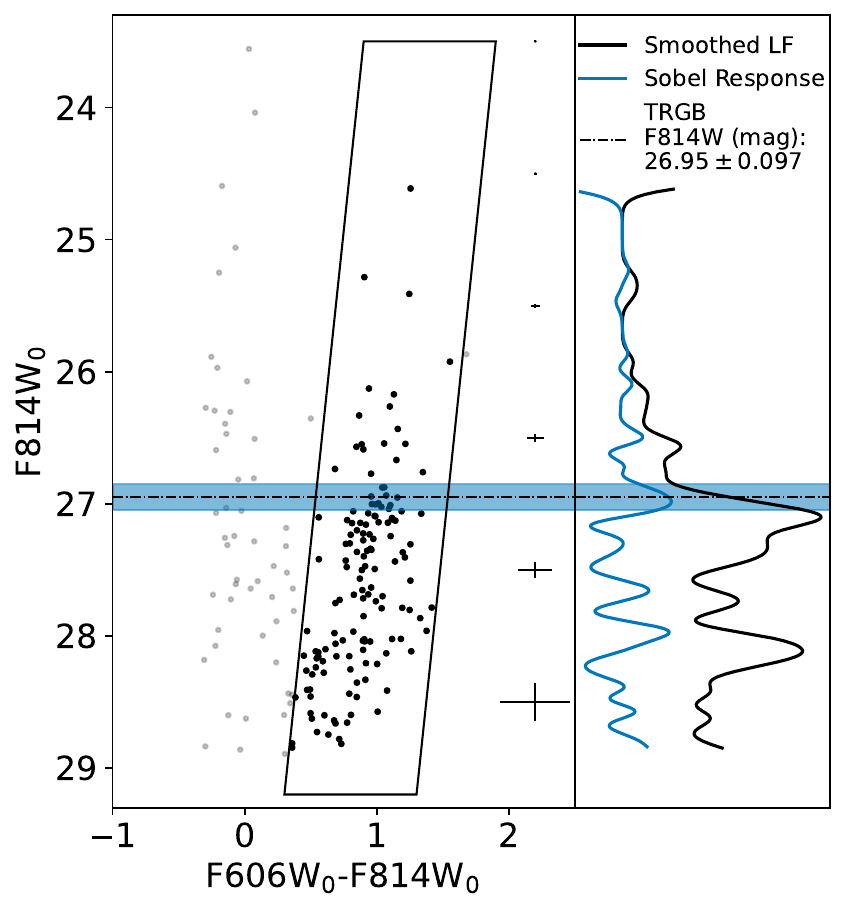}
\caption{\textit{Left}: CMD highlighting the stars used for measuring the TRGB (outlined by the polygon). The TRGB magnitude is traced by a dash-dot line, with total error indicated by the blue-shaded regions above and below the TRGB. \textit{Right}: Here we present the LF (solid black curve) and the smoothed sobel response (solid blue curve) for Leonessa. The TRGB stands out clearly as the peak of the smoothed sobel response at an F814W magnitude of $26.95 \pm 0.097$, corresponding to a distance of $15.86 \pm 0.78$ Mpc; see text for details.} 
\label{fig:cmd_histy}
\end{figure}

Before measuring the TRGB, we culled our photometric data to include only stars in the region of the RGB to eliminate any LF contamination from MS stars. The left panel of \autoref{fig:cmd_histy} displays a CMD highlighting the population of stars used for measuring the TRGB magnitude. This cut captures the RGB, TRGB, and some likely asymptotic giant branch (AGB) stars.

After smoothing the LF we applied a simple Sobel kernel to detect the discontinuity at the TRGB magnitude. When the LF is convolved with the kernel $\left[-2,0,+2\right]$ \citep{lee1993, brunker2019}, the output Sobel response is sensitive to the changes in star counts between neighboring bins in the LF. The highest peak in the Sobel response indicates the largest discontinuity in the LF, and thus the magnitude of the TRGB. The right panel of \autoref{fig:cmd_histy} shows the resultant smoothed LF (black line), and the Sobel response (blue line). 

The TRGB magnitude and statistical uncertainties are determined by implementing a Monte Carlo (MC) technique using the systematic errors determined from ASTs as standard deviations away from the output magnitude. We performed 5,000 iterations using this technique. The MC technique results in a normal distribution centered on the TRGB magnitude. We take the mean of the Gaussian in this distribution as the TRGB magnitude and the standard deviations are representative of the statistical errors. The measured, extinction corrected TRGB magnitude: F814W$_{0,TRGB}=26.95 \pm 0.1$ mag (dashed-dot line) is also shown in \autoref{fig:cmd_histy}.

The TRGB has a zero-point calibration magnitude of $M_V \sim -4.0$ mag, as well as a color-based metallicity dependence (lower metallicity means lower opacity in stellar atmosphere, which leads to higher temperatures and hence a bluer color) \citep{sakai1996, ferrarese2000, rizzi2007, jang2017}. For colors (F606W$_0$-F814W$_0$) $\le 1.5$, the color-magnitude relation for the TRGB is approximately constant requiring no corrections \citep{jang2017}. Only one (likely AGB) star on our CMD have a color $> 1.5$, therefore we did not apply a color correction to our data.  

Combining the TRGB magnitude with the zero-point calibration \cite[$F814W_{0,TRGB}+4.049(\pm 0.038)$;][]{freedman2021} results in a distance modulus value of $31.00 \pm 0.10$ mag, corresponding to a distance of $15.86 \pm 0.78$ Mpc. This is $\sim 6$ times farther than the distance initially reported in \citet{senchyna2019} of $2.7$ Mpc, estimated using a local-flow model from \citet{tonry2000}. The new, more robust, distance places Leonessa in an isolated region outside of the Local Volume.

The farther distance to Leonessa results in a brighter absolute luminosity (M$_g$), and a larger stellar mass. We used our TRGB-based distance to calculate an absolute $g$-band luminosity of M$_g = -11.91 \pm 0.11$ \cite[based on SDSS apparent $g$-band;][]{ahumada2020}. We then determined a stellar mass based on a mass-to-light ratio ($\rm M_\star/L$; \autoref{eq:m/l}), using the $r$- and $i$-band calibration from \citet{bell2003}, and \citet{hsyu2018}:
\begin{equation}
\label{eq:m/l}
\rm log\left(\frac{M_\star}{L}\right)=0.006+(1.114 \times (r-i))
\end{equation}
where $r$ and $i$ are SDSS magnitudes \citep{ahumada2020}. This resulted in a stellar mass of M$_\star = (1.31 \pm 0.24) \times 10^6$ M$_{\odot}$. This is considerably larger than the initial estimate for stellar mass of M$_\star = 5.01 \times 10^4$ M$_{\odot}$ that used a local flow model-based distance \citep{senchyna2019}.

\subsection{Environment Surrounding Leonessa}
\label{ssec:env}
Characterizing the environment surrounding a galaxy is critical for quantifying the density of the surrounding region and exploring possible past interactions with neighboring systems. Several XMP galaxies (SBS 0335-52W, IZw18A/C, DDO 68/B) are observed to be closely interacting or merging with other systems \citep{lelli2012a,annibali2023,correnti2025}. These interactions and mergers are thought to be responsible for triggering the recent SF in these systems that leads to increased luminosities, thus contributing to their outlier status on the LZR. Conversely, the two galaxies that agree with the LZR (Leo A, and Leo P) are observed to be isolated systems, not interacting, and not in a state of starburst. These examples outline the importance of identifying a galaxy's proximity to nearby systems.

The first step in investigating Leonessa's environment is to determine its distance to the nearest neighboring systems ($\rm D_{NN}$), which can offer insight into the likelihood of whether the galaxy has experienced any recent interaction. We compare Leonessa's supergalactic position (SGX/Y/Z; \autoref{tab:properties}) to systems in the $Cosmicflows$-4 (CF-4) database \citep{tully2023}, void galaxies \cite[P19-voids;][]{pustilnik2016,pustilnik2019}, as well as galaxies from the ALFALFA survey \citep{haynes2018, durbala2020}. Cartesian supergalactic coordinates for Leonessa were calculated using equations found in \citet{mcquinn2020}, where distance, supergalactic latitude (SGB), and supergalactic longitude (SGL) are used to determine the 3-dimensional location of a system relative to an origin centered on the Milky Way (SGX, \, SGY, \, SGZ = (0, 0, 0) in units of Mpc, with the SGX-SGY plane correlating with the super-galactic plane at SGZ = 0). 

The supergalactic position for Leonessa place it within the known void Cancer-Canis-Minor-Hydra \cite[\cch, centered at SGX, SGY, SGZ = (3.35, 9.42, $-14.36$);][]{pustilnik2019}. Void region boundaries can vary significantly depending on the method used to define a region as a void \citep{colberg2008, pustilnik2019}. We adopt the \citet{pustilnik2024} boundary condition which defines two regions within voids where galaxies can reside: the `inner' void region ($\rm D_{NN}\ge 1.7$ Mpc) and the `outer' void region ($\rm D_{NN} < 1.7$ Mpc). Inner void regions are inherently less populated. 

From the three galaxy catalogs (CF-4, P19-voids, and ALFALFA) no galaxies are found within a $\sim 1.8$ Mpc radius of Leonessa, making it an isolated `inner' void system. The nearest neighboring galaxy is AGC 5351 (NGC 3067), separated by a distance of $1.84$ Mpc \cite[ALFALFA;][]{haynes2018, durbala2020}. AGC 5351 is an intermediate-mass \cite[log(M$_\star/$M$_{\odot}) = 10.0$;][]{chang2015}, gas-poor (M$_{\rm HI}/$M$_\star = 0.04$) system. The next nearest galaxy to Leonessa is NGC 2964, with an approximate physical separation of 2.21 Mpc based on a Tully-Fisher (TF) distance \citep{tully2023}. NGC 2964 is an intermediate-mass, barred-spiral galaxy \cite[stellar mass of log(M$_\star/$M$_{\odot}) = 10.3$;][]{castignani2022}. Uncertainties in distances to neighboring galaxies can significantly change their calculated distance from Leonessa, and could affect how the galaxy's environment is characterized, i.e., isolated or interacting. Robust follow-up measurements of galaxies in the ALFALFA survey have found that they can be up to 73\% farther than the ALFALFA estimates \citep{mcquinn2014}. \citet{tully2023} report errors of 25\% for TF distances in the CF-4 catalog (TRGB, Cepheid Period, surface brightness fluctuation, and SN-Ia based distances were reported at $\sim 5 - 7$\% error). 

To investigate the effect that distance errors might have on an interpretation of environment, we also explored galaxies close to Leonessa on the sky. We considered all galaxies from the CF-4, P19-voids, and ALFALFA catalogs with reported distances between 10 and 22 Mpc (to account for maximum error on reported distances) to be at an equidistance of 15.86 Mpc. Calculating the angular separation between systems on the sky we find two low-mass systems that could be within $\sim 0.5$ Mpc proximity to Leonessa. UGC 5540 (CF-4 catalog) has an angular distance corresponding to a separation of 615 kpc from Leonessa. For comparison, the TF distance to UGC 5540 is 12.78 Mpc, where the assumed 15.86 Mpc is $\sim 25$\% farther than the estimated TF distance. UGC 5540 was also found in the P19-voids catalog at a distance of 19.37 Mpc, residing in ``Void 12" \citep{pustilnik2019}. \citet{pustilnik2019} found UGC 5540 to a have a $\rm D_{NN} = 3.41$, qualifying it as an `inner' void galaxy. AGC 198712 (ALFALFA catalog) is found 492 kpc from Leonessa, based on angular separation and assuming both systems are at the same distance (15.86 Mpc). The reported distance for AGC 198712 is 8.3 Mpc, making the assumed 15.86 Mpc $\sim 92$\% further than the reported distance. Given the two separate distance values for UGC 5540, the relatively high respective errors in distance differences to Leonessa, and the large difference in their heliocentric velocities with respect to Leonessa ($\Delta v_{Helio} > 100$ km s$^{-1}$), it is unlikely that these galaxies are in close physical proximity to Leonessa.

Assessing the environment surrounding Leonessa has revealed no galaxies within $\sim 1.8$ Mpc, based on their reported distances. As such, we categorize Leonessa as an isolated galaxy residing deep within a void region, and rule out Leonessa as a satellite system of a more massive galaxy. The environment surrounding Leonessa suggests that its current SF is unlikely to have been triggered by interactions with neighboring systems, and, thus, environment is likely not responsible for its enhanced luminosity on the LZR.

\section{Nebular \& Atomic Gas Measurements}
\label{sec:atom_neb}
\subsection{Optical Emission-Line Fluxes}
\label{sec:neb_meas}
For these calculations we use the \texttt{PyNeb} \texttt{python} package with atomic data. The underlying continuum of the optical spectra were fit by the STARLIGHT spectral synthesis code \citep{fernandes2005} using stellar models from \citet{bruzual2003}. Next, emission lines were fit in the continuum-subtracted spectrum with Gaussian profiles and allowing for an additional nebular continuum component. To determine the emission line fluxes and associated errors, we used the non-linear least-squares minimization and curve fitting \texttt{python} package \texttt{lmfit}. We fit each line with a Gaussian and fit nearby lines simultaneously, constraining the velocity widths to each other. 

The initial electron temperature estimate was derived from the ratio of the [\ion{O}{3}]\W4363 auroral line to the [\ion{O}{3}]\W5007 nebular lines, serving as input for the reddening determination. 
Dust reddening values, $E(B-V)$, were obtained from flux ratios of H$\gamma$/H$\beta$ and H$\delta$/H$\beta$ Balmer lines using the \citet{cardelli1989} reddening law with $\rm R_V=3.1$ and combined as an error-weighted average. We iterated this process until the temperature changed by $<10$K. Reddening corrected line intensities are reported in \autoref{tab:emlin}.

\subsection{Temperature and Density}
\label{ssec:T_e}
The direct method for measuring 12+log(O/H) uses ratios of collisionally-excited lines from different levels of the same ion (e.g., [\ion{O}{3}]\W\W4363, 5007) to determine the electron temperature ($T_e$) and from a single split level to determine density within the observed \ion{H}{2} region \citep{dinerstein1990, izotov2006a, ly2014}.
We use the reddening-corrected [\ion{O}{3}]I(\W5007)/I(\W4363) ratio to determine the high-ionization zone temperature. The low-ionization zone temperature is then determined using the
photoionization model based $T_e-T_e$ relationship from \citet{garnett1992}:
\begin{equation}
T_e[O II] = T_e[N II] = 0.70 \times T_e[O III] + 3000K
\end{equation}
Electron density was calculated using the [\ion{O}{2}]\W3726/\W3728 line ratio, $T_e[NII]$, and assuming a uniform density.

\subsection{Oxygen Abundance}
Ionic abundances relative to hydrogen are calculated using:
\begin{equation}
\label{eq:fracoh}
    \frac{N(X^i)}{N(H^+)} = \frac{I_{\lambda(i)}}{I_{H\beta}}\frac{j_{H\beta}}{j_{\lambda(i)}}
\end{equation}
The emissivity coefficients, $j_{\lambda(i)}$, which are functions of both
temperature and density, were calculated using \texttt{PyNeb} with $T_e[OIII]$ for O$^{+2}$/H$^+$, and $T_e[OII]$ for O$^{+}$/H$^+$. The total oxygen abundance (O/H) was thus found from the
simple sum of O$^+$/H$^+$ and O$^{+2}$/H$^+$, as contributions from O$^0$ and O$^{+3}$ (requiring ionization energies of 54.9 eV) are typically
negligible in H II regions \cite[e.g.,][]{berg2021}. We arrive at a total oxygen abundance of 12+log(O/H) $= 7.32 \pm 0.04$.

\subsection{N/O Ratio}
The ionic abundance of nitrogen relative to oxygen was calculated using the same ratio as \autoref{eq:fracoh}, adopting the convention that $\rm N/O = N^+/O^+$ \citep{peimbert1967}.

\begin{equation}
\label{eq:fracno}
    \frac{N^+}{O^+} = \frac{I_{\lambda6584}}{I_{\lambda3726,3728}}\frac{j_{\lambda3726,3728}}{j_{\lambda6584}}
\end{equation}

Emissivity functions were calculated using \texttt{PyNeb} with $T_e[OII]$ and $T_e[NII]$ for N$^{+}$/O$^+$, resulting in a value of $\rm log(N/O) = -1.41 \pm 0.2$ for Leonessa

\begin{table}[ht]
    \centering
    \caption{Optical Emission-Lines from HET Observations.}
    \label{tab:emlin}
    \begin{tabular}{lr} 
        \hline
        \hline
        Ion & $I(\lambda)/I(\mbox{H}\beta)$ \\ 
        \hline
        {[\ion{O}{2}]} $\lambda$3726.04 & 15.34$\pm$0.60 \\ 
        {[\ion{O}{2}]} $\lambda$3728.80 & 22.46$\pm$0.64 \\
        {[\ion{Ne}{3}]} $\lambda$3868.76 & 4.26$\pm$0.17  \\
        H$\delta$ $\lambda$4101.71 & 25.59$\pm$0.42 \\
        H$\gamma$ $\lambda$4340.44 & 10.58$\pm$0.17 \\
        {[\ion{O}{3}]} $\lambda$4363.86 & 1.20$\pm$0.11  \\
        H$\beta$ $\lambda$4861.35 & 100.00$\pm$1.40 \\
        {[\ion{O}{3}]} $\lambda$4958.91 & 17.06$\pm$0.46 \\
        {[\ion{O}{3}]} $\lambda$5006.84 & 51.81$\pm$0.52 \\
        {[\ion{N}{2}]} $\lambda$6548.51 & 1.15$\pm$1.13  \\
        {[\ion{N}{2}]} $\lambda$6584.24 & 1.69$\pm$1.42  \\
        \hline
        E(B$-$V) & 0.052$\pm$0.023 \\
        F$_{H\beta}$ & 41.86$\pm$0.24  \\
        \hline
    \end{tabular}
    \tablecomments{The [\ion{N}{2}] \W6548 line is affected in many exposures, see \S~\ref{ssec:het_deets}.}
\end{table}

\subsection{Atomic Hydrogen Flux}
The \mbox{\ion{H}{1}} 21cm line profile in \autoref{fig:hi_21} was detected with a $\sim 4.4 \, \sigma$ significance. From the smoothed radio spectrum, we measured the integrated flux density over the velocity channels containing the 21cm line detection using the \texttt{SciPy} Simpson's rule integration method, resulting in a total integrated flux of $35.51 \pm 8.08$ mJy km  s$^{-1}$. This corresponds to a line width of $\sim 13.3$ km s$^{-1}$. The line exhibits a symmetric Gaussian form, indicating a low velocity dispersion in the \mbox{\ion{H}{1}} disk. 

To determine the atomic hydrogen mass we used the TRGB distance ($\rm D_{\rm TRGB}$; Mpc) and integrated flux ($\rm S_{HI}$; Jy km s$^{-1}$) in conjunction with the standard formula for the \mbox{\ion{H}{1}} mass \citep{haynes2018}: $\rm M_{\rm HI} = 2.356 \times 10^5D_{\rm TRGB}^2S_{HI}$, resulting in an M$_{\rm HI}= (2.10 \pm 0.56) \times 10^{6} \, $M$_{\odot}$ for Leonessa. Uncertainties on the \mbox{\ion{H}{1}} mass were calculated following the error propagation approach outlined in \citet[][and references therein]{haynes2018}.

\section{Comparison Sample}
\label{sec:props_leonessa_comp}
To provide additional context for the properties of the XMP galaxy Leonessa, we identified and compiled a comparison sample of low-mass galaxies from the literature. The fundamental criterion for including a galaxy in the comparison sample was having a robust measurement of the gas-phase oxygen abundance using the direct method. We chose the direct method as a requirement because it is more accurate than the empirical strong line method. 

The comparison sample is comprised of 150 low-mass systems that we separate into three sub-groups: 
(1) LVL: 53 field galaxies from the \citet{berg2012} sample (\S~\ref{sssec:b12}), mostly comprised of \textit{Local Volume Legacy} galaxies \citep{dale2009}; (2) void: 44 galaxies found in underdense regions of the nearby Universe \cite[][\S~\ref{sssec:void}]{kreckel2015,pustilnik2016,kniazev2018,pustilnik2021}; (3) XMP: 53 galaxies with 12+log(O/H) $\le 7.35$ (\S~\ref{sssec:xmp}). Seven galaxies categorized as XMP overlap with the LVL (Leo A, UGCA 292) or void (Leoncino, Leonessa, SDSS 081239.53+483645.4, SDSS 085233.75+135028.3, SDSS 092609.45+334304.1) groups, and one galaxy (DDO 68/B) falls into all three categories. There is evidence that large scale environment plays a role in the chemical evolution of a galaxy \citep[e.g,][]{petropoulou2012,pilyugin2017,pustilnik2024}. However, we do not investigate the role void regions play in chemical evolution here, so while we make the distinction between void and LVL for systems with higher gas-phase oxygen abundances, we do not make that same distinction for XMP systems.

\autoref{tab:dwarf_gals} contains the list of galaxy names (Column 1) and coordinates (R.A. and Decl. (J2000): Columns 2 and 3, respectively). The remaining columns report gas-phase oxygen abundance, stellar and gas mass, luminosity, distance, galaxy subgroup, and references. We note that DDO 68 and IZw18 are listed as systems with two (separate) components due to their unique morphology and how they are presented in the literature. Certain values for Leonessa from \autoref{tab:properties} are repeated in \autoref{tab:dwarf_gals} for convenience.

\subsection{LVL Galaxies}
\label{sssec:b12}
LVL, and similar field dIrr galaxies, were used by \citet{berg2012} to establish MZR, LZR, and N/O-O/H trends for low-mass field galaxies in the nearby universe. LVL galaxy distances reported in \autoref{tab:dwarf_gals} were updated to more recent TRGB measurements where applicable \citep{tully2023}. \citet{berg2012} reported stellar mass uniformly estimated using M$_\star/L_K$ ratios and a $B-K$ color \citep{lee2006}. M$_{\rm HI}$ values for 29 of the LVL galaxies were calculated using \mbox{\ion{H}{1}} fluxes taken from the Galaxies Losing Oxygen via Winds project (GLOW; McQuinn et al. in prep; see \autoref{tab:dwarf_gals}).

\subsection{Void Galaxies}
\label{sssec:void}
As established above (\S~\ref{ssec:env}), Leonessa is located within the \cch\ void \citep{pustilnik2019}. Void galaxies are suspected to undergo a slower chemical evolution process than counterpart galaxies in regions with a mean cosmic density \citep{kniazev2018, pustilnik2019,pustilnik2022,pustilnik2024}. Therefore, we include a population of void galaxies in the comparison sample. 

The void galaxies in the sample were taken from \citet{kreckel2015,pustilnik2016,pustilnik2021}; and \citet{kniazev2018}. $B$-band magnitudes are reported for all void galaxies (see \autoref{tab:dwarf_gals}). \citet{kniazev2018} calculated distances using a Local Flow model.\footnote{$\rm D=V_{LG}/73$; $\rm V_{LG}$ is recession velocity in Local Group coordinate system} \citet{pustilnik2016,pustilnik2021} adopts TF relation distances \citep{tully2009}. A majority of void galaxies do not have stellar masses reported and are therefore included in our analysis of the LZR, but are not considered in the MZR.

\subsection{XMP Galaxies}
\label{sssec:xmp}
The last sub-group in the comparison sample is the XMP group, including well studied XMP galaxies: DDO 68/B, Leo A, Leo P, SBS 0335-052W, IZw18A/B, Leoncino (see \autoref{tab:dwarf_gals} for references). We gathered sky coordinates and gas-phase oxygen abundances for the XMP galaxy comparison sample from 16 different sources (see \autoref{tab:dwarf_gals}). The galaxy with the lowest gas-phase oxygen abundance is SDSS J081152.11+473026.2 \cite[J0811+4730;][]{izotov2019} with 12+log(O/H)$= 6.97$.

While the majority of the LVL and void galaxies have reported $B$-band magnitudes, many of the XMP galaxies are reported with $g$-band magnitudes instead of $B$-band. $g$-band luminosities cover a similar passband as $B$-band luminosities, making the $g$-band filter a comparable substitute for star forming galaxies when observations in $B$-band are not available \citep{papaderos2008}. Stellar masses for XMP galaxies, when reported in the literature, were determined using several different spectral energy distribution (SED) fitting models, or from calibrated M$_\star/L$ ratios. We found distances reported for a majority of the XMP systems, most of which were determined using galaxy velocities and local flow models. For gas masses, if more than one value was reported, we adopt the value with the largest signal-to-noise ratio in the \ion{H}{1} flux measurement.

\begin{figure*}
\epsscale{1.25}
\plotone{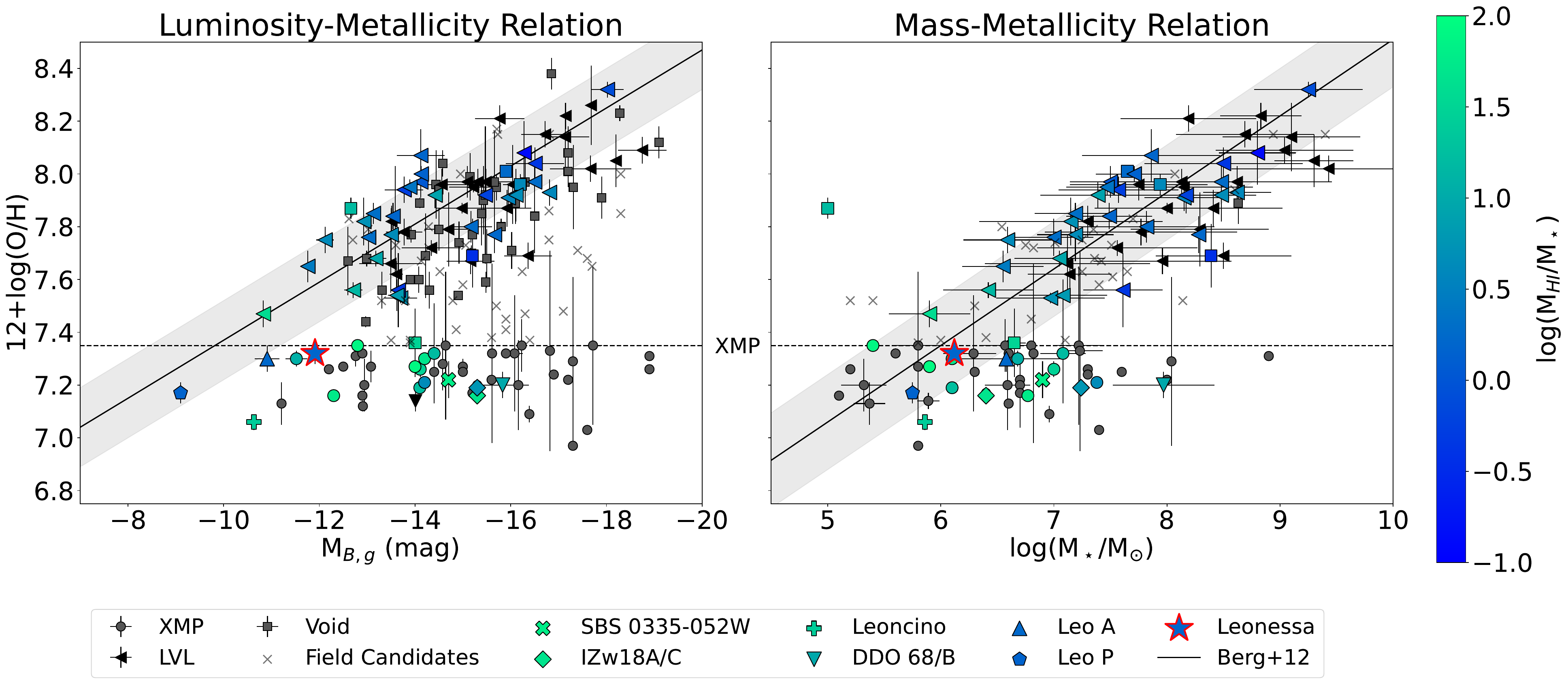}
\caption{\textit{Left Panel}: 12+log(O/H) vs Absolute B-band or $g$-band magnitude (luminosity; M$_B$ and M$_g$, respectively) for LVL (arrows), void (squares), XMP (circles) galaxies, including 6 well-studied XMP galaxies with special markers (DDO 68/B, IZw18A/C, Leo A, Leo P, Leoncino, and SBS 0335-052W), and field candidate galaxies (x). The black line indicates the best-fit line for the LZR (1$\sigma$ shaded in grey) \citep{berg2012}. M$_g$, or M$_B$ mags for our sample are taken from the literature (\autoref{tab:dwarf_gals}). Log of gas-to-stellar mass ratios (log(M$_{\rm HI}$/M$_\star$)) are represented with the colorbar. Higher-metallicity (12+log(O/H) $\gtrsim 7.6$) galaxies exhibit a correlation with magnitudes, while the XMP galaxies are scattered over 10 orders of magnitude and over $\lesssim 0.45$ dex in 12+log(O/H), with only 2 galaxies (Leo P and Leo A) confirmed to agree with the best-fitting LZR line. \textit{Right Panel}: Shows the 12+log(O/H) vs log(M$_\star$/M$_{\odot}$) for the sample. The MZR from \citet{berg2012} is represented by the black line, with 1$\sigma$ error shaded in grey. Leonessa, and many other XMP galaxies agree with the MZR found by \citet{berg2012}. However, there is a distinct scatter (though less-so than the LZR) away from the established MZR. We also see a feature on both the LZR and MZR where gas-to-stellar mass ratios anti-correlate with gas-phase oxygen abundances; XMP galaxies are all gas-rich systems (where the data is available). \newline
Note: While the systems labeled `Field Candidate' (represented with an `x') did have gas-phase oxygen abundances measured using the direct method, they were not classified in the literature as a field galaxy. As such these systems were not used for our comparison analysis with Leonessa. We include them in the figure for completeness.}
\label{fig:MZR_LZR}
\end{figure*}

\section{Galaxy Evolution Trends}
\label{sec:gal_evo_trend}
Here, we present the composite galaxy sample, including Leonessa, on the LZR and MZR planes. First, we discuss the LZR as it contains the most galaxies from the comparison sample. Next, we investigate the MZR, where we consider only galaxies from the LVL and the XMP groups; as previously stated, stellar masses were not reported for the void sub-group. Where \mbox{\ion{H}{1}} masses are available, we also consider the gas-to-stellar mass ratios with respect to the LZR and MZR.

\subsection{LZR}
The left panel of \autoref{fig:MZR_LZR} presents 146 galaxies from our sample on the LZR plane. The LVL and void galaxy sub-samples are both fully represented. Four XMP galaxies did not have absolute magnitudes published: J0216+1715; J1046+4047; J2313+2935; from \citet{yang2017}, and 1631+4426 from \citet{kojima2020}. We overlay the best-fitting line to the field dwarfs from \citet{berg2012} (12+log(O/H)$= (6.27 \pm 0.21) + (-0.11 \pm 0.01)$M$_B$) as a solid black line, with 1$\sigma$ ($\sigma = 0.15$) uncertainty shaded in grey. We also indicate with a horizontal dashed line the ISM oxygen abundance value below which a system is classified as XMP. 

It should be noted that on the LZR plane (and similarly on the MZR plane) there are systems labeled `Field Candidates' (represented with an `x'). These are systems that have been found in the literature, and have gas-phase oxygen abundances determined using the direct method, and are field dIrr candidate galaxies. However, for the purposes of this study, focusing on XMP galaxies in particular, we have not included them in our analysis. We have included these field candidate systems here for completeness; their chemical evolution pathways will be explored in future work (Breneman et al.\ in prep.).

Most void galaxies with 12+log(O/H) $\gtrsim 7.6$ in \autoref{fig:MZR_LZR} agree with the LZR, although they exhibit a slightly lower metallicity than their LVL counterparts. There is evidence that void galaxies may follow a trend similar to the LVL galaxies, albeit offset to lower metallicities and having a steeper slope \citep{pustilnik2016,kniazev2018,pustilnik2021,pustilnik2024}. \citet{mcquinn2020} investigated this using only void galaxies from \citet{pustilnik2016} with direct method measurement oxygen abundances and did not find any significant ($> 1 \sigma$) deviation from the \citet{berg2012} trend. However, \citet{kniazev2018} investigated a sample of Eridanus void galaxies with gas-phase metallicity measured using the direct method, which we include here, and found a trend that was consistent with \citet{pustilnik2016} in being below the \citet{berg2012} trend. \citet{pustilnik2024}, adding more direct method measurements to their previous sample, found evidence that the new values also supported a unique trend for void galaxies that deviates from the \citet{berg2012} relation. It should be noted that these void outliers at low-metallicity comprise only a small population of systems, and that a significant increase in void galaxy populations would be needed to confirm any deviation \citep{pustilnik2024}.

XMP galaxies in the left panel of \autoref{fig:MZR_LZR} are spread over a large parameter space in luminosity, with all but two systems (Leo A, Leo P) disagreeing with the LZR. This could be due in-part to a Malmquist bias in recovering the brightest XMP systems when searching for them. The overwhelming majority of XMP galaxies were discovered by searching for SF galaxies in survey data. Leo P, the faintest galaxy in the sample \cite[M$_B = -8.85 \pm 0.1$;][]{mcquinn2015}, is unlike most XMP galaxies in that it was discovered via its \mbox{\ion{H}{1}} content; it is also one of the few XMP systems that does not exhibit signs of intense SF, containing only one compact SF region. Leo A is also a faint system, though its close proximity inside the Local Group made it easier to find. J0808+3244 \cite[M$_g = -18.9$, 12+log(O/H) $= 7.31$;][]{izotov2019}, the farthest outlier galaxy from the LZR, and the brightest XMP system, is approximately $8$ magnitudes brighter (in the $g$-band) than the LZR value at the corresponding oxygen abundance. J0808+3244 was discovered through a search for SF galaxies in the SDSS DR14 catalog \citep{izotov2019}. The remaining XMP systems ranging from $6.97 \le$ 12+log(O/H) $\le 7.35$ provide a good covering fraction over the $\sim 10$ magnitude range separating Leo P and J0808+3244. In addition to Leo P and Leo A we also highlight several other well-studied XMP galaxies from the literature in \autoref{fig:MZR_LZR}: DDO 68/B, SBS 0335-052W, Leoncino, and IZw18A/C. 

We include information on gas content in \autoref{fig:MZR_LZR} by representing log of the gas-to-stellar mass (log(M$_{\rm HI}/$M$_\star$)) ratio using a colorbar. Galaxies with higher gas-phase metallicity show lower gas-to-stellar mass ratios. XMP systems, with an average $\left\langle \rm M_{\rm HI}/M_\star \right\rangle \sim 35$ within the comparison sample, are considerably more gas rich than the LVL galaxies ($\left\langle \rm M_{\rm HI}/M_\star \right\rangle \sim 7$). All but four XMP galaxies have M$_{\rm HI}$/M$_\star > 10$. The three least gas-rich XMP systems are Leo P, Leo A, and Leonessa. Notice in \autoref{fig:MZR_LZR} the XMP systems with the lowest gas-to-stellar mass ratios (Leo P, Leo A, Leonessa) are also closer to the LZR than their extremely gas-rich counterparts (Leoncino, IZw18C, SBS 0335-052W, DDO 68) that exhibit enhanced luminosities and/or the possible dilution of their O/H ratio via infall of pristine gas. All three of these systems either agree with the LZR (Leo A, Leo P) or have a small deviation away from the 1$\sigma$ error (Leonessa). 

The position of Leonessa on the LZR has significantly changed from calculations using the previous, highly uncertain, distance \citep{senchyna2019,mcquinn2020}. Whereas previous estimates had its luminosity agreeing with the LZR, Leonessa (represented as the star outlined in red) is now slightly offset from the LZR in \autoref{fig:MZR_LZR}, with a luminosity $\sim 2.4$ mags brighter than the previous estimate. This new position on the LZR joins Leonessa with a growing population of XMP galaxies that disagree with the established \citet{berg2012} LZR. XMP systems that are outliers on the LZR either have enhanced luminosity from their intense SF, and/or exhibit a lower measured chemical abundance compared to their counterparts at similar luminosities. We explore potential processes that are likely responsible for making Leonessa an outlier from the LZR in \S~\ref{sec:dis}.

\subsection{MZR}
The right panel of \autoref{fig:MZR_LZR} presents 100 galaxies from the sample on the the MZR plane. Only one void galaxy (above the XMP line) is represented since stellar masses were not reported for the rest of the void sample considered here. 48 of the XMP galaxies represented on the LZR also appear on the the MZR. We again overlay the best-fitting line from \citet{berg2012} (12+log(O/H) $= [(5.61 \pm 0.24) + (0.29 \pm 0.03)$log(M$_\star$)]) as a black line with 1$\sigma$ ($\sigma = 0.18$) uncertainty shaded in grey. The XMP metallicity value is marked with a dashed black line at 12+log(O/H) = 7.35. A majority of XMP galaxies are outliers from the MZR, diverging to higher stellar masses/lower measured oxygen abundances than their counterparts that agree with the trend. The scatter away from the trend in the XMP region of the MZR is smaller than that on the LZR, although the divergence is still significant, with only 19 out of the 48 systems agreeing with the trend.

SED fitting was used for approximating most of the stellar masses reported for the XMP galaxies. SED modeling has been found to under-estimate stellar masses for star-forming galaxies (like BCDs) with a bias of 25\% for galaxies with sSFR on the scale of XMP galaxies \cite[sSFR $\ge 0.1$ Gyr$^{-1}$;][]{sorba2015}. This is attributed to SED models fitting the younger (massive) stellar populations that have a lower M$_\star/L$ ratio, while also dominating the spectrum at optical wavelengths, outshining the older stellar populations that account for a majority of the stellar mass \citep{sorba2018}. We investigated the effects that outshining could have on our XMP sample by assuming the maximum reported bias from outshining (25\%), which would increase the log(M$_\star$/M$_{\odot}$) values for XMP galaxies by 0.12 dex. This extreme case scenario only results in a 6\% decrease of the number of XMP systems that agree with the MZR. The 0.12 dex increase from a maximum outshining bias is less than the stellar mass error on $\sim 68$\% of XMP galaxies with reported errors, and we therefore do not consider outshining to be a significant contributor to an XMP galaxy's position on the MZR plane. 

Only a quarter of the reported XMP galaxy distances in the sample were determined using robust methods. Investigating how distances affect SED-based calibrated stellar mass estimates, we find that the XMP systems that show a significant offset from the MZR would require distances to be $\sim 7$ times closer, on average, for the systems to be within the $1 \sigma$ errors of the MZR at their respective gas-phase oxygen abundance. We only need to consider overestimates on distances since XMP galaxies are only found to scatter below the MZR and not above.

When adding gas-to-stellar mass ratios to the MZR in \autoref{fig:MZR_LZR} (represented by the colorbar) we see some similarities and differences to how these systems are represented on the LZR plane. Similarly to the LZR plane, the least gas-rich galaxies in the XMP regime fall within, or very close to, the \citet{berg2012} trend. However, unlike on the LZR plane, on the MZR plane we find that a few very gas-rich XMP systems agree with the trend. This is in contrast to previous studies that found the most gas-rich systems to also be the farthest outliers away from the MZR \citep{ekta2010}.

Leonessa has the second lowest gas-to-stellar mass ratio (M$_{\rm HI}$/M$_\star = 1.61 \pm 0.52$) of the XMP galaxies in the comparison sample. This places its position in the $\rm log(M_{HI}/M_\star)$ vs. 12+log(O/H) parameter space close to Leo P (M$_{\rm HI}$/M$_\star = 1.45$) and Leo A (M$_{\rm HI}$/M$_\star = 1.66$).

\subsection{N/O vs.\ O/H}
\label{ssec:no_oh}
Here, we investigate how Leonessa, and the comparison sample, relate to N/O-O/H trends found in \citet{berg2012,berg2019}. Oxygen is predominantly produced as a primary\footnote{Primary and secondary elements are both produced during stellar nucleosynthesis. The distinction between them lies in whether or not the process is sensitive to initial stellar metallicity. Primary elemental yields are independent of the initial seed metallicity when the star formed, while secondary elemental yields depend on the initial stellar metallicity.} element through the triple$-\alpha$ process, and enriched into the ISM on Myr timescales by core-collapse supernovae of massive stars ($> 8 M_{\odot}$). In contrast, at solar metallicity nitrogen is produced predominantly as a secondary element through the CNO cycle and is mainly synthesized and enriched on Gyr timescales through low- and intermediate-mass AGB stars shedding their outer layers. However, at lower metallicities secondary nitrogen production yields are largely hindered by the lack of seed carbon and nitrogen abundances in the ISM when stars are formed. Because of this, primary nitrogen yields are thought to be dominant in very low-metallicity stars (like those found in XMP galaxies) where the seed carbon and nitrogen come from the helium burning phase of the star, and therefore are independent of initial carbon and nitrogen abundances \citep{vanzee2006,vincenzo2016,berg2020}. Previous studies have found that N/O ratios have a bimodal behavior as a function of their gas-phase oxygen abundance, indicating primary nitrogen dominating in low-metallicity and secondary nitrogen yields linearly increasing as metallicity, and carbon and nitrogen abundances increase \citep{izotov1999,henry2000,vanzee2006,guseva2011,berg2012,berg2019,vincenzo2016,vincenzo2018a,kojima2021,kobayashi2023,hunter2024}. 

\begin{figure}[htb!]
\epsscale{1.2}
\plotone{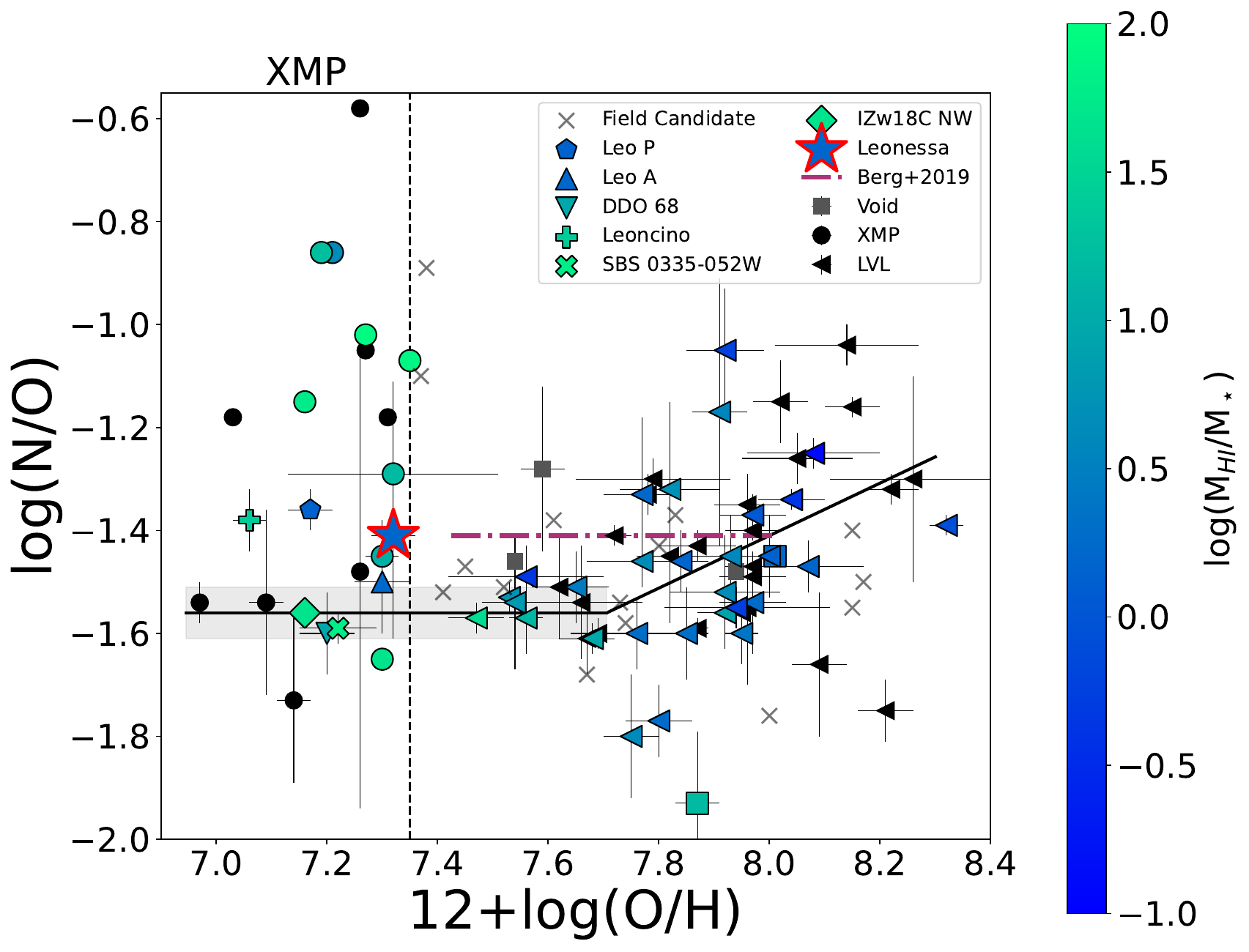}
\caption{Log(N/O) vs 12+log(O/H) for the comparison sample of dwarf galaxies. We use the same identifying markers as in \autoref{fig:MZR_LZR}. The XMP value is indicated with a dashed vertical line at 12+log(O/H)$=7.35$. We identify the trend(s) established in \citet{berg2012} with solid black lines; we also include an updated plateau from \citet{berg2019} represented by a purple dash-dot line. Log of the gas-to-stellar mass ratios is represented by the colorbar to the right of the figure. Note that XMP galaxies exhibit a significant scatter away from the log(N/O) plateau for low-metallicity systems. We do not find a correlation between gas-to-stellar mass ratios and N/O ratios. Leonessa does not stand out as a significant outlier in comparison to other XMP systems.}
\label{fig:NO_Z}
\end{figure}

\autoref{fig:NO_Z} presents 83 galaxies from the comparison sample with their N/O ratios (log(N/O)) shown as a function of gas-phase oxygen abundance (12+log(O/H)), or N/O-O/H. The bimodality of N/O is traced out by the overlaid trends (black and purple lines) established in \citet{berg2012,berg2019}. \citet{berg2012} found that at low-metallicity (12+log(O/H) $< 7.7$), N/O is constant at log(N/O) $=-1.56 \pm 0.05$, while at higher metallicity the log(N/O) ratio increases as $-5.49+0.51[12+$log(O/H)]. \citet{berg2019} updated the constant N/O (plateau) to log(N/O) $=-1.41 \pm 0.09$ (represented by a purple dash-dot line in \autoref{fig:NO_Z}) for systems with $\rm 7.50 \lesssim 12+log(O/H) \lesssim 8.00$. When adding more recently discovered XMP systems (see \autoref{tab:dwarf_gals}) to the N/O-O/H plane we find that a majority of them disagree with the N/O plateau at lower 12+log(O/H) values ($\lesssim 7.5$). There is a large scatter present above the trend in N/O values for XMP systems, with J0314-0108 \citep[log(N/O) $= -0.58$, 12+log(O/H) $= 7.26$;][]{izotov2019} having the largest enhanced N/O value at $\sim 20 \sigma$ away from the trend.

The measured log(N/O) value of $-1.41 \pm 0.20$ for Leonessa agrees with the plateau value for low-metallicity systems \citep{berg2012,berg2019}. The system does not stand out as an outlier like many other XMP systems do.

Similar to the LZR and MZR planes we also consider gas-to-stellar mass ratios on N/O-O/H. We do not find any correlation between log(M$_{\rm HI}$/M$_\star$) and N/O ratios for XMP systems; instead there is a large range ($\sim 0.5$ dex) of N/O ratios found in extremely gas-rich galaxies. We find that while all XMP galaxies are gas-rich, the most gas-rich systems are not the farthest outliers above the N/O-O/H plateau. We discuss possible scenarios responsible for the N/O scatter in \S~\ref{ssec:nitro_diss}. 

XMP galaxies show a large scatter in \autoref{fig:NO_Z},

\section{Chemical Enrichment of XMP Systems}
\label{sec:dis}
XMP galaxies standout in \autoref{fig:MZR_LZR} as outliers from the established \citet{berg2012} MZR and LZR for low-mass dwarf galaxies, and show a noticeably larger scatter away from the trend on the LZR than the MZR. When investigating the nitrogen content in XMP systems we also find that they exhibit a large scatter in N/O ratios. Here, we consider what different chemical evolution pathways XMP systems might follow that result in covering such a small range in gas-phase metallicity ($\Delta Z_{gas} \sim 0.4$) over such a wide parameter space in luminosity ($\Delta $M$_{B,g} \sim 10$ mag), stellar mass ($\Delta$log$($M$_\star/$M$_{\odot}) \sim 4$), and N/O ratios ($\Delta$log(N/O) $\sim 1.2$).

\subsection{MZR \& LZR}
XMP galaxies that agree with the MZR and LZR found in \autoref{fig:MZR_LZR} likely follow an evolutionary path that is expected of typical field dwarf galaxies, marked by secular evolution and dominated by stellar feedback. Due to dwarf galaxies having a propensity to host ``bursty" SF \citep{mcquinn2010b}, stellar feedback from clusters of massive stars can inject enough energy into the ISM to accelerate baryonic matter beyond the escape velocity of their shallow potential wells \cite[e.g.][and references therein.]{tremonti2004,mcquinn2019,collins2022}. The most metal-enhanced component (the `hot' phase of the wind) is also the most energetic, and therefore the most likely to be expelled from the system \citep{strickland2009}. 

Multiple studies have explored the effects of stellar-feedback driven outflows on ISM enrichment, finding that galactic winds can be considerably more chemically enriched than the surrounding ISM \citep{garnett2002,dalcanton2007,chisholm2018,collins2022}. \citet{peeples2011} and \citet{chisholm2018} both found steep inverse power-law correlations between their metal-loading fractions ($\zeta_w$) and virial velocity ($v_{vir}$; related to baryonic mass through the virial theorem). \citet{dalcanton2007} found that systems with M$_{bary} \le 10^7$ M$_{\odot}$ can achieve up to 50\% metal concentrations in their outflows; they also found that metal-biased winds were the only mechanism capable of driving effective yields down to those found in observations (log$(y_{\rm eff}) \le -2.4$). While modeled and observed outflows are found to be very effective at removing metal-enriched masses from the ISM, recovering these galactic winds in observations of low-mass galaxies is quite difficult. The diffuse X-ray emission extending beyond the gas disk used to identify the `hot' phase of these outflows is usually faint, and short lived \cite[fading away after $\sim 25$ Myr;][]{mcquinn2018}, making direct detections of `hot' outflows likely to be a rare occurrence in observations.

Leo P is an example of a galaxy that does not have outflows observed directly, but is suspected of having experienced preferential depletion of metals from its ISM through galactic winds \citep{skillman2013,mcquinn2015}. Leo P is found in an isolated environment where it shows no signs of recent interactions or mergers \citep{mcquinn2015}. \citet{mcquinn2015Z} investigated the metal retention fraction in Leo P, finding that most (95\%) of the oxygen produced over all previous star forming epochs was missing. Leo P also has an extremely low effective yield (log$(y_{\rm eff}) = -3.4$), another indicator that it has expelled its metals through metal-biased galactic winds \citep{dalcanton2007}. Leonessa is similar to Leo P in this respect, and the low effective yield observed in Leonessa (log$(y_{\rm eff}) = -3.2$) may serve as indirect evidence that it has expelled a significant amount of its metal yields.

Many XMP galaxies, including Leonessa, currently host intense star forming regions. The emission from these regions will skew the measured $B$- and $g$-band absolute magnitudes brighter, and move the galaxy's position away from the LZR in \autoref{fig:MZR_LZR} \citep{salzer2005,izotov2012,guseva2017,hsyu2018}. \citet{izotov2018b} investigated how much a star forming region in J0811+4730 was contributing to the system's M$_g$ mag. They found that the nebular emission increased the $g$-band luminosity by 2 magnitudes. \citet{mcquinn2020} similarly estimated that a population of young, massive stars in Leoncino increased the galaxy's M$_g$ by $\sim 1.3$ mags. We identified 25 stars on Leonessa's CMD ($\sim 13$\% of the the recovered stellar population) belonging to the upper MS, and corresponding to a 100 Myr PARSEC isochrone ($\le 27.5$ mag in F814W), likely making them O- and B-class stars. Removing the luminosity contribution of these upper MS stars reduces Leonessa's M$_g$ value by $\sim 1.1$ mag, putting it within the 1$\sigma$ error of the LZR. This indicates that the current SF in Leonessa is likely responsible for increasing the system's luminosity and moving its position on the LZR plane away from the trend.

For XMP galaxies that lie off both the LZR and the MZR, external mechanisms (such as inflows) are thought to be the dominant contributors to their status as outliers. Many of these systems are interacting with or merging with other dIrr galaxies. These galaxy interactions create a scenario where systems exchange gas components and experience heavy turbulence and mixing of their ISM. \citet{ekta2010} found their sample of XMP galaxies were very gas-rich and in most cases showed signs of disturbed morphologies, indicating external interaction as the most likely cause for the current SF. They also attributed lower metallicities in the ISM of their sample to dilution from accretion of pristine gas. \citet{annibali2023} were able to resolve a third component (DDO 68C) of the iconic DDO 68 system, and found a faint \mbox{\ion{H}{1}} bridge connecting it to other components, noting that while DDO 68C does not have a large gas mass it is possible the main components of DDO 68 have stripped gas from this system in the past, causing the XMP characteristics observed in DDO 68's main body. Their findings further corroborate previous \mbox{\ion{H}{1}} studies where extremely gas-rich XMP galaxies were often found to be interacting or merging with another system, and are expected to have accreted pristine gas, which lowers their measured gas-phase oxygen abundances \citep{vanzee2006,chengalur2006,ekta2006,ekta2008,hsyu2018}.

\subsection{\mbox{\ion{H}{1}} Content}
\begin{figure}[htb!]
\epsscale{1.2}
\plotone{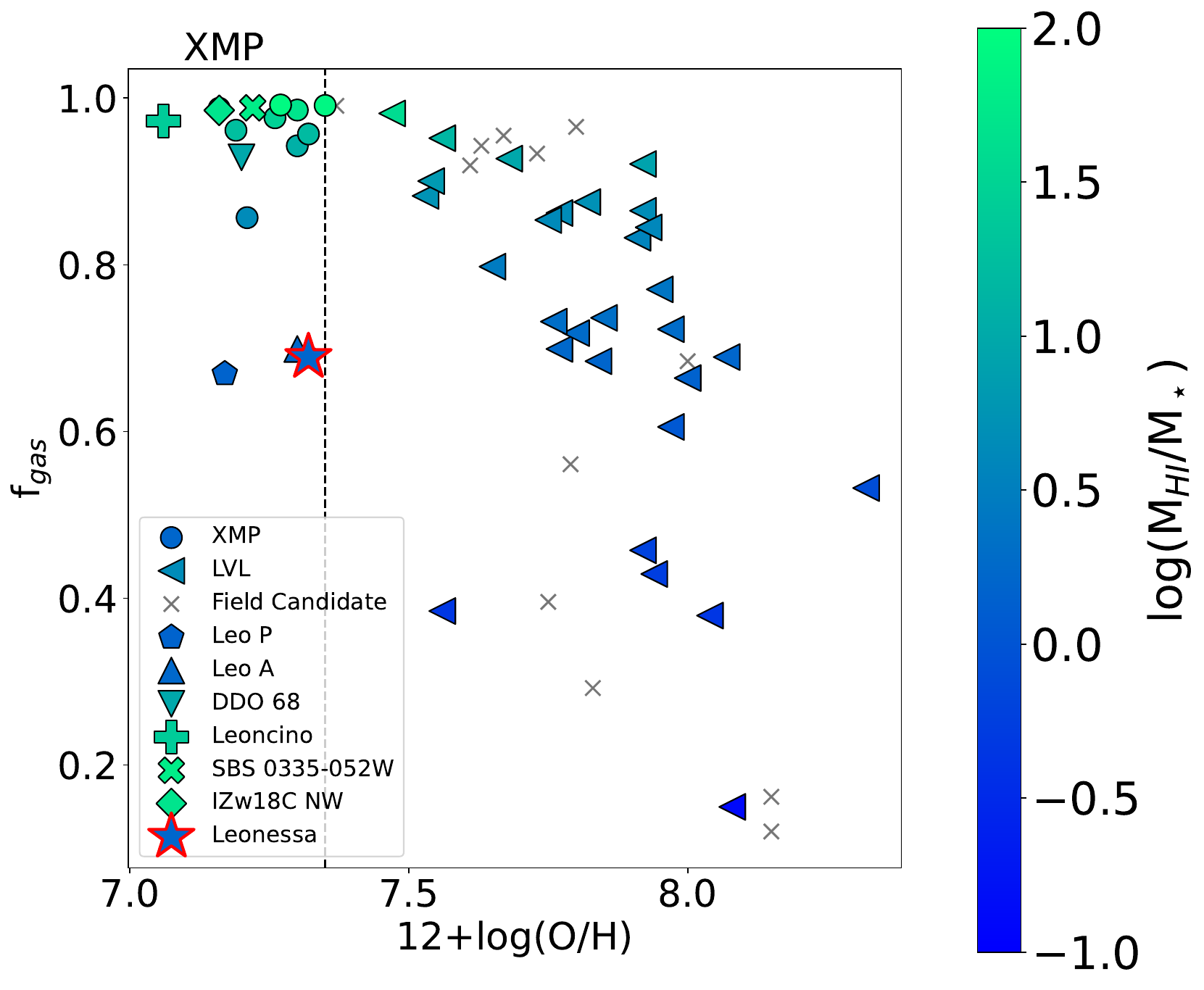}
\caption{$\mu$ vs. 12+log(O/H) for the comparison sample, including Leonessa. All markers and labels are identical to \autoref{fig:MZR_LZR} and \autoref{fig:NO_Z}. The colorbar is now representative of the log of the number of sigma a system is away from the LZR in \autoref{fig:MZR_LZR} ($\Delta LZR$). Notice that the majority of the XMP systems are concentrated in the upper left corner, also indicated by the highest $\mu$ values. These systems are also the farthest outliers from the LZR. Leonessa has a much lower $\mu$ value than most other XMP systems, except Leo A, and Leo P. Leo A, and Leo P are also the only two XMP systems to agree with both the LZR and MZR.}
\label{fig:Z_fgas_delzr}
\end{figure}

Adding an \mbox{\ion{H}{1}} gas-to-stellar mass ratio axis to the MZR and LZR planes reveals a distinct difference in the gas content of known XMP galaxies compared to their chemically enriched counterparts, finding a majority of XMP galaxies to be extremely gas-rich ($\mu \ge 0.95 \sim $M$_{\rm HI}/$M$_\star \ge 10$). \autoref{fig:Z_fgas_delzr} compares gas-mass fraction to gas-phase metallicity for the comparison sample, including Leonessa; here we indicate the log of the magnitude offset from the LZR with a colorbar. Most of the XMP galaxies are concentrated in the upper-left corner of the plot where very low metallicity correlates to very large gas-mass fractions. Some of the most gas-rich systems in the XMP regime are also some of the farthest outliers from the LZR. Although there are very few LVL systems with $\mu > 0.93$, the transition from a position of a significant outlier to a system that agrees with the LZR happens abruptly as gas-phase metallicity increases. Leo A, Leo P, and Leonessa have significantly less abundant atomic hydrogen disks ($\mu < 0.7$) compared to all other XMP systems, exhibiting $\mu$ values that are typical for dIrr systems.

We previously discussed how current SF can enhance the luminosity of a system and be responsible for its status as an outlier on the LZR; we now examine how dilution of the gas-phase metallicity through the accretion of atomic hydrogen could also be contributing to the scatter XMP systems exhibit on the LZR and MZR. \citet{mcquinn2020} explored how much gas would be needed to significantly dilute ISM metallicities to those observed in XMP galaxies. They reported that it would take a doubling of pristine \mbox{\ion{H}{1}} gas abundance in the star-forming region to lower 12+log(O/H) values by 0.3 dex. \autoref{fig:Z_fgas_delzr} demonstrates just how much more gas rich most XMP systems are overall in relation to non-XMP galaxies, and that the majority of them have gas-to-stellar mass ratios $> 4$, and dilution can easily be a contributing factor for lowering measured 12+log(O/H) values in a majority of XMP systems. However, Leonessa has a gas-to-stellar mass ratios $<2$, therefore we do not suspect recent accretion of pristine gas to be a large contributor to its gas-phase metallicity.

While XMP galaxies are rich in atomic \ion{H}{1}, SF is known to correlate with the presence of molecular hydrogen (H\textsubscript{2}). However, simulations have suggested that at very low metallicity stars may form directly from atomic \ion{H}{1} \citep{glover2012}. Recent observations of the \ion{H}{2} region in Leo P (at $\sim 3\%$ Z$_\odot$) have resulted in a detection of warm molecular hydrogen \citep{telford2025}. This implies that molecular gas plays a role in SF even in XMP galaxies, and that offsets of XMP systems from the LZR are therefore unlikely to be driven by unique SF processes.

\subsection{Nitrogen Content}
\label{ssec:nitro_diss}
When adding recently discovered XMP galaxies to the N/O-O/H plane shown in \autoref{fig:NO_Z}, a significant scatter appears. Several scenarios and mechanisms could possibly explain the large scatter in N/O at low-metallicity, such as length and efficiency of star formation, enrichment timescales, amplitude and mass loading of outflows, and primary nitrogen yields \cite[e.g.,][]{garnett1990,henry2000,meynet2002,vanzee2006,vincenzo2018}.

As stated in \S~\ref{ssec:no_oh}, at higher gas-phase metallicities oxygen and nitrogen will enrich on differing timescales. Combining this with the bursty nature of dwarf galaxies and the heightened amplitude of their starbursts, the SFR at the time of observation, along with the star formation history (SFH) of the observed system can effect the position of a system in the N/O-O/H parameter space. Current SF will be enriching the ISM with oxygen through massive stars going supernova, driving N/O values down, while at a later time low- and intermediate-mass stars created during a previous starburst will start enriching the ISM with nitrogen as they enter their AGB phase and shed their outer layers during periods of quiescence in the system, driving N/O values up \citep{garnett1990,vanzee2006}. Not only can the length of a starburst effect this N/O dynamic, the amplitude and mass-loading of the outflows can also have an impact, as the outflows will occur during oxygen enrichment from supernovae, but not during nitrogen enrichment from AGB stars. \citet{vanzee2006} investigated oxygen and nitrogen in dIrr galaxies (with $0.4 \le \mu \le 0.95$) and found that differential outflows (SNeII driven outflows rich in oxygen exclusively) will elevate N/O ratios by depleting oxygen abundances in the ISM through preferential oxygen expulsion.

Star formation efficiency (SFE) and SFH contribute to N/O scatter by establishing a pace and order by which the ISM is enriched, which is especially important when considering the XMP end of the abundance scale. \citet{berg2020} investigated a range of SFEs for chemical evolution models from \citet{henry2000}. They found that high SFRs during early SF epochs lead to a system having a lower N/O ratio at low-metallicity that would increase rapidly, while low SFRs at early epochs lead to a higher N/O ratio at lower metallicity that would increase slowly or in unison with O/H ratios. 

However, at the lowest metallicities massive metal-poor stars that have high rotational velocities could also be contributing to the nitrogen enhancement in their ISM. Metal-poor stars can experience better mixing in their outer shells due to decreased opacity increasing convection and dredge-up; they can also exhibit very large rotational velocities ($\gtrsim 100$ km s$^{-1}$) which contributes significantly to mixing of CNO produced nitrogen into the outer shells of the star \citep{vangioni2018,telford2021,brauer2024}. \citet{meynet2002} found an order of magnitude difference in nitrogen yields for rapidly-rotating ($\sim 300$ km s$^{-1}$) compared to non-rotating stars. \citet{schootemeijer2022} found that the percentage of massive stars that were rapid-rotators increased with decreasing metallicity, from $\sim 22$\% in galaxies with $0.5 Z_{\odot}$ to $\sim 31$\% for systems with $0.2 Z_{\odot}$. They concluded that extremely rapidly rotating stars were commonplace in low-metallicity dwarf galaxies. Their conclusion suggests that star-forming XMP galaxies may harbor significant populations of extremely rapidly rotating stars. Ultimately, a deeper investigation of nitrogen production in XMP systems is needed to determine what mechanisms are contributing the most to the enhanced N/O ratios observed in the ISM of these galaxies.

As previously stated \autoref{fig:NO_Z} displays XMP galaxies as having a wide range of N/O values including systems with enhanced nitrogen levels, similar to those found in systems at higher redshift. \citet{stiavelli2025} investigated N/O in 12 higher redshift systems ($3 \le z \le 6$) with metal poor ISMs ($ \rm 7.5 \le 12+log(O/H) \le 8.4$) and compared them to local dwarf galaxies and \ion{H}{2} regions from \citet{berg2019,berg2020}, respectively. They found that their systems also displayed a large range of rest-frame optical N/O values, with 3 systems above solar value \citep[$\rm log(N/O) \ge -0.85$;][]{asplund2021}. They concluded that Wolf-Rayet stars could be causing some enhancement, consistent with what is suspected in nearby XMP dIrr. They also found that dilution of the ISM with atomic \ion{H}{1} could be responsible for making the galaxies in their sample appear nitrogen enhanced, where a dilution factor of 5 corresponded to a 0.6 dex change in gas-phase oxygen abundance. This is also in accordance with the majority of extremely gas rich XMP systems in the comparison sample we have compiled. \citet{stiavelli2025} also included 8 systems at high redshift including GN-z11 \cite[$z=10.6$;][]{oesch2016}, but excluded these systems from their analysis for homogeneity, as the N/O calculation was based on rest-frame UV lines ([\ion{N}{3}]$\lambda\lambda 1747,1749$, and \ion{N}{4}]$\lambda1486$). \citet{senchyna2024} explored abundance ratios in GN-z11, finding that the suspected star forming environments responsible for producing the higher ionization states were more compact and had a higher critical density (of order $n_e \sim 10^6$ cm$^{-3}$) than the compact environments in nearby blue compact dwarfs where the collisionally ionized [\ion{O}{2}]$\lambda3727$ and [\ion{N}{2}]$\lambda6584$ is measured (densities of order $10^4$). These studies highlight that further investigations are needed to find the degeneracies between nitrogen enrichment over all redshift ranges, and that star forming environments at the highest redshifts are not a straightforward comparison to nearby star forming dwarfs, like XMP galaxies.

\section{Conclusion}
\label{sec:conc}
Leonessa (J1005+3722) was first introduced in a study aiming to identify XMP systems from the SDSS catalog based upon a color-color selection method \citep{senchyna2019}. Obtaining follow-up MMT spectroscopic observations of candidate SDSS systems \citet{senchyna2019} measured the gas-phase oxygen abundances of their sample confirming Leonessa as an XMP system. Their initial estimate for the distance to Leonessa (2.7 Mpc) made it a prime target for gathering deeper-field optical imaging, as well as optical and radio spectral observations. Here, we have presented follow-up HST optical imaging, HET optical spectroscopic, and GBT radio spectroscopic observations of Leonessa. From analysis of the reduced data we present improved estimates of Leonessa's properties:

\begin{itemize}
    \item The HST imaging of Leonessa shows an asymmetric morphology with an intense centralized star forming region, as well as an older stellar population that extends to large radii (see \autoref{fig:rgb}).
    \item We performed PSF photometry on the HST data for Leonessa. After quality cuts this resulted in a CMD of 194 recovered stellar sources. We identified and measured the TRGB magnitude, determining a distance to the system of $D= 15.86 \pm 0.78$ Mpc (see \autoref{fig:cmd_histy}). From the TRGB distance and SDSS apparent magnitude we measured an absolute $g$-band magnitude of M$_g = -11.91 \pm 0.11$ mag. Using the TRGB distance and SDSS $r$- and $i$-band magnitudes we calculated a stellar mass of M$_\star = (1.31 \pm 0.24) \times 10^6$ M$_{\odot}$.
    \item Supergalactic coordinates for Leonessa place it in an isolated ($\rm D_{NN} \sim 1.8$ Mpc) region within the \cch\ void.
    \item Detection of the [\ion{O}{3}]$\lambda 4363$ emission line from the HET optical spectra allowed us to make a direct method gas-phase oxygen abundance measurement for Leonessa. We measured the gas-phase metallicity in the galaxy to be 12+log(O/H) $= 7.32 \pm 0.04$. From the HET data we also measured an N/O ratio of log(N/O)$= -1.41 \pm 0.2$ using the [\ion{N}{2}] \W6584 spectral line (see \autoref{fig:HET_spec}).
    \item From the GBT L-band data we identified and measured the \mbox{\ion{H}{1}} 21cm radio spectral line, resulting in an integrated flux value of S$_{\rm HI} = 35.51 \pm 8.08$ mJy km s$^{-1}$ (see \autoref{fig:hi_21}). From the GBT data and TRGB distance we calculated an \mbox{\ion{H}{1}} gas mass of M$_{\rm HI} = (2.10 \pm 0.56) \times 10^6$ M$_{\odot}$. Combining our \mbox{\ion{H}{1}} gas mass and stellar mass measurements we determine that Leonessa is a gas-rich system with a gas-to-stellar mass ratio of M$_{\rm HI}$/M$_\star = 1.61 \pm 0.52$.
\end{itemize}

We used the derived properties (see \autoref{tab:properties}) from these observations to explore possible chemical evolution pathways for Leonessa. We compared properties of Leonessa to the well-established MZR, LZR and N/O-O/H trends found by \citet{berg2012} for LVL galaxies. In order to ascertain a better understanding of how Leonessa compares to other systems, we compiled a sample of XMP, and void, galaxies from the literature (see \autoref{tab:dwarf_gals}). From this analysis and comparison we find:

\begin{itemize}
    \item XMP galaxies show a large scatter away from the LZR, spanning a range of over 10 magnitudes in luminosity. All but two XMP systems (Leo A and Leo P) disagree with the LZR.
    \item A majority of XMP galaxies also disagree with the MZR. While the scatter away from the MZR is not as extreme as is shown on the LZR plane, it is still significant, covering over 4 dex in stellar mass.
    \item An anti-correlation is observed between gas-phase oxygen abundance and \mbox{\ion{H}{1}} gas-to-stellar mass ratios (see \autoref{fig:MZR_LZR}). All XMP galaxies are gas-rich, with a majority of them being extremely gas-rich with $\mu > 0.93$ (M$_{\rm HI}/$M$_\star \ge 10.0$). The farthest outliers from the LZR are also some of the most gas-rich systems.
    \item XMP galaxies also show a significant scatter above the N/O-O/H plateau at low metallicity, in contrast to findings in previous studies where metal-poor systems showed a smaller dispersion. However, previous studies lacked significant XMP populations due to observational constraints, making extrapolations into the XMP regime not robust. 
    \item Leonessa agrees with the MZR, but disagrees with the LZR. Bright O- and B-class stars in the central star forming region are enhancing the galactic $g$-band luminosity for Leonessa by $\sim 1.1$ mags. Removing the luminosity contribution from this population of stars places Leonessa in agreement with the LZR. \item Leonessa is a gas-rich system ((M$_{\rm HI}/$M$_\star = 1.61 \pm 0.52$). However, it is significantly less gas-rich than all but two other XMP systems, Leo A and Leo P. All three systems have comparatively similar $\mu$, and subsequently $y_{\rm eff}$ values. \\
\end{itemize}

Leonessa has likely followed a typical chemical evolution pathway similar both to field dwarfs and XMP systems like Leo A and Leo P, experiencing secular evolution, and whose present-day low gas phase oxygen abundance is likely due to poor metal retention as a result of stellar feedback fueling strong galactic outflows. Leonessa currently harbors an intense, compact star-forming region that is contributing to an increase in luminosity. We attribute the system's status as an outlier on the LZR to this enhanced luminosity from current SF. XMP galaxies in general are outliers on the well-established MZR and LZR from \citet{berg2012}, with only two systems agreeing with the LZR. Multiple scenarios and mechanisms are contributors to the large scatter we find for XMP systems on the LZR and MZR planes. Further investigation of the interplay between metal-enriched galactic outflows, pristine gas inflows, SF, and \mbox{\ion{H}{1}} gas content will offer us deeper insight on how these systems have chemically evolved over time.

\section{Acknowledgments}
We thank the referee for their insights, recommendations, and thoughtful inquiries that improved the quality of our work. This work was supported by NASA through Grant HST-GO-16048 (PI: McQuinn) from the Space Telescope Science Institute. This research has made use of the NASA/IPAC Extragalactic Database, which is funded by the National Aeronautics and Space Administration and operated by the California Institute of Technology. We would like to acknowledge that the HET is built on Indigenous land. Moreover, we would like to acknowledge and pay our respects to the Carrizo \& Comecrudo, Coahuiltecan, Caddo, Tonkawa, Comanche, Lipan Apache, Alabama-Coushatta, Kickapoo, Tigua Pueblo, and all the American Indian and Indigenous Peoples and communities who have been or have become a part of these lands and territories in Texas, here on Turtle Island. O.\ G.\ T.\ acknowledges support from a Carnegie-Princeton Fellowship through Princeton University and the Carnegie Observatories. The National Radio Astronomy Observatory and Green Bank Observatory are facilities of the U.S. National Science Foundation operated under cooperative agreement by Associated Universities, Inc. This research has made use of NASA's Astrophysics Data System Bibliographic Services. This research used NASA's Astrophysics Data System and the arXiv preprint server.

\facilities{HST (COS), HET (LRS2), GBO (GBT)} \software{This paper makes use of \texttt{AstroPy} \texttt{python} software package, we thank \citet{astropy1,astropy2,astropy3} for their contribution, and making their code publicly available; \textit{AstroDrizzle v3.2.1} (STScI), \texttt{jupyter} \citep{kluyver2016}, \texttt{PyNeb} \citep{luridiana2012,luridiana2015}, \texttt{SciPy} \citep{virtanen2020}, \texttt{AplPy} \citep{robitaille2012}, \texttt{python}.}

\startlongtable
\begin{deluxetable*}{lcccccccccc}
\setlength{\tabcolsep}{3.5pt} 
\tablecaption{Properties of Galaxies \label{tab:dwarf_gals}}
\tablewidth{0pt} 
\tablehead{\colhead{Name} & \colhead{RA} & \colhead{Dec} & \colhead{M$_{B,g}$} & \colhead{log} & \colhead{log} & \colhead{Dist} & \colhead{12+} & \colhead{log} & \colhead{Group} & \colhead{Ref.} \\ 
\colhead{} & \colhead{deg} & \colhead{deg} & \colhead{mag} & \colhead{($\rm \frac{M_\star}{M_{\odot}}$)} & \colhead{($\rm \frac{M_{HI}}{M_{\odot}}$)} & \colhead{Mpc} & \colhead{log(O/H)} & \colhead{(N/O)} & \colhead{} & \colhead{} \\
\colhead{(1)} & \colhead{(2)} & \colhead{(3)} & \colhead{(4)} & \colhead{(5)} & \colhead{(6)} & \colhead{(7)} & \colhead{(8)} & \colhead{(9)} & \colhead{(10)} & \colhead{(11)} }
\startdata
WLM & 0.49387 & -15.45756 & -13.5$\pm$0.05 & 7.19$\pm$0.34 & 7.84 & 0.98 & 7.77$\pm$0.1 & -1.46$\pm$0.05 & LVL & 1;20;52 \\
J0006+2413 & 1.704 & 24.21859 & -15.0$_g$ & 7.6 & \nodata & \nodata & 7.25 & \nodata & XMP & 7 \\
NGC 55 & 3.72333 & -39.19667 & -18.2$\pm$0.11 & 9.3$\pm$0.35 & \nodata & 2.11 & 8.05$\pm$0.1 & -1.26$\pm$0.05 & LVL & 1 \\
UM 240 & 6.28083 & 0.31278 & -16.1 & \nodata & \nodata & 46.5 & 7.89$\pm$0.08 & \nodata & void & 9;36 \\
UM 40 & 7.11083 & 5.00444 & -16.3 & \nodata & \nodata & 20.8 & 7.97$\pm$0.1 & \nodata & void & 9 \\
UM 38 & 7.21458 & 3.48972 & -15.7 & \nodata & 8.52 & 21.1 & 7.95$\pm$0.05 & \nodata & void & 9;31 \\
J0029-0025 & 7.45625 & -0.42775 & -14.4$_g$ & 6.3 & \nodata & 67.77 & 7.25 & \nodata & XMP & 7;26 \\
UGC 300 & 7.51708 & 3.51306 & -15.8 & \nodata & \nodata & 20.8 & 7.8$\pm$0.03 & \nodata & void & 9 \\
J0041-01b & 10.42958 & -1.98833 & -15.5 & \nodata & \nodata & 28.6 & 7.68$\pm$0.11 & \nodata & void & 9 \\
J0042+3247 & 10.63904 & 32.78917 & -17.2$_g$ & 8.0 & \nodata & \nodata & 7.22 & \nodata & XMP & 7 \\
IC 52 & 12.09917 & 4.09194 & -17.2 & \nodata & \nodata & 29.1 & 8.01$\pm$0.06 & \nodata & void & 9 \\
UGC 521 & 12.80042 & 12.02389 & -15.16$\pm$0.5 & 7.96$\pm$0.61 & \nodata & 10.9 & 7.67$\pm$0.05 & -1.61$\pm$0.07 & LVL & 1 \\
UGC 527 & 12.95792 & 3.10583 & -16.5 & \nodata & 8.84 & 28.8 & 7.84$\pm$0.14 & \nodata & void & 9;31 \\
UM 285 & 12.99458 & -1.67194 & -15.2 & \nodata & 8.27 & 28.0 & 7.77$\pm$0.06 & \nodata & void & 9;46 \\
UM 286 & 12.99875 & -0.48917 & -17.2 & \nodata & 9.26 & 24.1 & 8.08$\pm$0.1 & \nodata & void & 9;51 \\
SMC & 13.18333 & -72.82833 & -16.04$\pm$0.2 & \nodata & \nodata & 0.06 & 7.96$\pm$0.15 & -1.55$\pm$0.15 & LVL & 1 \\
J0057-0021 & 14.3025 & -0.36611 & -13.9 & \nodata & \nodata & 41.1 & 7.6$\pm$0.01 & \nodata & void & 9 \\
MCG-01-03-072 & 15.59542 & -4.50861 & -16.5 & \nodata & \nodata & 25.8 & 7.84$\pm$0.09 & \nodata & void & 9 \\
UGC 00668 & 16.20458 & 2.12528 & -13.61$\pm$0.14 & 7.14$\pm$0.37 & \nodata & 0.75 & 7.62$\pm$0.05 & -1.51$\pm$0.1 & LVL & 1 \\
J0106+2345 & 16.53821 & 23.75935 & -15.2$_g$ & 6.4 & \nodata & \nodata & 7.17 & \nodata & XMP & 7 \\
UGC 00685 & 16.845 & 16.68389 & -14.13$\pm$0.11 & 7.71$\pm$0.34 & 7.86 & 4.81 & 8.0$\pm$0.03 & -1.45$\pm$0.08 & LVL & 1;31;52 \\
UGC 695 & 16.94333 & 1.06361 & -15.2 & \nodata & 7.93 & 10.5 & 7.69$\pm$0.12 & \nodata & void & 9;36 \\
UGC 711 & 17.15375 & 1.64167 & -17.9 & \nodata & 9.57 & 29.0 & 7.91$\pm$0.08 & \nodata & void & 9;31 \\
NGC 428 & 18.23208 & 0.98167 & -19.1 & \nodata & \nodata & 17.6 & 8.12$\pm$0.06 & \nodata & void & 9 \\
J0113+0052 & 18.4185 & 0.87754 & -12.3$_g$ & 6.77 & 8.53 & 15.8 & 7.16 & -1.15 & XMP & 7;9;30;31 \\
MCG+00-04-049 & 18.58542 & 0.91722 & -14.3 & \nodata & 8.21 & 17.2 & 7.56$\pm$0.07 & \nodata & void & 9;31 \\
J0122+0048 & 20.67337 & 0.81167 & -16.16$\pm$0.06$_g$ & 6.59$\pm$0.2 & \nodata & \nodata & 7.2$\pm$0.17 & \nodata & XMP & 4 \\
UM 323 & 21.69375 & -0.64583 & -16.2 & 7.94 & 8.55 & 27.8 & 7.96$\pm$0.04 & -1.5$\pm$-0.73 & void & 9;50 \\
UGC 1056 & 22.19833 & 16.68917 & -15.09$\pm$0.52 & 8.62$\pm$0.61 & \nodata & 10.32 & 7.97$\pm$0.06 & -1.49$\pm$0.02 & LVL & 1 \\
NGC 625 & 23.76625 & -41.43722 & -16.28$\pm$0.4 & 8.8$\pm$0.34 & 7.9 & 4.02 & 8.08$\pm$0.12 & -1.25$\pm$0.03 & LVL & 1;25;52 \\
J0137+1810 & 24.47683 & 18.17666 & -16.08$\pm$0.17$_g$ & 6.29$\pm$0.2 & \nodata & \nodata & 7.32$\pm$0.22 & \nodata & XMP & 4 \\
UGC 1176 & 25.04125 & 15.90556 & -15.48$\pm$0.93 & 8.48$\pm$0.98 & \nodata & 9.04 & 7.97$\pm$0.05 & -1.4$\pm$0.02 & LVL & 1 \\
J0141+2124 & 25.38842 & 21.41398 & -17.72$\pm$0.19$_g$ & 7.22$\pm$0.2 & \nodata & \nodata & 7.35$\pm$0.3 & \nodata & XMP & 4 \\
J0153+0104 & 28.29983 & 1.07781 & -16.23$\pm$0.06$_g$ & 6.57$\pm$0.2 & \nodata & \nodata & 7.35$\pm$0.1 & \nodata & XMP & 4 \\
NGC 784 & 30.32042 & 28.83583 & -16.5$\pm$0.12 & 8.48$\pm$0.36 & 8.75 & 5.37 & 7.97$\pm$0.06 & -1.54$\pm$0.1 & LVL & 1;31;52 \\
J0216+1715 & 34.09903 & 17.2572 & \nodata & 6.8 & \nodata & \nodata & 7.35$\pm$0.01 & \nodata & XMP & 16 \\
J0222-0935 & 35.66062 & -9.59311 & -17.3$_g$ & 8.04 & \nodata & \nodata & 7.29$\pm$0.33 & \nodata & XMP & 3 \\
AGC 124609 & 42.36833 & 34.55806 & -14.1 & \nodata & \nodata & 25.0 & 7.89$\pm$0.02 & \nodata & void & 15 \\
J0314-0108 & 48.60879 & -1.14626 & -18.9$_g$ & 7.3 & \nodata & \nodata & 7.26 & -0.58 & XMP & 7 \\
UGC 2716 & 51.03375 & 17.75417 & -15.31$\pm$0.5 & 8.13$\pm$0.61 & \nodata & 6.2 & 7.97$\pm$0.05 & -1.47$\pm$0.02 & LVL & 1 \\
SBS 0335-052W & 54.43358 & -5.0445 & -14.7 & 6.9 & 8.68 & 58.0 & 7.22$\pm$0.07 & -1.59$\pm$0.03 & XMP & 6;21;30;44 \\
NGC 1705 & 73.55708 & -53.36139 & -15.77$\pm$0.52 & 8.19$\pm$0.6 & \nodata & 5.11 & 8.21$\pm$0.05 & -1.75$\pm$0.06 & LVL & 1 \\
LMC & 80.89417 & -69.75367 & -17.68$\pm$0.05 & \nodata & \nodata & 0.05 & 8.26$\pm$0.15 & -1.3$\pm$0.2 & LVL & 1 \\
UGC 3476 & 97.62175 & 33.302 & -16.02 & \nodata & \nodata & 9.84 & 7.71$\pm$0.07 & \nodata & void & 14 \\
UGC 3501 & 99.66 & 49.25833 & 13.31 & \nodata & \nodata & 10.07 & 7.56$\pm$0.07 & \nodata & void & 14 \\
UGC 3600 & 103.91667 & 39.09522 & -14.22 & \nodata & \nodata & 9.3 & 7.69$\pm$0.16 & \nodata & void & 14 \\
UGC 3672 & 106.61483 & 30.32206 & -15.43 & \nodata & \nodata & 16.93 & 7.9$\pm$0.13 & \nodata & void & 14 \\
UGC 3698 & 107.32 & 44.38 & -14.92 & \nodata & \nodata & 9.6 & 7.74$\pm$0.08 & \nodata & void & 14 \\
NGC 2337 & 107.55667 & 44.45694 & -16.85 & \nodata & \nodata & 9.79 & 8.38$\pm$0.06 & \nodata & void & 14 \\
UGC 3817 & 110.68533 & 45.10853 & -14.44 & \nodata & \nodata & 9.82 & 7.96$\pm$0.13 & \nodata & void & 14 \\
UGC 3860 & 112.07167 & 40.77028 & -14.5 & \nodata & \nodata & 7.81 & 7.79$\pm$0.08 & \nodata & void & 14 \\
NGC 2366 & 112.20667 & 69.20889 & -15.95$\pm$0.11 & 8.15$\pm$0.35 & 8.7 & 3.28 & 7.91$\pm$0.05 & -1.17$\pm$0.26 & LVL & 1;25;52 \\
UGC 3876 & 112.32287 & 27.90053 & -17.31 & \nodata & \nodata & 15.01 & 7.95$\pm$0.16 & \nodata & void & 14 \\
J0730+4109 & 112.74542 & 41.16661 & -14.58 & \nodata & \nodata & 15.7 & 8.04$\pm$0.05 & \nodata & void & 14 \\
J0739+4434 & 114.76079 & 44.57394 & -14.64$\pm$0.09$_g$ & 5.8$\pm$0.2 & \nodata & \nodata & 7.35$\pm$0.28 & \nodata & XMP & 4 \\
MCG 9-13-56 & 116.88375 & 51.19139 & -15.18 & \nodata & \nodata & 10.0 & 7.7$\pm$0.03 & \nodata & void & 14 \\
J0808+3244 & 122.23475 & 32.73867 & -18.9$_g$ & 8.9 & \nodata & \nodata & 7.31 & -1.18 & XMP & 7 \\
MCG 7-17-19 & 122.40042 & 41.59444 & -15.39 & \nodata & \nodata & 13.37 & 7.85$\pm$0.12 & \nodata & void & 14 \\
J0810+1837 & 122.62771 & 18.61781 & -13.59 & \nodata & \nodata & 23.05 & 7.83$\pm$0.2 & \nodata & void & 14 \\
J0811+4730 & 122.96717 & 47.50729 & -17.3$_g$ & 5.8 & \nodata & \nodata & 6.97 & -1.54$\pm$0.04 & XMP & 7 \\
J0812+4836 & 123.16471 & 48.61261 & -13.08 & \nodata & \nodata & 11.05 & 7.27$\pm$0.06 & \nodata & XMP & 14 \\
NGC 2537 & 123.31083 & 45.99167 & -17.14$\pm$0.5 & 9.1$\pm$0.61 & \nodata & 6.88 & 8.14$\pm$0.13 & -1.04$\pm$0.04 & LVL & 1 \\
UGC 4278 & 123.49542 & 45.74361 & -16.36$\pm$0.5 & 8.5$\pm$0.6 & \nodata & 7.6 & 7.69$\pm$0.05 & -1.6$\pm$0.03 & LVL & 1 \\
NGC 2541 & 123.66742 & 49.06169 & -18.28 & \nodata & \nodata & 12.0 & 8.23$\pm$0.03 & \nodata & void & 14 \\
UGC 4305 & 124.7875 & 70.72444 & -16.11$\pm$0.12 & 8.48$\pm$0.31 & 9.14 & 3.47 & 7.92$\pm$0.1 & -1.52$\pm$0.11 & LVL & 1;42;52 \\
NGC 2552 & 124.83 & 50.01028 & -16.72$\pm$0.5 & 8.69$\pm$0.61 & \nodata & 7.7 & 8.15$\pm$0.05 & -1.16$\pm$0.02 & LVL & 1 \\
AGC 189201 & 125.85667 & 17.91583 & -12.76 & \nodata & \nodata & 23.4 & 7.31$\pm$0.04 & \nodata & XMP & 15 \\
HS 0822+3542 & 126.48096 & 35.54219 & -12.97 & \nodata & \nodata & 13.49 & 7.44$\pm$0.02 & \nodata & void & 14 \\
UGC 4393 & 126.51833 & 45.96778 & -17.67$\pm$0.85 & 9.43$\pm$0.86 & \nodata & 16.8 & 8.02$\pm$0.05 & -1.15$\pm$0.08 & LVL & 1 \\
UGC 4459 & 128.53167 & 66.1775 & -12.93$\pm$0.12 & 7.15$\pm$0.81 & 7.85 & 3.68 & 7.82$\pm$0.09 & -1.32$\pm$0.17 & LVL & 1;33;52 \\
J0834+5905 & 128.65496 & 59.09333 & -15.6$_g$ & 6.7 & \nodata & \nodata & 7.22 & \nodata & XMP & 7 \\
UGC 4483 & 129.26375 & 69.77611 & -12.71$\pm$0.19 & 6.42$\pm$0.4 & 7.57 & 3.58 & 7.56$\pm$0.03 & -1.57$\pm$0.07 & LVL & 1;33;52 \\
J0852+1350 & 133.14062 & 13.84119 & -14.58 & \nodata & \nodata & 23.08 & 7.28$\pm$0.1 & \nodata & XMP & 14 \\
UGC 4704 & 134.75117 & 39.20992 & -15.66 & \nodata & \nodata & 11.74 & 7.97$\pm$0.19 & \nodata & void & 14 \\
J0859+3923 & 134.94554 & 39.3849 & -12.8$_g$ & 5.4 & 7.3 & 10.2 & 7.35 & -1.07 & XMP & 7;48 \\
KKH 46 & 137.15225 & 5.29078 & -12.99 & \nodata & \nodata & 10.0 & 7.68$\pm$0.03 & \nodata & void & 14 \\
J0911+3135 & 137.99758 & 31.59331 & -12.9$_g$ & 5.6 & \nodata & \nodata & 7.32 & \nodata & XMP & 7 \\
J0921+4038 & 140.40229 & 40.64829 & -15.9$_g$ & 6.6 & \nodata & \nodata & 7.32 & \nodata & XMP & 7 \\
J0926+3343 & 141.53937 & 33.71781 & -12.91 & \nodata & \nodata & 10.63 & 7.12$\pm$0.02 & \nodata & XMP & 14 \\
IZw18C NW & 143.50846 & 55.24107 & -15.3$_g$ & 6.4 & 8.08 & 18.2 & 7.16$\pm$0.01 & -1.56$\pm$0.02 & XMP & 6;17;34;38;39 \\
IZw18A SE & 143.50996 & 55.23978 & -15.3$_g$ & 7.24 & 8.0 & 18.2 & 7.19$\pm$0.02 & -1.53$\pm$0.04 & XMP & 6;17;34;38;39 \\
KISSB 23 & 145.05279 & 29.59147 & -14.08 & \nodata & \nodata & 10.21 & 7.6$\pm$0.05 & \nodata & void & 14 \\
UGC 5139 & 145.125 & 71.18472 & -14.42$\pm$0.12 & 7.39$\pm$0.51 & 8.31 & 4.02 & 7.92$\pm$0.05 & -1.56$\pm$0.05 & LVL & 1;47;52 \\
Leoncino & 145.885 & 33.44944 & -10.63 & 5.86 & 7.26 & 12.1 & 7.06$\pm$0.03 & -1.38$\pm$0.06 & XMP & 13 \\
IC 559 & 146.1825 & 9.615 & -14.12$\pm$0.5 & 7.86$\pm$0.61 & 7.4 & 4.9 & 8.07$\pm$0.1 & -1.47$\pm$0.05 & LVL & 1;31;43 \\
J0945+3835 & 146.33146 & 38.59803 & -16.82$\pm$0.06$_g$ & 7.23$\pm$0.2 & \nodata & \nodata & 7.33$\pm$0.38 & \nodata & XMP & 4 \\
J0947+4138 & 146.82646 & 41.63789 & -13.92 & \nodata & \nodata & 22.56 & 7.77$\pm$0.03 & \nodata & void & 14 \\
UGC 5272B & 147.58121 & 31.45619 & -12.6 & \nodata & \nodata & 10.27 & 7.67$\pm$0.13 & \nodata & void & 14 \\
UGC 5272 & 147.59333 & 31.48778 & -14.98$\pm$0.55 & 8.0$\pm$0.64 & \nodata & 7.11 & 7.87$\pm$0.02 & -1.59$\pm$0.02 & LVL & 1 \\
J0955+6442 & 148.88104 & 64.71391 & -12.9$_g$ & 5.1 & \nodata & \nodata & 7.16 & \nodata & XMP & 7 \\
DDO 68 & 149.19083 & 28.82556 & -15.83$\pm$0.55 & 7.97$\pm$0.45 & 8.94 & 12.1 & 7.2$\pm$0.05 & -1.6$\pm$0.08 & XMP & 1;31 \\
DDO 68B & 149.19875 & 28.82306 & -14.01 & \nodata & \nodata & 12.75 & 7.14$\pm$0.04 & \nodata & XMP & 14 \\
J0959+4626 & 149.774 & 46.44736 & -12.2$_g$ & 5.2 & \nodata & \nodata & 7.26 & -1.48$\pm$0.46 & XMP & 7;28 \\
Leo A & 149.85333 & 30.74694 & -10.91$\pm$0.26 & 6.58$\pm$0.43 & 6.8 & 0.74 & 7.3$\pm$0.05 & -1.5$\pm$0.1 & XMP & 1;33;52 \\
Sex B & 149.99896 & 5.33267 & -13.54$\pm$0.16 & 7.49$\pm$0.38 & 7.68 & 1.43 & 7.84$\pm$0.05 & -1.46$\pm$0.06 & LVL & 1;31;52 \\
NGC 3109 & 150.79025 & -26.158 & -15.66 & 8.28 & 8.5 & 1.34 & 7.77$\pm$0.07 & -1.33$\pm$0.15 & LVL & 1;24 \\
UGC 5427 & 151.17104 & 29.36533 & -15.15 & \nodata & \nodata & 9.79 & 7.99$\pm$0.09 & \nodata & void & 14 \\
Leonessa & 151.30064 & 37.3671 & -11.92$\pm$0.09$_g$ & 6.12$\pm$0.08 & 6.32 & 15.86 & 7.32$\pm$0.04 & -1.41$\pm$0.2 & XMP & \nodata \\
UGC 5423 & 151.3775 & 70.36444 & -13.77$\pm$0.38 & 7.77$\pm$0.85 & \nodata & 5.27 & 7.78$\pm$0.05 & -1.33$\pm$0.04 & LVL & 1 \\
UGC 5464 & 152.03208 & 29.54289 & -15.47 & \nodata & \nodata & 16.9 & 7.92$\pm$0.26 & \nodata & void & 14 \\
Sex A & 152.75292 & -4.69361 & -13.62$\pm$0.19 & 7.08$\pm$0.39 & 7.89 & 1.45 & 7.54$\pm$0.06 & -1.54$\pm$0.13 & LVL & 1;20;52 \\
HS 1013+3809 & 154.10208 & 37.91278 & -15.48 & \nodata & \nodata & 19.3 & 7.59$\pm$0.04 & -1.28$\pm$0.16 & void & 14 \\
Leo P & 155.43792 & 18.08806 & -9.1$\pm$0.1 & 5.75 & 5.91 & 1.62 & 7.17$\pm$0.04 & -1.36$\pm$0.04 & XMP & 12;31;36;45 \\
UGC 5666 & 157.14708 & 68.43139 & -16.81$\pm$0.13 & 8.62$\pm$0.3 & 9.21 & 3.93 & 7.93$\pm$0.05 & -1.45$\pm$0.08 & LVL & 1;35;52 \\
J1036+2036 & 159.16446 & 20.60439 & -15.61$_g$ & 6.82 & \nodata & \nodata & 7.32$\pm$0.34 & \nodata & XMP & 3 \\
UGC 5797 & 159.855 & 1.71806 & -14.56$\pm$0.51 & 7.75$\pm$0.61 & \nodata & 6.8 & 7.96$\pm$0.06 & -1.35$\pm$0.06 & LVL & 1 \\
Little Cub & 161.17775 & 63.10056 & -11.21 & 5.37$\pm$0.14 & \nodata & 20.6 & 7.13$\pm$0.08 & \nodata & XMP & 5 \\
J1046+4047 & 161.53849 & 40.78532 & \nodata & 6.6 & \nodata & \nodata & 7.13$\pm$0.01 & \nodata & XMP & 16 \\
UGC 5923 & 162.28167 & 6.9175 & -14.7$\pm$0.5 & 8.29$\pm$0.61 & \nodata & 7.2 & 7.79$\pm$0.14 & -1.3$\pm$0.04 & LVL & 1 \\
J1053+4713 & 163.33887 & 47.22246 & -16.9$_g$ & 7.3 & \nodata & \nodata & 7.24 & \nodata & XMP & 7 \\
J1109+2007 & 167.28971 & 20.12493 & -14.1$_g$ & 6.1 & 7.43 & 17.7 & 7.19 & -0.86 & XMP & 7;22;54 \\
J1119+0935 & 169.86704 & 9.59563 & -14.0$_g$ & 5.9 & 7.83 & \nodata & 7.27 & -1.02 & XMP & 7;28 \\
UGC 6541 & 173.37 & 49.23972 & -13.51$\pm$0.06 & 7.3$\pm$0.34 & \nodata & 3.89 & 7.82$\pm$0.06 & -1.45$\pm$0.13 & LVL & 1 \\
NGC 3738 & 173.9525 & 54.52472 & -16.51$\pm$0.61 & 8.5$\pm$0.7 & 8.14 & 5.3 & 8.04$\pm$0.06 & -1.34$\pm$0.02 & LVL & 1;35;52 \\
NGC 3741 & 174.02417 & 45.28639 & -13.18$\pm$0.22 & 7.05$\pm$0.4 & 8.01 & 3.22 & 7.68$\pm$0.03 & -1.61$\pm$0.03 & LVL & 1;47;52 \\
UGC 6817 & 177.72542 & 38.88083 & -13.7$\pm$0.34 & 6.97$\pm$0.48 & 7.7 & 2.65 & 7.53$\pm$0.02 & -1.53$\pm$0.03 & LVL & 1;35;52 \\
J1206+5007 & 181.53554 & 50.12255 & -15.3$_g$ & 6.7 & \nodata & \nodata & 7.2 & \nodata & XMP & 7 \\
NGC 4163 & 183.03833 & 36.16944 & -13.65$\pm$0.12 & 7.61$\pm$0.35 & 7.26 & 2.99 & 7.56$\pm$0.14 & -1.49$\pm$0.06 & LVL & 1;33;52 \\
NGC 4214 & 183.9125 & 36.32639 & -17.15$\pm$0.1 & 8.83$\pm$0.36 & \nodata & 3.03 & 8.22$\pm$0.05 & -1.32$\pm$0.03 & LVL & 1 \\
CGCG 269-049 & 183.94667 & 52.38806 & -10.83$\pm$0.14 & 5.9$\pm$0.36 & 7.48 & 4.61 & 7.47$\pm$0.02 & -1.57$\pm$0.03 & LVL & 1;48;52 \\
J1220+4915 & 185.21504 & 49.26541 & -12.94$\pm$0.04$_g$ & 5.32$\pm$0.2 & \nodata & \nodata & 7.2$\pm$0.1 & \nodata & XMP & 4 \\
J1226+0952 & 186.73217 & 9.8823 & -14.2$_g$ & 7.38 & 8.02 & 16.18 & 7.21 & -0.86 & XMP & 7;22;54 \\
UGC 7577 & 186.92042 & 43.49556 & -14.12$\pm$0.14 & 7.5$\pm$0.36 & 7.54 & 2.61 & 7.97$\pm$0.06 & -1.37$\pm$0.04 & LVL & 1;33;52 \\
NGC 4449 & 187.04208 & 44.09194 & -18.02$\pm$0.34 & 9.25$\pm$0.48 & 9.16 & 4.27 & 8.32$\pm$0.03 & -1.39$\pm$0.02 & LVL & 1;19;52 \\
UGC 7605 & 187.16042 & 35.71611 & -13.49$\pm$0.66 & 7.12$\pm$0.73 & \nodata & 4.43 & 7.66$\pm$0.11 & -1.54$\pm$0.11 & LVL & 1 \\
J1234+3901 & 188.56542 & 39.02122 & -17.6$_g$ & 7.4 & \nodata & \nodata & 7.03 & -1.18 & XMP & 7 \\
UGCA 292 & 189.66958 & 32.76139 & -11.52$\pm$0.23 & 6.68$\pm$0.4 & 7.75 & 3.85 & 7.3$\pm$0.03 & -1.45$\pm$0.07 & XMP & 1;47;52 \\
NGC 4656 & 190.99042 & 32.16806 & -18.75$\pm$0.51 & 9.04$\pm$0.61 & \nodata & 8.6 & 8.09$\pm$0.05 & -1.66$\pm$0.14 & LVL & 1 \\
UGC 8091 & 194.66583 & 14.21833 & -11.76$\pm$0.07 & 6.55$\pm$0.36 & 7.0 & 2.19 & 7.65$\pm$0.06 & -1.51$\pm$0.07 & LVL & 1;33;52 \\
J1258+1413 & 194.6675 & 14.21689 & -12.5$_g$ & 5.8 & \nodata & \nodata & 7.27 & -1.05 & XMP & 7 \\
UGC 8201 & 196.60208 & 67.70778 & -15.17$\pm$0.44 & 7.82$\pm$0.6 & 8.08 & 4.83 & 7.8$\pm$0.06 & -1.77$\pm$0.07 & LVL & 1;36 \\
UGC 8508 & 202.68375 & 54.91111 & -13.03$\pm$0.07 & 7.0$\pm$0.35 & 7.29 & 2.67 & 7.76$\pm$0.07 & -1.6$\pm$0.07 & LVL & 1;36;52 \\
UGC 8638 & 204.83 & 24.77667 & -13.77$\pm$0.4 & 7.57$\pm$0.53 & 7.3 & 4.29 & 7.94$\pm$0.05 & -1.55$\pm$0.04 & LVL & 1;31;52 \\
UGC 8651 & 204.97458 & 40.74056 & -13.13$\pm$0.11 & 7.19$\pm$0.36 & 7.49 & 3.1 & 7.85$\pm$0.04 & -1.6$\pm$0.09 & LVL & 1;47;52 \\
UGC 8837 & 208.69042 & 53.90083 & -15.92$\pm$0.51 & 8.41$\pm$0.61 & \nodata & 8.3 & 7.87$\pm$0.07 & -1.43$\pm$0.03 & LVL & 1 \\
VGS 38b & 210.10417 & 55.22167 & -14.5 & \nodata & 8.1 & 55.0 & 7.94 & -1.48 & void & 11;37 \\
VGS 38a & 210.13333 & 55.24611 & -15.9 & 7.65$\pm$0.05 & 7.9 & 54.0 & 8.01 & -1.45 & void & 11;27;37 \\
VGS 39a & 210.8375 & 32.75833 & -14.9 & \nodata & 8.4 & \nodata & 7.54 & -1.46 & void & 11 \\
NGC 5477 & 211.38792 & 54.46083 & -15.22$\pm$0.51 & 8.15$\pm$0.61 & \nodata & 7.7 & 7.95$\pm$0.02 & -1.56$\pm$0.02 & LVL & 1 \\
J1414-0208 & 213.72554 & -2.13971 & -14.2$_g$ & 6.1 & 7.79 & \nodata & 7.3 & -1.65 & XMP & 7;53 \\
UGC 9128 & 213.98667 & 23.05611 & -12.12$\pm$0.18 & 6.59$\pm$0.39 & 7.21 & 2.3 & 7.75$\pm$0.05 & -1.8$\pm$0.12 & LVL & 1;47;52 \\
UGC 9240 & 216.17958 & 44.52694 & -13.89$\pm$0.09 & 7.47$\pm$0.36 & 7.85 & 2.83 & 7.95$\pm$0.03 & -1.6$\pm$0.06 & LVL & 1;32;52 \\
UGC 9497 & 221.05333 & 42.62889 & -14.4$\pm$0.05$_g$ & 7.08$\pm$0.2 & 8.28 & 11.04 & 7.32$\pm$0.19 & -1.29$\pm$0.18 & XMP & 4;22;28 \\
J1631+4426 & 247.80933 & 44.43456 & \nodata & 5.89$\pm$0.1 & \nodata & \nodata & 7.14$\pm$0.03 & -1.73$\pm$0.16 & XMP & 10;49 \\
J2104-0035 & 316.23042 & -0.58944 & -14.11 & 7.0 & 8.47 & 24.97 & 7.26$\pm$0.03 & \nodata & XMP & 2;23;26 \\
IC 5152 & 330.67333 & -51.29444 & -15.47$\pm$0.03 & 8.17$\pm$0.34 & 7.95 & 1.96 & 7.92$\pm$0.07 & -1.05$\pm$0.12 & LVL & 1;35;52 \\
J2229+2725 & 337.38829 & 27.42378 & -16.39$\pm$0.06$_g$ & 6.96 & \nodata & 347.0 & 7.09$\pm$0.03 & -1.54$\pm$0.18 & XMP & 8 \\
J2313+2935 & 348.48918 & 29.58686 & \nodata & 6.7 & \nodata & \nodata & 7.17$\pm$0.13 & \nodata & XMP & 16 \\
UGCA 442 & 355.94292 & -31.95694 & -14.34$\pm$0.62 & 7.56$\pm$0.71 & \nodata & 4.27 & 7.72$\pm$0.03 & -1.41$\pm$0.02 & LVL & 1 \\
J2354-0005 & 358.65542 & -0.08389 & -14.0 & \nodata & 8.14 & 34.1 & 7.36$\pm$0.13 & \nodata & void & 9;29;40 \\
\enddata
\tablecomments{Column 1 is the galaxy name. Column 2 and Column 3 report the R.A. and the Decl. (J2000), respectively. Column 4 is the luminosity (subscripts of $B$ and $g$ indicate absolute $B$- or $g$-band mag, respectively). Column 5 is log of the stellar mass. Column 6 is log of the \mbox{\ion{H}{1}} gas mass. Column 7 is the distance, given in units of Mpc. Column 8 is the gas-phase oxygen abundance measurement, determined using the `direct' method. Column 9 is the log of the N/O ratio. Column 10 is the respective group that we have put the galaxy into for the purposes of this study. Please see the References for details in column 11 (Ref.). \\
A majority of the values in \autoref{tab:dwarf_gals} are taken from the first source cited in Column 16. When additional values for a system are added, those citations are listed after the first source in Column 16; the References for \autoref{tab:dwarf_gals} reflect what specific values are obtained from the corresponding citation. This table gives the values for the sample used to make the MZR, LZR, and log(N/O) figures.\\
Table 3 is published in machine-readable format in the online journal.}
\tablerefs{(1) \citet{berg2012}; (2) \citet{ekta2010}; (3) \citet{guseva2015}; (4) \citet{guseva2017}; (5) \citet{hsyu2018}; (6) \citet{izotov2018}; (7) \citet{izotov2019}; (8) \citet{izotov2021}; (9) \citet{kniazev2018}; (10) \citet{kojima2020}; (11) \citet{kreckel2015}; (12) \citet{mcquinn2015}; (13) \citet{mcquinn2020}; (14) \citet{pustilnik2016}; (15) \citet{pustilnik2021}; (16) \citet{yang2017}; (17) \citet{annibali2013}: Stellar mass; (18) \citet{aver2022}: O/H, N/O; (19) \citet{bajaja1994}: \mbox{\ion{H}{1}}; (20) \citet{barnes2004}: \mbox{\ion{H}{1}}; (21) \citet{berg2022}: Dist; (22) \citet{cattorini2023}: Dist; (23) \citet{chang2015}: Stellar mass; (24) \citet{cook2014}: M$_B$, Stellar Mass; (25) \citet{courtois2009}: \mbox{\ion{H}{1}}; (26) \citet{dalya2018}: Dist, Abs. mag; (27) \citet{dominguez2022}: Stellar mass; (28) \citet{douglass2018}: N/O; (29) \citet{durbala2020}: Stellar mass; (30) \citet{filho2013}: \mbox{\ion{H}{1}}; (31) \citet{haynes2018}: \mbox{\ion{H}{1}}; (32) \citet{hogg2007}: \mbox{\ion{H}{1}}; (33) \citet{huchtmeier1986}: \mbox{\ion{H}{1}}; (34) \citet{izotov1998}: N/O; (35) \citet{karachentsev2004}: \mbox{\ion{H}{1}}; (36) \citet{karachentsev2013}: Abs. mag; (37) \citet{kreckel2011a}: Dist.; (38) \citet{lelli2012}: \mbox{\ion{H}{1}}; (39) \citet{lelli2014}: Dist.; (40) \citet{pustilnik2013}: \mbox{\ion{H}{1}}; (41) \citet{pustilnik2020a}: \mbox{\ion{H}{1}}; (42) \citet{rots1980}: \mbox{\ion{H}{1}}; (43) \citet{sabbi2018}: Dist.; (44) \citet{schneider2016}; (45) \citet{skillman1993}: N/O; (46) \citet{smoker2000}: \mbox{\ion{H}{1}}; (47) \citet{springob2005}: \mbox{\ion{H}{1}}; (48) \citet{thuan2016}: \mbox{\ion{H}{1}}; (49) \citet{thuan2022}: \mbox{\ion{H}{1}}; (50) \citet{trujillo2020}: Stellar mass; (51) \citet{tully2009}: \mbox{\ion{H}{1}}; (52) \citet{tully2013}: Dist.; (53) \citet{vandriel2016}: \mbox{\ion{H}{1}}; (54) \citet{yu2022}: \mbox{\ion{H}{1}}}
\label{tab:dwarf_gals}
\end{deluxetable*}

\bibliography{bib}

\begin{thebibliography}{}
\expandafter\ifx\csname natexlab\endcsname\relax\def\natexlab#1{#1}\fi
\providecommand{\url}[1]{\href{#1}{#1}}
\providecommand{\dodoi}[1]{doi:~\href{http://doi.org/#1}{\nolinkurl{#1}}}
\providecommand{\doeprint}[1]{\href{http://ascl.net/#1}{\nolinkurl{http://ascl.net/#1}}}
\providecommand{\doarXiv}[1]{\href{https://arxiv.org/abs/#1}{\nolinkurl{https://arxiv.org/abs/#1}}}

\bibitem[{{Ahumada} {et~al.}(2020){Ahumada}, {Allende Prieto}, {Almeida},
  {Anders}, {Anderson}, {Andrews}, {Anguiano}, {Arcodia}, {Armengaud},
  {Aubert}, {Avila}, {Avila-Reese}, {Badenes}, {Balland}, {Barger},
  {Barrera-Ballesteros}, {Basu}, {Bautista}, {Beaton}, {Beers}, {Benavides},
  {Bender}, {Bernardi}, {Bershady}, {Beutler}, {Bidin}, {Bird}, {Bizyaev},
  {Blanc}, {Blanton}, {Boquien}, {Borissova}, {Bovy}, {Brandt}, {Brinkmann},
  {Brownstein}, {Bundy}, {Bureau}, {Burgasser}, {Burtin}, {Cano-D{\'\i}az},
  {Capasso}, {Cappellari}, {Carrera}, {Chabanier}, {Chaplin}, {Chapman},
  {Cherinka}, {Chiappini}, {Doohyun Choi}, {Chojnowski}, {Chung}, {Clerc},
  {Coffey}, {Comerford}, {Comparat}, {da Costa}, {Cousinou}, {Covey}, {Crane},
  {Cunha}, {Ilha}, {Dai}, {Damsted}, {Darling}, {Davidson}, {Davies}, {Dawson},
  {De}, {de la Macorra}, {De Lee}, {Queiroz}, {Deconto Machado}, {de la Torre},
  {Dell'Agli}, {du Mas des Bourboux}, {Diamond-Stanic}, {Dillon}, {Donor},
  {Drory}, {Duckworth}, {Dwelly}, {Ebelke}, {Eftekharzadeh}, {Davis Eigenbrot},
  {Elsworth}, {Eracleous}, {Erfanianfar}, {Escoffier}, {Fan}, {Farr},
  {Fern{\'a}ndez-Trincado}, {Feuillet}, {Finoguenov}, {Fofie},
  {Fraser-McKelvie}, {Frinchaboy}, {Fromenteau}, {Fu}, {Galbany}, {Garcia},
  {Garc{\'\i}a-Hern{\'a}ndez}, {Garma Oehmichen}, {Ge}, {Geimba Maia},
  {Geisler}, {Gelfand}, {Goddy}, {Gonzalez-Perez}, {Grabowski}, {Green},
  {Grier}, {Guo}, {Guy}, {Harding}, {Hasselquist}, {Hawken}, {Hayes}, {Hearty},
  {Hekker}, {Hogg}, {Holtzman}, {Horta}, {Hou}, {Hsieh}, {Huber}, {Hunt}, {Ider
  Chitham}, {Imig}, {Jaber}, {Jimenez Angel}, {Johnson}, {Jones},
  {J{\"o}nsson}, {Jullo}, {Kim}, {Kinemuchi}, {Kirkpatrick}, {Kite}, {Klaene},
  {Kneib}, {Kollmeier}, {Kong}, {Kounkel}, {Krishnarao}, {Lacerna}, {Lan},
  {Lane}, {Law}, {Le Goff}, {Leung}, {Lewis}, {Li}, {Lian}, {Lin}, {Long},
  {Longa-Pe{\~n}a}, {Lundgren}, {Lyke}, {Mackereth}, {MacLeod}, {Majewski},
  {Manchado}, {Maraston}, {Martini}, {Masseron}, {Masters}, {Mathur},
  {McDermid}, {Merloni}, {Merrifield}, {M{\'e}sz{\'a}ros}, {Miglio}, {Minniti},
  {Minsley}, {Miyaji}, {Mohammad}, {Mosser}, {Mueller}, {Muna},
  {Mu{\~n}oz-Guti{\'e}rrez}, {Myers}, {Nadathur}, {Nair}, {Nandra}, {Correa do
  Nascimento}, {Nevin}, {Newman}, {Nidever}, {Nitschelm}, {Noterdaeme},
  {O'Connell}, {Olmstead}, {Oravetz}, {Oravetz}, {Osorio}, {Pace}, {Padilla},
  {Palanque-Delabrouille}, {Palicio}, {Pan}, {Pan}, {Parker}, {Paviot},
  {Peirani}, {Ram{\'r}ez}, {Penny}, {Percival}, {Perez-Fournon},
  {P{\'e}rez-R{\`a}fols}, {Petitjean}, {Pieri}, {Pinsonneault}, {Poovelil},
  {Povick}, {Prakash}, {Price-Whelan}, {Raddick}, {Raichoor}, {Ray}, {Rembold},
  {Rezaie}, {Riffel}, {Riffel}, {Rix}, {Robin}, {Roman-Lopes},
  {Rom{\'a}n-Z{\'u}{\~n}iga}, {Rose}, {Ross}, {Rossi}, {Rowlands}, {Rubin},
  {Salvato}, {S{\'a}nchez}, {S{\'a}nchez-Menguiano}, {S{\'a}nchez-Gallego},
  {Sayres}, {Schaefer}, {Schiavon}, {Schimoia}, {Schlafly}, {Schlegel},
  {Schneider}, {Schultheis}, {Schwope}, {Seo}, {Serenelli}, {Shafieloo},
  {Shamsi}, {Shao}, {Shen}, {Shetrone}, {Shirley}, {Silva Aguirre}, {Simon},
  {Skrutskie}, {Slosar}, {Smethurst}, {Sobeck}, {Sodi}, {Souto}, {Stark},
  {Stassun}, {Steinmetz}, {Stello}, {Stermer}, {Storchi-Bergmann},
  {Streblyanska}, {Stringfellow}, {Stutz}, {Su{\'a}rez}, {Sun},
  {Taghizadeh-Popp}, {Talbot}, {Tayar}, {Thakar}, {Theriault}, {Thomas},
  {Thomas}, {Tinker}, {Tojeiro}, {Toledo}, {Tremonti}, {Troup}, {Tuttle},
  {Unda-Sanzana}, {Valentini}, {Vargas-Gonz{\'a}lez}, {Vargas-Maga{\~n}a},
  {V{\'a}zquez-Mata}, {Vivek}, {Wake}, {Wang}, {Weaver}, {Weijmans}, {Wild},
  {Wilson}, {Wilson}, {Wolthuis}, {Wood-Vasey}, {Yan}, {Yang}, {Y{\`e}che},
  {Zamora}, {Zarrouk}, {Zasowski}, {Zhang}, {Zhao}, {Zhao}, {Zheng}, {Zheng},
  {Zhu}, \& {Zou}}]{ahumada2020}
{Ahumada}, R., {Allende Prieto}, C., {Almeida}, A., {et~al.} 2020, \apjs, 249,
  3, \dodoi{10.3847/1538-4365/ab929e}

\bibitem[{{Annibali} {et~al.}(2013){Annibali}, {Cignoni}, {Tosi}, {van der
  Marel}, {Aloisi}, {Clementini}, {Contreras Ramos}, {Fiorentino}, {Marconi},
  \& {Musella}}]{annibali2013}
{Annibali}, F., {Cignoni}, M., {Tosi}, M., {et~al.} 2013, \aj, 146, 144,
  \dodoi{10.1088/0004-6256/146/6/144}

\bibitem[{{Annibali} {et~al.}(2023){Annibali}, {Pinna}, {Hunt}, {Paris},
  {Cusano}, {Bellazzini}, {Cannon}, {Pascale}, {Tosi}, \&
  {Rossi}}]{annibali2023}
{Annibali}, F., {Pinna}, E., {Hunt}, L.~K., {et~al.} 2023, \apjl, 942, L23,
  \dodoi{10.3847/2041-8213/acab63}

\bibitem[{{Asplund} {et~al.}(2021){Asplund}, {Amarsi}, \&
  {Grevesse}}]{asplund2021}
{Asplund}, M., {Amarsi}, A.~M., \& {Grevesse}, N. 2021, \aap, 653, A141,
  \dodoi{10.1051/0004-6361/202140445}

\bibitem[{{Astropy Collaboration} {et~al.}(2013){Astropy Collaboration},
  {Robitaille}, {Tollerud}, {Greenfield}, {Droettboom}, {Bray}, {Aldcroft},
  {Davis}, {Ginsburg}, {Price-Whelan}, {Kerzendorf}, {Conley}, {Crighton},
  {Barbary}, {Muna}, {Ferguson}, {Grollier}, {Parikh}, {Nair}, {Unther},
  {Deil}, {Woillez}, {Conseil}, {Kramer}, {Turner}, {Singer}, {Fox}, {Weaver},
  {Zabalza}, {Edwards}, {Azalee Bostroem}, {Burke}, {Casey}, {Crawford},
  {Dencheva}, {Ely}, {Jenness}, {Labrie}, {Lim}, {Pierfederici}, {Pontzen},
  {Ptak}, {Refsdal}, {Servillat}, \& {Streicher}}]{astropy2}
{Astropy Collaboration}, {Robitaille}, T.~P., {Tollerud}, E.~J., {et~al.} 2013,
  \aap, 558, A33, \dodoi{10.1051/0004-6361/201322068}

\bibitem[{{Astropy Collaboration} {et~al.}(2018){Astropy Collaboration},
  {Price-Whelan}, {Sip{\H{o}}cz}, {G{\"u}nther}, {Lim}, {Crawford}, {Conseil},
  {Shupe}, {Craig}, {Dencheva}, {Ginsburg}, {Vand erPlas}, {Bradley},
  {P{\'e}rez-Su{\'a}rez}, {de Val-Borro}, {Aldcroft}, {Cruz}, {Robitaille},
  {Tollerud}, {Ardelean}, {Babej}, {Bach}, {Bachetti}, {Bakanov}, {Bamford},
  {Barentsen}, {Barmby}, {Baumbach}, {Berry}, {Biscani}, {Boquien}, {Bostroem},
  {Bouma}, {Brammer}, {Bray}, {Breytenbach}, {Buddelmeijer}, {Burke},
  {Calderone}, {Cano Rodr{\'\i}guez}, {Cara}, {Cardoso}, {Cheedella}, {Copin},
  {Corrales}, {Crichton}, {D'Avella}, {Deil}, {Depagne}, {Dietrich}, {Donath},
  {Droettboom}, {Earl}, {Erben}, {Fabbro}, {Ferreira}, {Finethy}, {Fox},
  {Garrison}, {Gibbons}, {Goldstein}, {Gommers}, {Greco}, {Greenfield},
  {Groener}, {Grollier}, {Hagen}, {Hirst}, {Homeier}, {Horton}, {Hosseinzadeh},
  {Hu}, {Hunkeler}, {Ivezi{\'c}}, {Jain}, {Jenness}, {Kanarek}, {Kendrew},
  {Kern}, {Kerzendorf}, {Khvalko}, {King}, {Kirkby}, {Kulkarni}, {Kumar},
  {Lee}, {Lenz}, {Littlefair}, {Ma}, {Macleod}, {Mastropietro}, {McCully},
  {Montagnac}, {Morris}, {Mueller}, {Mumford}, {Muna}, {Murphy}, {Nelson},
  {Nguyen}, {Ninan}, {N{\"o}the}, {Ogaz}, {Oh}, {Parejko}, {Parley}, {Pascual},
  {Patil}, {Patil}, {Plunkett}, {Prochaska}, {Rastogi}, {Reddy Janga},
  {Sabater}, {Sakurikar}, {Seifert}, {Sherbert}, {Sherwood-Taylor}, {Shih},
  {Sick}, {Silbiger}, {Singanamalla}, {Singer}, {Sladen}, {Sooley},
  {Sornarajah}, {Streicher}, {Teuben}, {Thomas}, {Tremblay}, {Turner},
  {Terr{\'o}n}, {van Kerkwijk}, {de la Vega}, {Watkins}, {Weaver}, {Whitmore},
  {Woillez}, {Zabalza}, \& {Astropy Contributors}}]{astropy1}
{Astropy Collaboration}, {Price-Whelan}, A.~M., {Sip{\H{o}}cz}, B.~M., {et~al.}
  2018, \aj, 156, 123, \dodoi{10.3847/1538-3881/aabc4f}

\bibitem[{{Astropy Collaboration} {et~al.}(2022){Astropy Collaboration},
  {Price-Whelan}, {Lim}, {Earl}, {Starkman}, {Bradley}, {Shupe}, {Patil},
  {Corrales}, {Brasseur}, {N{"o}the}, {Donath}, {Tollerud}, {Morris},
  {Ginsburg}, {Vaher}, {Weaver}, {Tocknell}, {Jamieson}, {van Kerkwijk},
  {Robitaille}, {Merry}, {Bachetti}, {G{"u}nther}, {Aldcroft},
  {Alvarado-Montes}, {Archibald}, {B{'o}di}, {Bapat}, {Barentsen}, {Baz{'a}n},
  {Biswas}, {Boquien}, {Burke}, {Cara}, {Cara}, {Conroy}, {Conseil}, {Craig},
  {Cross}, {Cruz}, {D'Eugenio}, {Dencheva}, {Devillepoix}, {Dietrich},
  {Eigenbrot}, {Erben}, {Ferreira}, {Foreman-Mackey}, {Fox}, {Freij}, {Garg},
  {Geda}, {Glattly}, {Gondhalekar}, {Gordon}, {Grant}, {Greenfield}, {Groener},
  {Guest}, {Gurovich}, {Handberg}, {Hart}, {Hatfield-Dodds}, {Homeier},
  {Hosseinzadeh}, {Jenness}, {Jones}, {Joseph}, {Kalmbach}, {Karamehmetoglu},
  {Ka{l}uszy{'n}ski}, {Kelley}, {Kern}, {Kerzendorf}, {Koch}, {Kulumani},
  {Lee}, {Ly}, {Ma}, {MacBride}, {Maljaars}, {Muna}, {Murphy}, {Norman},
  {O'Steen}, {Oman}, {Pacifici}, {Pascual}, {Pascual-Granado}, {Patil},
  {Perren}, {Pickering}, {Rastogi}, {Roulston}, {Ryan}, {Rykoff}, {Sabater},
  {Sakurikar}, {Salgado}, {Sanghi}, {Saunders}, {Savchenko}, {Schwardt},
  {Seifert-Eckert}, {Shih}, {Jain}, {Shukla}, {Sick}, {Simpson},
  {Singanamalla}, {Singer}, {Singhal}, {Sinha}, {Sip{H{o}}cz}, {Spitler},
  {Stansby}, {Streicher}, {{{S}}umak}, {Swinbank}, {Taranu}, {Tewary},
  {Tremblay}, {Val-Borro}, {Van Kooten}, {Vasovi{'c}}, {Verma}, {de Miranda
  Cardoso}, {Williams}, {Wilson}, {Winkel}, {Wood-Vasey}, {Xue}, {Yoachim},
  {Zhang}, {Zonca}, \& {Astropy Project Contributors}}]{astropy3}
{Astropy Collaboration}, {Price-Whelan}, A.~M., {Lim}, P.~L., {et~al.} 2022,
  \apj, 935, 167, \dodoi{10.3847/1538-4357/ac7c74}

\bibitem[{{Aver} {et~al.}(2022){Aver}, {Berg}, {Hirschauer}, {Olive}, {Pogge},
  {Rogers}, {Salzer}, \& {Skillman}}]{aver2022}
{Aver}, E., {Berg}, D.~A., {Hirschauer}, A.~S., {et~al.} 2022, \mnras, 510,
  373, \dodoi{10.1093/mnras/stab3226}

\bibitem[{{Bajaja} {et~al.}(1994){Bajaja}, {Huchtmeier}, \&
  {Klein}}]{bajaja1994}
{Bajaja}, E., {Huchtmeier}, W.~K., \& {Klein}, U. 1994, \aap, 285, 385

\bibitem[{{Barnes} \& {de Blok}(2004)}]{barnes2004}
{Barnes}, D.~G., \& {de Blok}, W.~J.~G. 2004, \mnras, 351, 333,
  \dodoi{10.1111/j.1365-2966.2004.07790.x}

\bibitem[{{Beaton} {et~al.}(2019){Beaton}, {Seibert}, {Hatt}, {Freedman},
  {Hoyt}, {Jang}, {Lee}, {Madore}, {Monson}, {Neeley}, {Rich}, \&
  {Scowcroft}}]{beaton2019}
{Beaton}, R.~L., {Seibert}, M., {Hatt}, D., {et~al.} 2019, \apj, 885, 141,
  \dodoi{10.3847/1538-4357/ab4263}

\bibitem[{{Bell} {et~al.}(2003){Bell}, {McIntosh}, {Katz}, \&
  {Weinberg}}]{bell2003}
{Bell}, E.~F., {McIntosh}, D.~H., {Katz}, N., \& {Weinberg}, M.~D. 2003, \apjs,
  149, 289, \dodoi{10.1086/378847}

\bibitem[{{Berg} {et~al.}(2021){Berg}, {Chisholm}, {Erb}, {Skillman}, {Pogge},
  \& {Olivier}}]{berg2021}
{Berg}, D.~A., {Chisholm}, J., {Erb}, D.~K., {et~al.} 2021, \apj, 922, 170,
  \dodoi{10.3847/1538-4357/ac141b}

\bibitem[{{Berg} {et~al.}(2019){Berg}, {Erb}, {Henry}, {Skillman}, \&
  {McQuinn}}]{berg2019}
{Berg}, D.~A., {Erb}, D.~K., {Henry}, R. B.~C., {Skillman}, E.~D., \&
  {McQuinn}, K. B.~W. 2019, \apj, 874, 93, \dodoi{10.3847/1538-4357/ab020a}

\bibitem[{{Berg} {et~al.}(2020){Berg}, {Pogge}, {Skillman}, {Croxall},
  {Moustakas}, {Rogers}, \& {Sun}}]{berg2020}
{Berg}, D.~A., {Pogge}, R.~W., {Skillman}, E.~D., {et~al.} 2020, \apj, 893, 96,
  \dodoi{10.3847/1538-4357/ab7eab}

\bibitem[{{Berg} {et~al.}(2012){Berg}, {Skillman}, {Marble}, {van Zee},
  {Engelbracht}, {Lee}, {Kennicutt}, {Calzetti}, {Dale}, \&
  {Johnson}}]{berg2012}
{Berg}, D.~A., {Skillman}, E.~D., {Marble}, A.~R., {et~al.} 2012, \apj, 754,
  98, \dodoi{10.1088/0004-637X/754/2/98}

\bibitem[{{Berg} {et~al.}(2022){Berg}, {James}, {King}, {McDonald}, {Chen},
  {Chisholm}, {Heckman}, {Martin}, {Stark}, {Aloisi}, {Amor{\'\i}n},
  {Arellano-C{\'o}rdova}, {Bayliss}, {Bordoloi}, {Brinchmann}, {Charlot},
  {Chevallard}, {Clark}, {Erb}, {Feltre}, {Gronke}, {Hayes}, {Henry},
  {Hernandez}, {Jaskot}, {Jones}, {Kewley}, {Kumari}, {Leitherer}, {Llerena},
  {Maseda}, {Mingozzi}, {Nanayakkara}, {Ouchi}, {Plat}, {Pogge},
  {Ravindranath}, {Rigby}, {Sanders}, {Scarlata}, {Senchyna}, {Skillman},
  {Steidel}, {Strom}, {Sugahara}, {Wilkins}, {Wofford}, {Xu}, \& {Classy
  Team}}]{berg2022}
{Berg}, D.~A., {James}, B.~L., {King}, T., {et~al.} 2022, \apjs, 261, 31,
  \dodoi{10.3847/1538-4365/ac6c03}

\bibitem[{{Brauer} {et~al.}(2024){Brauer}, {Emerick}, {Mead}, {Ji}, {Wise},
  {Bryan}, {Mac Low}, {Cote}, {Andersson}, \& {Frebel}}]{brauer2024}
{Brauer}, K., {Emerick}, A., {Mead}, J., {et~al.} 2024, arXiv e-prints,
  arXiv:2410.16366, \dodoi{10.48550/arXiv.2410.16366}

\bibitem[{{Bressan} {et~al.}(2012){Bressan}, {Marigo}, {Girardi}, {Salasnich},
  {Dal Cero}, {Rubele}, \& {Nanni}}]{bressan2012}
{Bressan}, A., {Marigo}, P., {Girardi}, L., {et~al.} 2012, \mnras, 427, 127,
  \dodoi{10.1111/j.1365-2966.2012.21948.x}

\bibitem[{{Brooks} {et~al.}(2009){Brooks}, {Governato}, {Quinn}, {Brook}, \&
  {Wadsley}}]{brooks2009}
{Brooks}, A.~M., {Governato}, F., {Quinn}, T., {Brook}, C.~B., \& {Wadsley}, J.
  2009, \apj, 694, 396, \dodoi{10.1088/0004-637X/694/1/396}

\bibitem[{{Brown} {et~al.}(2008){Brown}, {Kewley}, \& {Geller}}]{brown2008}
{Brown}, W.~R., {Kewley}, L.~J., \& {Geller}, M.~J. 2008, \aj, 135, 92,
  \dodoi{10.1088/0004-6256/135/1/92}

\bibitem[{{Brunker} {et~al.}(2019){Brunker}, {McQuinn}, {Salzer}, {Cannon},
  {Janowiecki}, {Leisman}, {Rhode}, {Adams}, {Ball}, {Dolphin}, {Giovanelli},
  \& {Haynes}}]{brunker2019}
{Brunker}, S.~W., {McQuinn}, K. B.~W., {Salzer}, J.~J., {et~al.} 2019, \aj,
  157, 76, \dodoi{10.3847/1538-3881/aafb39}

\bibitem[{{Bruzual} \& {Charlot}(2003)}]{bruzual2003}
{Bruzual}, G., \& {Charlot}, S. 2003, \mnras, 344, 1000,
  \dodoi{10.1046/j.1365-8711.2003.06897.x}

\bibitem[{{Cardelli} {et~al.}(1989){Cardelli}, {Clayton}, \&
  {Mathis}}]{cardelli1989}
{Cardelli}, J.~A., {Clayton}, G.~C., \& {Mathis}, J.~S. 1989, \apj, 345, 245,
  \dodoi{10.1086/167900}

\bibitem[{{Castignani} {et~al.}(2022){Castignani}, {Combes}, {Jablonka},
  {Finn}, {Rudnick}, {Vulcani}, {Desai}, {Zaritsky}, \&
  {Salom{\'e}}}]{castignani2022}
{Castignani}, G., {Combes}, F., {Jablonka}, P., {et~al.} 2022, \aap, 657, A9,
  \dodoi{10.1051/0004-6361/202040141}

\bibitem[{{Cattorini} {et~al.}(2023){Cattorini}, {Gavazzi}, {Boselli}, \&
  {Fossati}}]{cattorini2023}
{Cattorini}, F., {Gavazzi}, G., {Boselli}, A., \& {Fossati}, M. 2023, \aap,
  671, A118, \dodoi{10.1051/0004-6361/202244738}

\bibitem[{{Chang} {et~al.}(2015){Chang}, {van der Wel}, {da Cunha}, \&
  {Rix}}]{chang2015}
{Chang}, Y.-Y., {van der Wel}, A., {da Cunha}, E., \& {Rix}, H.-W. 2015, \apjs,
  219, 8, \dodoi{10.1088/0067-0049/219/1/8}

\bibitem[{{Chengalur} {et~al.}(2006){Chengalur}, {Pustilnik}, {Martin}, \&
  {Kniazev}}]{chengalur2006}
{Chengalur}, J.~N., {Pustilnik}, S.~A., {Martin}, J.~M., \& {Kniazev}, A.~Y.
  2006, \mnras, 371, 1849, \dodoi{10.1111/j.1365-2966.2006.10816.x}

\bibitem[{{Chisholm} {et~al.}(2018){Chisholm}, {Tremonti}, \&
  {Leitherer}}]{chisholm2018}
{Chisholm}, J., {Tremonti}, C., \& {Leitherer}, C. 2018, \mnras, 481, 1690,
  \dodoi{10.1093/mnras/sty2380}

\bibitem[{{Chonis} {et~al.}(2014){Chonis}, {Hill}, {Lee}, {Tuttle}, \&
  {Vattiat}}]{chonis2014}
{Chonis}, T.~S., {Hill}, G.~J., {Lee}, H., {Tuttle}, S.~E., \& {Vattiat}, B.~L.
  2014, in Society of Photo-Optical Instrumentation Engineers (SPIE) Conference
  Series, Vol. 9147, Ground-based and Airborne Instrumentation for Astronomy V,
  ed. S.~K. {Ramsay}, I.~S. {McLean}, \& H.~{Takami}, 91470A,
  \dodoi{10.1117/12.2056005}

\bibitem[{{Chonis} {et~al.}(2016){Chonis}, {Hill}, {Lee}, {Tuttle}, {Vattiat},
  {Drory}, {Indahl}, {Peterson}, \& {Ramsey}}]{chonis2016}
{Chonis}, T.~S., {Hill}, G.~J., {Lee}, H., {et~al.} 2016, in Society of
  Photo-Optical Instrumentation Engineers (SPIE) Conference Series, Vol. 9908,
  Ground-based and Airborne Instrumentation for Astronomy VI, ed. C.~J.
  {Evans}, L.~{Simard}, \& H.~{Takami}, 99084C, \dodoi{10.1117/12.2232209}

\bibitem[{{Cid Fernandes} {et~al.}(2005){Cid Fernandes}, {Mateus}, {Sodr{\'e}},
  {Stasi{\'n}ska}, \& {Gomes}}]{fernandes2005}
{Cid Fernandes}, R., {Mateus}, A., {Sodr{\'e}}, L., {Stasi{\'n}ska}, G., \&
  {Gomes}, J.~M. 2005, \mnras, 358, 363,
  \dodoi{10.1111/j.1365-2966.2005.08752.x}

\bibitem[{{Colberg} {et~al.}(2008){Colberg}, {Pearce}, {Foster}, {Platen},
  {Brunino}, {Neyrinck}, {Basilakos}, {Fairall}, {Feldman}, {Gottl{\"o}ber},
  {Hahn}, {Hoyle}, {M{\"u}ller}, {Nelson}, {Plionis}, {Porciani}, {Shandarin},
  {Vogeley}, \& {van de Weygaert}}]{colberg2008}
{Colberg}, J.~M., {Pearce}, F., {Foster}, C., {et~al.} 2008, \mnras, 387, 933,
  \dodoi{10.1111/j.1365-2966.2008.13307.x}

\bibitem[{{Cole} {et~al.}(2007){Cole}, {Skillman}, {Tolstoy}, {Gallagher},
  {Aparicio}, {Dolphin}, {Gallart}, {Hidalgo}, {Saha}, {Stetson}, \&
  {Weisz}}]{cole2007}
{Cole}, A.~A., {Skillman}, E.~D., {Tolstoy}, E., {et~al.} 2007, \apjl, 659,
  L17, \dodoi{10.1086/516711}

\bibitem[{{Collins} \& {Read}(2022)}]{collins2022}
{Collins}, M. L.~M., \& {Read}, J.~I. 2022, Nature Astronomy, 6, 647,
  \dodoi{10.1038/s41550-022-01657-4}

\bibitem[{{Cook} {et~al.}(2014){Cook}, {Dale}, {Johnson}, {Van Zee}, {Lee},
  {Kennicutt}, {Calzetti}, {Staudaher}, \& {Engelbracht}}]{cook2014}
{Cook}, D.~O., {Dale}, D.~A., {Johnson}, B.~D., {et~al.} 2014, \mnras, 445,
  899, \dodoi{10.1093/mnras/stu1787}

\bibitem[{{Correnti} {et~al.}(2025){Correnti}, {Annibali}, {Bellazzini},
  {Marinelli}, {Aloisi}, {Cignoni}, {Tosi}, {Pascale}, {Cannon}, {Schisgal},
  {Hunt}, {Sacchi}, \& {Sohn}}]{correnti2025}
{Correnti}, M., {Annibali}, F., {Bellazzini}, M., {et~al.} 2025, arXiv
  e-prints, arXiv:2502.18171, \dodoi{10.48550/arXiv.2502.18171}

\bibitem[{{Courtois} {et~al.}(2009){Courtois}, {Tully}, {Fisher}, {Bonhomme},
  {Zavodny}, \& {Barnes}}]{courtois2009}
{Courtois}, H.~M., {Tully}, R.~B., {Fisher}, J.~R., {et~al.} 2009, \aj, 138,
  1938, \dodoi{10.1088/0004-6256/138/6/1938}

\bibitem[{{Dalcanton}(2007)}]{dalcanton2007}
{Dalcanton}, J.~J. 2007, \apj, 658, 941, \dodoi{10.1086/508913}

\bibitem[{{Dale} {et~al.}(2009){Dale}, {Cohen}, {Johnson}, {Schuster},
  {Calzetti}, {Engelbracht}, {Gil de Paz}, {Kennicutt}, {Lee}, {Begum},
  {Block}, {Dalcanton}, {Funes}, {Gordon}, {Johnson}, {Marble}, {Sakai},
  {Skillman}, {van Zee}, {Walter}, {Weisz}, {Williams}, {Wu}, \&
  {Wu}}]{dale2009}
{Dale}, D.~A., {Cohen}, S.~A., {Johnson}, L.~C., {et~al.} 2009, \apj, 703, 517,
  \dodoi{10.1088/0004-637X/703/1/517}

\bibitem[{{D{\'a}lya} {et~al.}(2018){D{\'a}lya}, {Galg{\'o}czi}, {Dobos},
  {Frei}, {Heng}, {Macas}, {Messenger}, {Raffai}, \& {de Souza}}]{dalya2018}
{D{\'a}lya}, G., {Galg{\'o}czi}, G., {Dobos}, L., {et~al.} 2018, \mnras, 479,
  2374, \dodoi{10.1093/mnras/sty1703}

\bibitem[{{Dinerstein}(1990)}]{dinerstein1990}
{Dinerstein}, H.~L. 1990, in Astrophysics and Space Science Library, Vol. 161,
  The Interstellar Medium in Galaxies, ed. J.~{Thronson}, Harley~A. \& J.~M.
  {Shull}, 257--285, \dodoi{10.1007/978-94-009-0595-5\_10}

\bibitem[{{Dolphin}(2016)}]{dolphin2016}
{Dolphin}, A. 2016, {DOLPHOT: Stellar photometry}, Astrophysics Source Code
  Library, record ascl:1608.013

\bibitem[{{Dolphin}(2000)}]{dolphin2000}
{Dolphin}, A.~E. 2000, \pasp, 112, 1383, \dodoi{10.1086/316630}

\bibitem[{{Dom{\'\i}nguez-G{\'o}mez} {et~al.}(2022){Dom{\'\i}nguez-G{\'o}mez},
  {Lisenfeld}, {P{\'e}rez}, {L{\'o}pez-S{\'a}nchez}, {Duarte Puertas},
  {Falc{\'o}n-Barroso}, {Kreckel}, {Peletier}, {Ruiz-Lara}, {van de Weygaert},
  {van der Hulst}, \& {Verley}}]{dominguez2022}
{Dom{\'\i}nguez-G{\'o}mez}, J., {Lisenfeld}, U., {P{\'e}rez}, I., {et~al.}
  2022, \aap, 658, A124, \dodoi{10.1051/0004-6361/202141888}

\bibitem[{{Douglass} {et~al.}(2018{\natexlab{a}}){Douglass}, {Vogeley}, \&
  {Cen}}]{douglass2018}
{Douglass}, K.~A., {Vogeley}, M.~S., \& {Cen}, R. 2018{\natexlab{a}}, \apj,
  864, 144, \dodoi{10.3847/1538-4357/aad86e}

\bibitem[{{Douglass} {et~al.}(2018{\natexlab{b}}){Douglass}, {Vogeley}, \&
  {Cen}}]{karachentsev2013}
---. 2018{\natexlab{b}}, \apj, 864, 144, \dodoi{10.3847/1538-4357/aad86e}

\bibitem[{{Durbala} {et~al.}(2020){Durbala}, {Finn}, {Crone Odekon}, {Haynes},
  {Koopmann}, \& {O'Donoghue}}]{durbala2020}
{Durbala}, A., {Finn}, R.~A., {Crone Odekon}, M., {et~al.} 2020, \aj, 160, 271,
  \dodoi{10.3847/1538-3881/abc018}

\bibitem[{{Ekta} {et~al.}(2006){Ekta}, {Chengalur}, \& {Pustilnik}}]{ekta2006}
{Ekta}, {Chengalur}, J.~N., \& {Pustilnik}, S.~A. 2006, \mnras, 372, 853,
  \dodoi{10.1111/j.1365-2966.2006.10904.x}

\bibitem[{{Ekta} {et~al.}(2008){Ekta}, {Chengalur}, \& {Pustilnik}}]{ekta2008}
---. 2008, \mnras, 391, 881, \dodoi{10.1111/j.1365-2966.2008.13928.x}

\bibitem[{{Ekta} \& {Chengalur}(2010)}]{ekta2010}
{Ekta}, B., \& {Chengalur}, J.~N. 2010, \mnras, 406, 1238,
  \dodoi{10.1111/j.1365-2966.2010.16756.x}

\bibitem[{{Ferrarese} {et~al.}(2000){Ferrarese}, {Ford}, {Huchra}, {Kennicutt},
  {Mould}, {Sakai}, {Freedman}, {Stetson}, {Madore}, {Gibson}, {Graham},
  {Hughes}, {Illingworth}, {Kelson}, {Macri}, {Sebo}, \&
  {Silbermann}}]{ferrarese2000}
{Ferrarese}, L., {Ford}, H.~C., {Huchra}, J., {et~al.} 2000, \apjs, 128, 431,
  \dodoi{10.1086/313391}

\bibitem[{{Filho} {et~al.}(2015){Filho}, {S{\'a}nchez Almeida},
  {Mu{\~n}oz-Tu{\~n}{\'o}n}, {Nuza}, {Kitaura}, \& {He{\ss}}}]{filho2015}
{Filho}, M.~E., {S{\'a}nchez Almeida}, J., {Mu{\~n}oz-Tu{\~n}{\'o}n}, C.,
  {et~al.} 2015, \apj, 802, 82, \dodoi{10.1088/0004-637X/802/2/82}

\bibitem[{{Filho} {et~al.}(2013){Filho}, {Winkel}, {S{\'a}nchez Almeida},
  {Aguerri}, {Amor{\'\i}n}, {Ascasibar}, {Elmegreen}, {Elmegreen}, {Gomes},
  {Humphrey}, {Lagos}, {Morales-Luis}, {Mu{\~n}oz-Tu{\~n}{\'o}n}, {Papaderos},
  \& {V{\'\i}lchez}}]{filho2013}
{Filho}, M.~E., {Winkel}, B., {S{\'a}nchez Almeida}, J., {et~al.} 2013, \aap,
  558, A18, \dodoi{10.1051/0004-6361/201322098}

\bibitem[{{Ford} {et~al.}(1998){Ford}, {Bartko}, {Bely}, {Broadhurst},
  {Burrows}, {Cheng}, {Clampin}, {Crocker}, {Feldman}, {Golimowski}, {Hartig},
  {Illingworth}, {Kimble}, {Lesser}, {Miley}, {Neff}, {Postman}, {Sparks},
  {Tsvetanov}, {White}, {Sullivan}, {Krebs}, {Leviton}, {La Jeunesse},
  {Burmester}, {Fike}, {Johnson}, {Slusher}, {Volmer}, \&
  {Woodruff}}]{ford1998}
{Ford}, H.~C., {Bartko}, F., {Bely}, P.~Y., {et~al.} 1998, in Society of
  Photo-Optical Instrumentation Engineers (SPIE) Conference Series, Vol. 3356,
  Space Telescopes and Instruments V, ed. P.~Y. {Bely} \& J.~B. {Breckinridge},
  234--248, \dodoi{10.1117/12.324464}

\bibitem[{{Freedman}(2021)}]{freedman2021}
{Freedman}, W.~L. 2021, \apj, 919, 16, \dodoi{10.3847/1538-4357/ac0e95}

\bibitem[{{Freedman} {et~al.}(2019){Freedman}, {Madore}, {Hatt}, {Hoyt},
  {Jang}, {Beaton}, {Burns}, {Lee}, {Monson}, {Neeley}, {Phillips}, {Rich}, \&
  {Seibert}}]{freedman2019}
{Freedman}, W.~L., {Madore}, B.~F., {Hatt}, D., {et~al.} 2019, \apj, 882, 34,
  \dodoi{10.3847/1538-4357/ab2f73}

\bibitem[{{Garnett}(1990)}]{garnett1990}
{Garnett}, D.~R. 1990, \apj, 363, 142, \dodoi{10.1086/169324}

\bibitem[{{Garnett}(1992)}]{garnett1992}
---. 1992, \aj, 103, 1330, \dodoi{10.1086/116146}

\bibitem[{{Garnett}(2002)}]{garnett2002}
---. 2002, \apj, 581, 1019, \dodoi{10.1086/344301}

\bibitem[{{Geha} {et~al.}(2006){Geha}, {Blanton}, {Masjedi}, \&
  {West}}]{geha2006}
{Geha}, M., {Blanton}, M.~R., {Masjedi}, M., \& {West}, A.~A. 2006, \apj, 653,
  240, \dodoi{10.1086/508604}

\bibitem[{{Glover} \& {Clark}(2012)}]{glover2012}
{Glover}, S. C.~O., \& {Clark}, P.~C. 2012, \mnras, 421, 9,
  \dodoi{10.1111/j.1365-2966.2011.19648.x}

\bibitem[{{Goddy} {et~al.}(2020){Goddy}, {Stark}, \& {Masters}}]{goddy2020}
{Goddy}, J., {Stark}, D.~V., \& {Masters}, K.~L. 2020, Research Notes of the
  American Astronomical Society, 4, 3, \dodoi{10.3847/2515-5172/ab66bd}

\bibitem[{{Guseva} {et~al.}(2015){Guseva}, {Izotov}, {Fricke}, \&
  {Henkel}}]{guseva2015}
{Guseva}, N.~G., {Izotov}, Y.~I., {Fricke}, K.~J., \& {Henkel}, C. 2015, \aap,
  579, A11, \dodoi{10.1051/0004-6361/201525697}

\bibitem[{{Guseva} {et~al.}(2017){Guseva}, {Izotov}, {Fricke}, \&
  {Henkel}}]{guseva2017}
---. 2017, \aap, 599, A65, \dodoi{10.1051/0004-6361/201629181}

\bibitem[{{Guseva} {et~al.}(2011){Guseva}, {Izotov}, {Stasi{\'n}ska}, {Fricke},
  {Henkel}, \& {Papaderos}}]{guseva2011}
{Guseva}, N.~G., {Izotov}, Y.~I., {Stasi{\'n}ska}, G., {et~al.} 2011, \aap,
  529, A149, \dodoi{10.1051/0004-6361/201016291}

\bibitem[{{Hatt} {et~al.}(2017){Hatt}, {Beaton}, {Freedman}, {Madore}, {Jang},
  {Hoyt}, {Lee}, {Monson}, {Rich}, {Scowcroft}, \& {Seibert}}]{hatt2017}
{Hatt}, D., {Beaton}, R.~L., {Freedman}, W.~L., {et~al.} 2017, \apj, 845, 146,
  \dodoi{10.3847/1538-4357/aa7f73}

\bibitem[{{Haynes} {et~al.}(2018){Haynes}, {Giovanelli}, {Kent}, {Adams},
  {Balonek}, {Craig}, {Fertig}, {Finn}, {Giovanardi}, {Hallenbeck}, {Hess},
  {Hoffman}, {Huang}, {Jones}, {Koopmann}, {Kornreich}, {Leisman}, {Miller},
  {Moorman}, {O'Connor}, {O'Donoghue}, {Papastergis}, {Troischt}, {Stark}, \&
  {Xiao}}]{haynes2018}
{Haynes}, M.~P., {Giovanelli}, R., {Kent}, B.~R., {et~al.} 2018, \apj, 861, 49,
  \dodoi{10.3847/1538-4357/aac956}

\bibitem[{{Henry} {et~al.}(2000){Henry}, {Edmunds}, \&
  {K{\"o}ppen}}]{henry2000}
{Henry}, R.~B.~C., {Edmunds}, M.~G., \& {K{\"o}ppen}, J. 2000, \apj, 541, 660,
  \dodoi{10.1086/309471}

\bibitem[{{Hill} {et~al.}(2021){Hill}, {Lee}, {MacQueen}, {Kelz}, {Drory},
  {Vattiat}, {Good}, {Ramsey}, {Kriel}, {Peterson}, {DePoy}, {Gebhardt},
  {Marshall}, {Tuttle}, {Bauer}, {Chonis}, {Fabricius}, {Froning},
  {H{\"a}user}, {Indahl}, {Jahn}, {Landriau}, {Leck}, {Montesano}, {Prochaska},
  {Snigula}, {Zeimann}, {Bryant}, {Damm}, {Fowler}, {Janowiecki}, {Martin},
  {Mrozinski}, {Odewahn}, {Rostopchin}, {Shetrone}, {Spencer}, {Mentuch
  Cooper}, {Armandroff}, {Bender}, {Dalton}, {Hopp}, {Komatsu}, {Nicklas},
  {Ramsey}, {Roth}, {Schneider}, {Sneden}, \& {Steinmetz}}]{hill2021}
{Hill}, G.~J., {Lee}, H., {MacQueen}, P.~J., {et~al.} 2021, \aj, 162, 298,
  \dodoi{10.3847/1538-3881/ac2c02}

\bibitem[{{Hoessel} \& {Mould}(1982)}]{hoessel1982}
{Hoessel}, J.~G., \& {Mould}, J.~R. 1982, \apj, 254, 38, \dodoi{10.1086/159702}

\bibitem[{{Hogg} {et~al.}(2007){Hogg}, {Roberts}, {Haynes}, \&
  {Maddalena}}]{hogg2007}
{Hogg}, D.~E., {Roberts}, M.~S., {Haynes}, M.~P., \& {Maddalena}, R.~J. 2007,
  \aj, 134, 1046, \dodoi{10.1086/520766}

\bibitem[{{Hsyu} {et~al.}(2017){Hsyu}, {Cooke}, {Prochaska}, \&
  {Bolte}}]{hsyu2017}
{Hsyu}, T., {Cooke}, R.~J., {Prochaska}, J.~X., \& {Bolte}, M. 2017, \apjl,
  845, L22, \dodoi{10.3847/2041-8213/aa821f}

\bibitem[{{Hsyu} {et~al.}(2018){Hsyu}, {Cooke}, {Prochaska}, \&
  {Bolte}}]{hsyu2018}
---. 2018, \apj, 863, 134, \dodoi{10.3847/1538-4357/aad18a}

\bibitem[{{Huchtmeier} \& {Richter}(1986)}]{huchtmeier1986}
{Huchtmeier}, W.~K., \& {Richter}, O.~G. 1986, \aaps, 63, 323

\bibitem[{{Hunter} {et~al.}(2024){Hunter}, {Elmegreen}, \&
  {Madden}}]{hunter2024}
{Hunter}, D.~A., {Elmegreen}, B.~G., \& {Madden}, S.~C. 2024, arXiv e-prints,
  arXiv:2402.17004.
\newblock \doarXiv{2402.17004}

\bibitem[{{Izotov} {et~al.}(2019){Izotov}, {Guseva}, {Fricke}, \&
  {Henkel}}]{izotov2019}
{Izotov}, Y.~I., {Guseva}, N.~G., {Fricke}, K.~J., \& {Henkel}, C. 2019, \aap,
  623, A40, \dodoi{10.1051/0004-6361/201834768}

\bibitem[{{Izotov} {et~al.}(2018{\natexlab{a}}){Izotov}, {Schaerer}, {Worseck},
  {Guseva}, {Thuan}, {Verhamme}, {Orlitov{\'a}}, \& {Fricke}}]{izotov2018}
{Izotov}, Y.~I., {Schaerer}, D., {Worseck}, G., {et~al.} 2018{\natexlab{a}},
  \mnras, 474, 4514, \dodoi{10.1093/mnras/stx3115}

\bibitem[{{Izotov} {et~al.}(2006){Izotov}, {Stasi{\'n}ska}, {Meynet}, {Guseva},
  \& {Thuan}}]{izotov2006a}
{Izotov}, Y.~I., {Stasi{\'n}ska}, G., {Meynet}, G., {Guseva}, N.~G., \&
  {Thuan}, T.~X. 2006, \aap, 448, 955, \dodoi{10.1051/0004-6361:20053763}

\bibitem[{{Izotov} \& {Thuan}(1998)}]{izotov1998}
{Izotov}, Y.~I., \& {Thuan}, T.~X. 1998, \apj, 497, 227, \dodoi{10.1086/305440}

\bibitem[{{Izotov} \& {Thuan}(1999)}]{izotov1999}
---. 1999, \apj, 511, 639, \dodoi{10.1086/306708}

\bibitem[{{Izotov} {et~al.}(2012){Izotov}, {Thuan}, \& {Guseva}}]{izotov2012}
{Izotov}, Y.~I., {Thuan}, T.~X., \& {Guseva}, N.~G. 2012, \aap, 546, A122,
  \dodoi{10.1051/0004-6361/201219733}

\bibitem[{{Izotov} {et~al.}(2021){Izotov}, {Thuan}, \& {Guseva}}]{izotov2021}
---. 2021, \mnras, 504, 3996, \dodoi{10.1093/mnras/stab1099}

\bibitem[{{Izotov} {et~al.}(2018{\natexlab{b}}){Izotov}, {Thuan}, {Guseva}, \&
  {Liss}}]{izotov2018b}
{Izotov}, Y.~I., {Thuan}, T.~X., {Guseva}, N.~G., \& {Liss}, S.~E.
  2018{\natexlab{b}}, \mnras, 473, 1956, \dodoi{10.1093/mnras/stx2478}

\bibitem[{{Jang} \& {Lee}(2017)}]{jang2017}
{Jang}, I.~S., \& {Lee}, M.~G. 2017, \apj, 835, 28,
  \dodoi{10.3847/1538-4357/835/1/28}

\bibitem[{{Karachentsev} {et~al.}(2004){Karachentsev}, {Karachentseva},
  {Huchtmeier}, \& {Makarov}}]{karachentsev2004}
{Karachentsev}, I.~D., {Karachentseva}, V.~E., {Huchtmeier}, W.~K., \&
  {Makarov}, D.~I. 2004, \aj, 127, 2031, \dodoi{10.1086/382905}

\bibitem[{{Kere{\v{s}}} {et~al.}(2005){Kere{\v{s}}}, {Katz}, {Weinberg}, \&
  {Dav{\'e}}}]{keres2005}
{Kere{\v{s}}}, D., {Katz}, N., {Weinberg}, D.~H., \& {Dav{\'e}}, R. 2005,
  \mnras, 363, 2, \dodoi{10.1111/j.1365-2966.2005.09451.x}

\bibitem[{{Kluyver} {et~al.}(2016){Kluyver}, {Ragan-Kelley}, {P{\'e}rez},
  {Granger}, {Bussonnier}, {Frederic}, {Kelley}, {Hamrick}, {Grout}, {Corlay},
  {Ivanov}, {Avila}, {Abdalla}, {Willing}, \& {Jupyter Development
  Team}}]{kluyver2016}
{Kluyver}, T., {Ragan-Kelley}, B., {P{\'e}rez}, F., {et~al.} 2016, in IOS
  Press, 87--90, \dodoi{10.3233/978-1-61499-649-1-87}

\bibitem[{{Kniazev} {et~al.}(2018){Kniazev}, {Egorova}, \&
  {Pustilnik}}]{kniazev2018}
{Kniazev}, A.~Y., {Egorova}, E.~S., \& {Pustilnik}, S.~A. 2018, \mnras, 479,
  3842, \dodoi{10.1093/mnras/sty1704}

\bibitem[{{Kobayashi} \& {Ferrara}(2023)}]{kobayashi2023}
{Kobayashi}, C., \& {Ferrara}, A. 2023, arXiv e-prints, arXiv:2308.15583,
  \dodoi{10.48550/arXiv.2308.15583}

\bibitem[{{Kojima} {et~al.}(2020){Kojima}, {Ouchi}, {Rauch}, {Ono}, {Nakajima},
  {Isobe}, {Fujimoto}, {Harikane}, {Hashimoto}, {Hayashi}, {Komiyama},
  {Kusakabe}, {Kim}, {Lee}, {Mukae}, {Nagao}, {Onodera}, {Shibuya}, {Sugahara},
  {Umemura}, \& {Yabe}}]{kojima2020}
{Kojima}, T., {Ouchi}, M., {Rauch}, M., {et~al.} 2020, \apj, 898, 142,
  \dodoi{10.3847/1538-4357/aba047}

\bibitem[{{Kojima} {et~al.}(2021){Kojima}, {Ouchi}, {Rauch}, {Ono}, {Nakajima},
  {Isobe}, {Fujimoto}, {Harikane}, {Hashimoto}, {Hayashi}, {Komiyama},
  {Kusakabe}, {Kim}, {Lee}, {Mukae}, {Nagao}, {Onodera}, {Shibuya}, {Sugahara},
  {Umemura}, \& {Yabe}}]{kojima2021}
---. 2021, \apj, 913, 22, \dodoi{10.3847/1538-4357/abec3d}

\bibitem[{{Kreckel} {et~al.}(2015){Kreckel}, {Croxall}, {Groves}, {van de
  Weygaert}, \& {Pogge}}]{kreckel2015}
{Kreckel}, K., {Croxall}, K., {Groves}, B., {van de Weygaert}, R., \& {Pogge},
  R.~W. 2015, \apjl, 798, L15, \dodoi{10.1088/2041-8205/798/1/L15}

\bibitem[{{Kreckel} {et~al.}(2011){Kreckel}, {Platen}, {Arag{\'o}n-Calvo}, {van
  Gorkom}, {van de Weygaert}, {van der Hulst}, {Kova{\v{c}}}, {Yip}, \&
  {Peebles}}]{kreckel2011a}
{Kreckel}, K., {Platen}, E., {Arag{\'o}n-Calvo}, M.~A., {et~al.} 2011, \aj,
  141, 4, \dodoi{10.1088/0004-6256/141/1/4}

\bibitem[{{Lee} {et~al.}(2006{\natexlab{a}}){Lee}, {Skillman}, {Cannon},
  {Jackson}, {Gehrz}, {Polomski}, \& {Woodward}}]{lee2006a}
{Lee}, H., {Skillman}, E.~D., {Cannon}, J.~M., {et~al.} 2006{\natexlab{a}},
  \apj, 647, 970, \dodoi{10.1086/505573}

\bibitem[{{Lee} {et~al.}(2006{\natexlab{b}}){Lee}, {Skillman}, \&
  {Venn}}]{lee2006}
{Lee}, H., {Skillman}, E.~D., \& {Venn}, K.~A. 2006{\natexlab{b}}, \apj, 642,
  813, \dodoi{10.1086/500568}

\bibitem[{{Lee} {et~al.}(1993{\natexlab{a}}){Lee}, {Freedman}, \&
  {Madore}}]{lee1993}
{Lee}, M.~G., {Freedman}, W.~L., \& {Madore}, B.~F. 1993{\natexlab{a}}, \apj,
  417, 553, \dodoi{10.1086/173334}

\bibitem[{{Lee} {et~al.}(1993{\natexlab{b}}){Lee}, {Freedman}, \&
  {Madore}}]{lee_freedman_madore1993}
---. 1993{\natexlab{b}}, \apj, 417, 553, \dodoi{10.1086/173334}

\bibitem[{{Lelli} {et~al.}(2014){Lelli}, {Verheijen}, \&
  {Fraternali}}]{lelli2014}
{Lelli}, F., {Verheijen}, M., \& {Fraternali}, F. 2014, \mnras, 445, 1694,
  \dodoi{10.1093/mnras/stu1804}

\bibitem[{{Lelli} {et~al.}(2012{\natexlab{a}}){Lelli}, {Verheijen},
  {Fraternali}, \& {Sancisi}}]{lelli2012a}
{Lelli}, F., {Verheijen}, M., {Fraternali}, F., \& {Sancisi}, R.
  2012{\natexlab{a}}, \aap, 537, A72, \dodoi{10.1051/0004-6361/201117867}

\bibitem[{{Lelli} {et~al.}(2012{\natexlab{b}}){Lelli}, {Verheijen},
  {Fraternali}, \& {Sancisi}}]{lelli2012}
---. 2012{\natexlab{b}}, \aap, 544, A145, \dodoi{10.1051/0004-6361/201219457}

\bibitem[{{Luridiana} {et~al.}(2012){Luridiana}, {Morisset}, \&
  {Shaw}}]{luridiana2012}
{Luridiana}, V., {Morisset}, C., \& {Shaw}, R.~A. 2012, in IAU Symposium, Vol.
  283, Planetary Nebulae: An Eye to the Future, 422--423,
  \dodoi{10.1017/S1743921312011738}

\bibitem[{{Luridiana} {et~al.}(2015){Luridiana}, {Morisset}, \&
  {Shaw}}]{luridiana2015}
{Luridiana}, V., {Morisset}, C., \& {Shaw}, R.~A. 2015, \aap, 573, A42,
  \dodoi{10.1051/0004-6361/201323152}

\bibitem[{{Ly} {et~al.}(2014){Ly}, {Malkan}, {Nagao}, {Kashikawa}, {Shimasaku},
  \& {Hayashi}}]{ly2014}
{Ly}, C., {Malkan}, M.~A., {Nagao}, T., {et~al.} 2014, \apj, 780, 122,
  \dodoi{10.1088/0004-637X/780/2/122}

\bibitem[{{Madore} \& {Freedman}(1995)}]{madore1995}
{Madore}, B.~F., \& {Freedman}, W.~L. 1995, \aj, 109, 1645,
  \dodoi{10.1086/117391}

\bibitem[{{Makarov} {et~al.}(2006){Makarov}, {Makarova}, {Rizzi}, {Tully},
  {Dolphin}, {Sakai}, \& {Shaya}}]{makarov2006}
{Makarov}, D., {Makarova}, L., {Rizzi}, L., {et~al.} 2006, \aj, 132, 2729,
  \dodoi{10.1086/508925}

\bibitem[{{McQuinn} {et~al.}(2018){McQuinn}, {Skillman}, {Heilman}, {Mitchell},
  \& {Kelley}}]{mcquinn2018}
{McQuinn}, K. B.~W., {Skillman}, E.~D., {Heilman}, T.~N., {Mitchell}, N.~P., \&
  {Kelley}, T. 2018, \mnras, 477, 3164, \dodoi{10.1093/mnras/sty839}

\bibitem[{{McQuinn} {et~al.}(2019){McQuinn}, {van Zee}, \&
  {Skillman}}]{mcquinn2019}
{McQuinn}, K. B.~W., {van Zee}, L., \& {Skillman}, E.~D. 2019, \apj, 886, 74,
  \dodoi{10.3847/1538-4357/ab4c37}

\bibitem[{{McQuinn} {et~al.}(2010){McQuinn}, {Skillman}, {Cannon}, {Dalcanton},
  {Dolphin}, {Hidalgo-Rodr{\'\i}guez}, {Holtzman}, {Stark}, {Weisz}, \&
  {Williams}}]{mcquinn2010b}
{McQuinn}, K. B.~W., {Skillman}, E.~D., {Cannon}, J.~M., {et~al.} 2010, \apj,
  724, 49, \dodoi{10.1088/0004-637X/724/1/49}

\bibitem[{{McQuinn} {et~al.}(2014){McQuinn}, {Cannon}, {Dolphin}, {Skillman},
  {Salzer}, {Haynes}, {Adams}, {Cave}, {Elson}, {Giovanelli}, {Ott}, \&
  {Saintonge}}]{mcquinn2014}
{McQuinn}, K. B.~W., {Cannon}, J.~M., {Dolphin}, A.~E., {et~al.} 2014, \apj,
  785, 3, \dodoi{10.1088/0004-637X/785/1/3}

\bibitem[{{McQuinn} {et~al.}(2015{\natexlab{a}}){McQuinn}, {Skillman},
  {Dolphin}, {Cannon}, {Salzer}, {Rhode}, {Adams}, {Berg}, {Giovanelli},
  {Girardi}, \& {Haynes}}]{mcquinn2015}
{McQuinn}, K. B.~W., {Skillman}, E.~D., {Dolphin}, A., {et~al.}
  2015{\natexlab{a}}, \apj, 812, 158, \dodoi{10.1088/0004-637X/812/2/158}

\bibitem[{{McQuinn} {et~al.}(2015{\natexlab{b}}){McQuinn}, {Skillman},
  {Dolphin}, {Cannon}, {Salzer}, {Rhode}, {Adams}, {Berg}, {Giovanelli}, \&
  {Haynes}}]{mcquinn2015Z}
---. 2015{\natexlab{b}}, \apjl, 815, L17, \dodoi{10.1088/2041-8205/815/2/L17}

\bibitem[{{McQuinn} {et~al.}(2020){McQuinn}, {Berg}, {Skillman}, {Adams},
  {Cannon}, {Dolphin}, {Salzer}, {Giovanelli}, {Haynes}, {Hirschauer},
  {Janoweicki}, {Klapkowski}, \& {Rhode}}]{mcquinn2020}
{McQuinn}, K. B.~W., {Berg}, D.~A., {Skillman}, E.~D., {et~al.} 2020, \apj,
  891, 181, \dodoi{10.3847/1538-4357/ab7447}

\bibitem[{{Meynet} \& {Maeder}(2002)}]{meynet2002}
{Meynet}, G., \& {Maeder}, A. 2002, \aap, 390, 561,
  \dodoi{10.1051/0004-6361:20020755}

\bibitem[{{Miller} {et~al.}(2023){Miller}, {Salzer}, {Janowiecki}, {Haynes}, \&
  {Hirschauer}}]{miller2023}
{Miller}, J.~H., {Salzer}, J.~J., {Janowiecki}, S., {Haynes}, M.~P., \&
  {Hirschauer}, A.~S. 2023, \apj, 943, 93, \dodoi{10.3847/1538-4357/aca89b}

\bibitem[{{Morales-Luis} {et~al.}(2011){Morales-Luis}, {S{\'a}nchez Almeida},
  {Aguerri}, \& {Mu{\~n}oz-Tu{\~n}{\'o}n}}]{morales-luis2011}
{Morales-Luis}, A.~B., {S{\'a}nchez Almeida}, J., {Aguerri}, J.~A.~L., \&
  {Mu{\~n}oz-Tu{\~n}{\'o}n}, C. 2011, \apj, 743, 77,
  \dodoi{10.1088/0004-637X/743/1/77}

\bibitem[{{Nishigaki} {et~al.}(2023){Nishigaki}, {Ouchi}, {Nakajima}, {Ono},
  {Rauch}, {Isobe}, {Harikane}, {Narita}, {Zahedy}, {Xu}, {Yajima},
  {Fukushima}, {Hirai}, {Kim}, {Inoue}, {Kusakabe}, {Lee}, {Nagao}, \&
  {Onodera}}]{nishigaki2023}
{Nishigaki}, M., {Ouchi}, M., {Nakajima}, K., {et~al.} 2023, \apj, 952, 11,
  \dodoi{10.3847/1538-4357/accf14}

\bibitem[{{Oesch} {et~al.}(2016){Oesch}, {Brammer}, {van Dokkum},
  {Illingworth}, {Bouwens}, {Labb{\'e}}, {Franx}, {Momcheva}, {Ashby}, {Fazio},
  {Gonzalez}, {Holden}, {Magee}, {Skelton}, {Smit}, {Spitler}, {Trenti}, \&
  {Willner}}]{oesch2016}
{Oesch}, P.~A., {Brammer}, G., {van Dokkum}, P.~G., {et~al.} 2016, \apj, 819,
  129, \dodoi{10.3847/0004-637X/819/2/129}

\bibitem[{{Papaderos} {et~al.}(2008){Papaderos}, {Guseva}, {Izotov}, \&
  {Fricke}}]{papaderos2008}
{Papaderos}, P., {Guseva}, N.~G., {Izotov}, Y.~I., \& {Fricke}, K.~J. 2008,
  \aap, 491, 113, \dodoi{10.1051/0004-6361:200810028}

\bibitem[{{Peeples} \& {Shankar}(2011)}]{peeples2011}
{Peeples}, M.~S., \& {Shankar}, F. 2011, \mnras, 417, 2962,
  \dodoi{10.1111/j.1365-2966.2011.19456.x}

\bibitem[{{Peimbert}(1967)}]{peimbert1967}
{Peimbert}, M. 1967, \apj, 150, 825, \dodoi{10.1086/149385}

\bibitem[{{Persson} {et~al.}(2004){Persson}, {Madore}, {Krzemi{\'n}ski},
  {Freedman}, {Roth}, \& {Murphy}}]{persson2004}
{Persson}, S.~E., {Madore}, B.~F., {Krzemi{\'n}ski}, W., {et~al.} 2004, \aj,
  128, 2239, \dodoi{10.1086/424934}

\bibitem[{{Petropoulou} {et~al.}(2012){Petropoulou}, {V{\'\i}lchez}, \&
  {Iglesias-P{\'a}ramo}}]{petropoulou2012}
{Petropoulou}, V., {V{\'\i}lchez}, J., \& {Iglesias-P{\'a}ramo}, J. 2012, \apj,
  749, 133, \dodoi{10.1088/0004-637X/749/2/133}

\bibitem[{{Pilyugin} {et~al.}(2017){Pilyugin}, {Grebel}, {Zinchenko},
  {Nefedyev}, \& {Mattsson}}]{pilyugin2017}
{Pilyugin}, L.~S., {Grebel}, E.~K., {Zinchenko}, I.~A., {Nefedyev}, Y.~A., \&
  {Mattsson}, L. 2017, \mnras, 465, 1358, \dodoi{10.1093/mnras/stw2831}

\bibitem[{{Pustilnik} {et~al.}(2021){Pustilnik}, {Egorova}, {Kniazev},
  {Perepelitsyna}, {Tepliakova}, {Burenkov}, \& {Oparin}}]{pustilnik2021}
{Pustilnik}, S.~A., {Egorova}, E.~S., {Kniazev}, A.~Y., {et~al.} 2021, \mnras,
  507, 944, \dodoi{10.1093/mnras/stab2084}

\bibitem[{{Pustilnik} {et~al.}(2020){Pustilnik}, {Egorova}, {Perepelitsyna}, \&
  {Kniazev}}]{pustilnik2020a}
{Pustilnik}, S.~A., {Egorova}, E.~S., {Perepelitsyna}, Y.~A., \& {Kniazev},
  A.~Y. 2020, \mnras, 492, 1078, \dodoi{10.1093/mnras/stz3417}

\bibitem[{{Pustilnik} {et~al.}(2024){Pustilnik}, {Kniazev}, {Tepliakova},
  {Perepelitsyna}, \& {Egorova}}]{pustilnik2024}
{Pustilnik}, S.~A., {Kniazev}, A.~Y., {Tepliakova}, A.~L., {Perepelitsyna},
  Y.~A., \& {Egorova}, E.~S. 2024, \mnras, 527, 11066,
  \dodoi{10.1093/mnras/stad3926}

\bibitem[{{Pustilnik} {et~al.}(2013){Pustilnik}, {Martin}, {Lyamina}, \&
  {Kniazev}}]{pustilnik2013}
{Pustilnik}, S.~A., {Martin}, J.~M., {Lyamina}, Y.~A., \& {Kniazev}, A.~Y.
  2013, \mnras, 432, 2224, \dodoi{10.1093/mnras/stt609}

\bibitem[{{Pustilnik} {et~al.}(2022){Pustilnik}, {Perepelitsyna}, {Tepliakova},
  {Kniazev}, {Egorova}, {Chengalur}, \& {Kurapati}}]{pustilnik2022}
{Pustilnik}, S.~A., {Perepelitsyna}, Y., {Tepliakova}, A., {et~al.} 2022, in
  The Multifaceted Universe: Theory and Observations - 2000, 26,
  \dodoi{10.48550/arXiv.2212.05640}

\bibitem[{{Pustilnik} {et~al.}(2016){Pustilnik}, {Perepelitsyna}, \&
  {Kniazev}}]{pustilnik2016}
{Pustilnik}, S.~A., {Perepelitsyna}, Y.~A., \& {Kniazev}, A.~Y. 2016, \mnras,
  463, 670, \dodoi{10.1093/mnras/stw2039}

\bibitem[{{Pustilnik} {et~al.}(2019){Pustilnik}, {Tepliakova}, \&
  {Makarov}}]{pustilnik2019}
{Pustilnik}, S.~A., {Tepliakova}, A.~L., \& {Makarov}, D.~I. 2019, \mnras, 482,
  4329, \dodoi{10.1093/mnras/sty2947}

\bibitem[{{Ramsey} {et~al.}(1998){Ramsey}, {Adams}, {Barnes}, {Booth},
  {Cornell}, {Fowler}, {Gaffney}, {Glaspey}, {Good}, {Hill}, {Kelton},
  {Krabbendam}, {Long}, {MacQueen}, {Ray}, {Ricklefs}, {Sage}, {Sebring},
  {Spiesman}, \& {Steiner}}]{ramsey1998}
{Ramsey}, L.~W., {Adams}, M.~T., {Barnes}, T.~G., {et~al.} 1998, in Society of
  Photo-Optical Instrumentation Engineers (SPIE) Conference Series, Vol. 3352,
  Advanced Technology Optical/IR Telescopes VI, ed. L.~M. {Stepp}, 34--42,
  \dodoi{10.1117/12.319287}

\bibitem[{{Reid} {et~al.}(1987){Reid}, {Mould}, \& {Thompson}}]{reid1987}
{Reid}, N., {Mould}, J., \& {Thompson}, I. 1987, \apj, 323, 433,
  \dodoi{10.1086/165841}

\bibitem[{{Rizzi} {et~al.}(2007){Rizzi}, {Tully}, {Makarov}, {Makarova},
  {Dolphin}, {Sakai}, \& {Shaya}}]{rizzi2007}
{Rizzi}, L., {Tully}, R.~B., {Makarov}, D., {et~al.} 2007, \apj, 661, 815,
  \dodoi{10.1086/516566}

\bibitem[{{Robitaille} \& {Bressert}(2012)}]{robitaille2012}
{Robitaille}, T., \& {Bressert}, E. 2012, {APLpy: Astronomical Plotting Library
  in Python}, Astrophysics Source Code Library, record ascl:1208.017

\bibitem[{{Rots}(1980)}]{rots1980}
{Rots}, A.~H. 1980, \aaps, 41, 189

\bibitem[{{Sabbi} {et~al.}(2018){Sabbi}, {Calzetti}, {Ubeda}, {Adamo},
  {Cignoni}, {Thilker}, {Aloisi}, {Elmegreen}, {Elmegreen}, {Gouliermis},
  {Grebel}, {Messa}, {Smith}, {Tosi}, {Dolphin}, {Andrews}, {Ashworth},
  {Bright}, {Brown}, {Chandar}, {Christian}, {Clayton}, {Cook}, {Dale}, {de
  Mink}, {Dobbs}, {Evans}, {Fumagalli}, {Gallagher}, {Grasha}, {Herrero},
  {Hunter}, {Johnson}, {Kahre}, {Kennicutt}, {Kim}, {Krumholz}, {Lee},
  {Lennon}, {Martin}, {Nair}, {Nota}, {{\"O}stlin}, {Pellerin}, {Prieto},
  {Regan}, {Ryon}, {Sacchi}, {Schaerer}, {Schiminovich}, {Shabani}, {Van Dyk},
  {Walterbos}, {Whitmore}, \& {Wofford}}]{sabbi2018}
{Sabbi}, E., {Calzetti}, D., {Ubeda}, L., {et~al.} 2018, \apjs, 235, 23,
  \dodoi{10.3847/1538-4365/aaa8e5}

\bibitem[{{Sakai} {et~al.}(1996){Sakai}, {Madore}, \& {Freedman}}]{sakai1996}
{Sakai}, S., {Madore}, B.~F., \& {Freedman}, W.~L. 1996, \apj, 461, 713,
  \dodoi{10.1086/177096}

\bibitem[{{Salzer} {et~al.}(2005){Salzer}, {Jangren}, {Gronwall}, {Werk},
  {Chomiuk}, {Caperton}, {Melbourne}, \& {McKinstry}}]{salzer2005}
{Salzer}, J.~J., {Jangren}, A., {Gronwall}, C., {et~al.} 2005, \aj, 130, 2584,
  \dodoi{10.1086/497365}

\bibitem[{{S{\'a}nchez Almeida} {et~al.}(2016){S{\'a}nchez Almeida},
  {P{\'e}rez-Montero}, {Morales-Luis}, {Mu{\~n}oz-Tu{\~n}{\'o}n},
  {Garc{\'\i}a-Benito}, {Nuza}, \& {Kitaura}}]{sanchez2016}
{S{\'a}nchez Almeida}, J., {P{\'e}rez-Montero}, E., {Morales-Luis}, A.~B.,
  {et~al.} 2016, \apj, 819, 110, \dodoi{10.3847/0004-637X/819/2/110}

\bibitem[{{Schlafly} \& {Finkbeiner}(2011)}]{schlafly2011}
{Schlafly}, E.~F., \& {Finkbeiner}, D.~P. 2011, \apj, 737, 103,
  \dodoi{10.1088/0004-637X/737/2/103}

\bibitem[{{Schlegel} {et~al.}(1998){Schlegel}, {Finkbeiner}, \&
  {Davis}}]{schlegal1998}
{Schlegel}, D.~J., {Finkbeiner}, D.~P., \& {Davis}, M. 1998, \apj, 500, 525,
  \dodoi{10.1086/305772}

\bibitem[{{Schneider} {et~al.}(2016){Schneider}, {Hunt}, \&
  {Valiante}}]{schneider2016}
{Schneider}, R., {Hunt}, L., \& {Valiante}, R. 2016, \mnras, 457, 1842,
  \dodoi{10.1093/mnras/stw114}

\bibitem[{{Schootemeijer} {et~al.}(2022){Schootemeijer}, {Lennon}, {Garcia},
  {Langer}, {Hastings}, \& {Sch{\"u}rmann}}]{schootemeijer2022}
{Schootemeijer}, A., {Lennon}, D.~J., {Garcia}, M., {et~al.} 2022, \aap, 667,
  A100, \dodoi{10.1051/0004-6361/202244730}

\bibitem[{{Senchyna} {et~al.}(2024){Senchyna}, {Plat}, {Stark}, {Rudie},
  {Berg}, {Charlot}, {James}, \& {Mingozzi}}]{senchyna2024}
{Senchyna}, P., {Plat}, A., {Stark}, D.~P., {et~al.} 2024, \apj, 966, 92,
  \dodoi{10.3847/1538-4357/ad235e}

\bibitem[{{Senchyna} \& {Stark}(2019)}]{senchyna2019}
{Senchyna}, P., \& {Stark}, D.~P. 2019, \mnras, 484, 1270,
  \dodoi{10.1093/mnras/stz058}

\bibitem[{{Skillman} \& {Kennicutt}(1993)}]{skillman1993}
{Skillman}, E.~D., \& {Kennicutt}, Robert~C., J. 1993, \apj, 411, 655,
  \dodoi{10.1086/172868}

\bibitem[{{Skillman} {et~al.}(1989){Skillman}, {Kennicutt}, \&
  {Hodge}}]{skillman1989}
{Skillman}, E.~D., {Kennicutt}, R.~C., \& {Hodge}, P.~W. 1989, \apj, 347, 875,
  \dodoi{10.1086/168178}

\bibitem[{{Skillman} {et~al.}(2013){Skillman}, {Salzer}, {Berg}, {Pogge},
  {Haurberg}, {Cannon}, {Aver}, {Olive}, {Giovanelli}, {Haynes}, {Adams},
  {McQuinn}, \& {Rhode}}]{skillman2013}
{Skillman}, E.~D., {Salzer}, J.~J., {Berg}, D.~A., {et~al.} 2013, \aj, 146, 3,
  \dodoi{10.1088/0004-6256/146/1/3}

\bibitem[{{Smoker} {et~al.}(2000){Smoker}, {Davies}, {Axon}, \&
  {Hummel}}]{smoker2000}
{Smoker}, J.~V., {Davies}, R.~D., {Axon}, D.~J., \& {Hummel}, E. 2000, \aap,
  361, 19

\bibitem[{{Sorba} \& {Sawicki}(2015)}]{sorba2015}
{Sorba}, R., \& {Sawicki}, M. 2015, \mnras, 452, 235,
  \dodoi{10.1093/mnras/stv1235}

\bibitem[{{Sorba} \& {Sawicki}(2018)}]{sorba2018}
---. 2018, \mnras, 476, 1532, \dodoi{10.1093/mnras/sty186}

\bibitem[{{Springob} {et~al.}(2005){Springob}, {Haynes}, {Giovanelli}, \&
  {Kent}}]{springob2005}
{Springob}, C.~M., {Haynes}, M.~P., {Giovanelli}, R., \& {Kent}, B.~R. 2005,
  \apjs, 160, 149, \dodoi{10.1086/431550}

\bibitem[{{Stiavelli} {et~al.}(2025){Stiavelli}, {Morishita}, {Chiaberge},
  {Leethochawalit}, {Norman}, {Ricotti}, {Roberts-Borsani}, {Treu}, {Vanzella},
  {Wyse}, {Zhang}, \& {Boyett}}]{stiavelli2025}
{Stiavelli}, M., {Morishita}, T., {Chiaberge}, M., {et~al.} 2025, \apj, 981,
  136, \dodoi{10.3847/1538-4357/adb5f3}

\bibitem[{{Strickland} \& {Heckman}(2009)}]{strickland2009}
{Strickland}, D.~K., \& {Heckman}, T.~M. 2009, \apj, 697, 2030,
  \dodoi{10.1088/0004-637X/697/2/2030}

\bibitem[{{STSCI Development Team}(2012)}]{drizzlepac}
{STSCI Development Team}. 2012, {DrizzlePac: HST image software}, Astrophysics
  Source Code Library, record ascl:1212.011.
\newblock \doeprint{1212.011}

\bibitem[{{Telford} {et~al.}(2021){Telford}, {Chisholm}, {McQuinn}, \&
  {Berg}}]{telford2021}
{Telford}, O.~G., {Chisholm}, J., {McQuinn}, K. B.~W., \& {Berg}, D.~A. 2021,
  \apj, 922, 191, \dodoi{10.3847/1538-4357/ac1ce2}

\bibitem[{{Telford} {et~al.}(2025){Telford}, {Sandstrom}, {McQuinn}, {Glover},
  {Tarantino}, {Bolatto}, \& {Rickards Vaught}}]{telford2025}
{Telford}, O.~G., {Sandstrom}, K.~M., {McQuinn}, K. B.~W., {et~al.} 2025, \nat,
  642, 900, \dodoi{10.1038/s41586-025-09115-7}

\bibitem[{{Thuan} {et~al.}(2016){Thuan}, {Goehring}, {Hibbard}, {Izotov}, \&
  {Hunt}}]{thuan2016}
{Thuan}, T.~X., {Goehring}, K.~M., {Hibbard}, J.~E., {Izotov}, Y.~I., \&
  {Hunt}, L.~K. 2016, \mnras, 463, 4268, \dodoi{10.1093/mnras/stw2259}

\bibitem[{{Thuan} {et~al.}(2022){Thuan}, {Guseva}, \& {Izotov}}]{thuan2022}
{Thuan}, T.~X., {Guseva}, N.~G., \& {Izotov}, Y.~I. 2022, \mnras, 516, L81,
  \dodoi{10.1093/mnrasl/slac095}

\bibitem[{{Tonry} {et~al.}(2000){Tonry}, {Blakeslee}, {Ajhar}, \&
  {Dressler}}]{tonry2000}
{Tonry}, J.~L., {Blakeslee}, J.~P., {Ajhar}, E.~A., \& {Dressler}, A. 2000,
  \apj, 530, 625, \dodoi{10.1086/308409}

\bibitem[{{Tremonti} {et~al.}(2004){Tremonti}, {Heckman}, {Kauffmann},
  {Brinchmann}, {Charlot}, {White}, {Seibert}, {Peng}, {Schlegel}, {Uomoto},
  {Fukugita}, \& {Brinkmann}}]{tremonti2004}
{Tremonti}, C.~A., {Heckman}, T.~M., {Kauffmann}, G., {et~al.} 2004, \apj, 613,
  898, \dodoi{10.1086/423264}

\bibitem[{{Trujillo} {et~al.}(2020){Trujillo}, {Chamba}, \&
  {Knapen}}]{trujillo2020}
{Trujillo}, I., {Chamba}, N., \& {Knapen}, J.~H. 2020, \mnras, 493, 87,
  \dodoi{10.1093/mnras/staa236}

\bibitem[{{Tully} {et~al.}(2009){Tully}, {Rizzi}, {Shaya}, {Courtois},
  {Makarov}, \& {Jacobs}}]{tully2009}
{Tully}, R.~B., {Rizzi}, L., {Shaya}, E.~J., {et~al.} 2009, \aj, 138, 323,
  \dodoi{10.1088/0004-6256/138/2/323}

\bibitem[{{Tully} {et~al.}(2013){Tully}, {Courtois}, {Dolphin}, {Fisher},
  {H{\'e}raudeau}, {Jacobs}, {Karachentsev}, {Makarov}, {Makarova},
  {Mitronova}, {Rizzi}, {Shaya}, {Sorce}, \& {Wu}}]{tully2013}
{Tully}, R.~B., {Courtois}, H.~M., {Dolphin}, A.~E., {et~al.} 2013, \aj, 146,
  86, \dodoi{10.1088/0004-6256/146/4/86}

\bibitem[{{Tully} {et~al.}(2023){Tully}, {Kourkchi}, {Courtois}, {Anand},
  {Blakeslee}, {Brout}, {Jaeger}, {Dupuy}, {Guinet}, {Howlett}, {Jensen},
  {Pomar{\`e}de}, {Rizzi}, {Rubin}, {Said}, {Scolnic}, \& {Stahl}}]{tully2023}
{Tully}, R.~B., {Kourkchi}, E., {Courtois}, H.~M., {et~al.} 2023, \apj, 944,
  94, \dodoi{10.3847/1538-4357/ac94d8}

\bibitem[{{van de Rydt} {et~al.}(1991){van de Rydt}, {Demers}, \&
  {Kunkel}}]{vanderydt1991}
{van de Rydt}, F., {Demers}, S., \& {Kunkel}, W.~E. 1991, \aj, 102, 130,
  \dodoi{10.1086/115861}

\bibitem[{{van Driel} {et~al.}(2016){van Driel}, {Butcher}, {Schneider},
  {Lehnert}, {Minchin}, {Blyth}, {Chemin}, {Hallet}, {Joseph}, {Kotze},
  {Kraan-Korteweg}, {Olofsson}, \& {Ramatsoku}}]{vandriel2016}
{van Driel}, W., {Butcher}, Z., {Schneider}, S., {et~al.} 2016, \aap, 595,
  A118, \dodoi{10.1051/0004-6361/201528048}

\bibitem[{{van Zee} \& {Haynes}(2006)}]{vanzee2006}
{van Zee}, L., \& {Haynes}, M.~P. 2006, \apj, 636, 214, \dodoi{10.1086/498017}

\bibitem[{{Vangioni} {et~al.}(2018){Vangioni}, {Dvorkin}, {Olive}, {Dubois},
  {Molaro}, {Petitjean}, {Silk}, \& {Kimm}}]{vangioni2018}
{Vangioni}, E., {Dvorkin}, I., {Olive}, K.~A., {et~al.} 2018, \mnras, 477, 56,
  \dodoi{10.1093/mnras/sty559}

\bibitem[{{Vincenzo} {et~al.}(2016){Vincenzo}, {Belfiore}, {Maiolino},
  {Matteucci}, \& {Ventura}}]{vincenzo2016}
{Vincenzo}, F., {Belfiore}, F., {Maiolino}, R., {Matteucci}, F., \& {Ventura},
  P. 2016, \mnras, 458, 3466, \dodoi{10.1093/mnras/stw532}

\bibitem[{{Vincenzo} \& {Kobayashi}(2018{\natexlab{a}})}]{vincenzo2018a}
{Vincenzo}, F., \& {Kobayashi}, C. 2018{\natexlab{a}}, \aap, 610, L16,
  \dodoi{10.1051/0004-6361/201732395}

\bibitem[{{Vincenzo} \& {Kobayashi}(2018{\natexlab{b}})}]{vincenzo2018}
---. 2018{\natexlab{b}}, \mnras, 478, 155, \dodoi{10.1093/mnras/sty1047}

\bibitem[{{Virtanen} {et~al.}(2020){Virtanen}, {Gommers}, {Oliphant},
  {Haberland}, {Reddy}, {Cournapeau}, {Burovski}, {Peterson}, {Weckesser},
  {Bright}, {van der Walt}, {Brett}, {Wilson}, {Millman}, {Mayorov}, {Nelson},
  {Jones}, {Kern}, {Larson}, {Carey}, {Polat}, {Feng}, {Moore}, {VanderPlas},
  {Laxalde}, {Perktold}, {Cimrman}, {Henriksen}, {Quintero}, {Harris},
  {Archibald}, {Ribeiro}, {Pedregosa}, {van Mulbregt}, \& {SciPy 1. 0
  Contributors}}]{virtanen2020}
{Virtanen}, P., {Gommers}, R., {Oliphant}, T.~E., {et~al.} 2020, Nature
  Methods, 17, 261, \dodoi{10.1038/s41592-019-0686-2}

\bibitem[{{Williams} {et~al.}(2014){Williams}, {Lang}, {Dalcanton}, {Dolphin},
  {Weisz}, {Bell}, {Bianchi}, {Byler}, {Gilbert}, {Girardi}, {Gordon},
  {Gregersen}, {Johnson}, {Kalirai}, {Lauer}, {Monachesi}, {Rosenfield},
  {Seth}, \& {Skillman}}]{williams2014}
{Williams}, B.~F., {Lang}, D., {Dalcanton}, J.~J., {et~al.} 2014, \apjs, 215,
  9, \dodoi{10.1088/0067-0049/215/1/9}

\bibitem[{{Yang} {et~al.}(2017){Yang}, {Malhotra}, {Rhoads}, \&
  {Wang}}]{yang2017}
{Yang}, H., {Malhotra}, S., {Rhoads}, J.~E., \& {Wang}, J. 2017, \apj, 847, 38,
  \dodoi{10.3847/1538-4357/aa8809}

\bibitem[{{Yu} {et~al.}(2022){Yu}, {Ho}, {Wang}, \& {Li}}]{yu2022}
{Yu}, N., {Ho}, L.~C., {Wang}, J., \& {Li}, H. 2022, \apjs, 261, 21,
  \dodoi{10.3847/1538-4365/ac626b}

\end{thebibliography}

\end{document}